\documentclass[12pt, draftclsnofoot, onecolumn]{IEEEtran}
\newtheorem{Theorem}{Theorem}
\newtheorem{Lemma}{Lemma}
\newtheorem{prop}{Proposition}
\newtheorem{Corollary}{Corollary}
\newtheorem{defi}{Definition}

\usepackage{algorithm}
\usepackage{algorithmic}
\usepackage{graphicx}
\usepackage{textcomp}
\usepackage{stfloats}
\usepackage{subfigure}
\usepackage{amssymb,amsmath,bm}
\usepackage{cite}
\usepackage{times}
\usepackage{latexsym}
\usepackage{array}
\usepackage{fancyhdr}
\usepackage{psfrag}
\usepackage{multirow}
\usepackage{xcolor}
\usepackage{color}
\usepackage{makecell}
\usepackage{threeparttable}
\usepackage{hyperref}
\usepackage[OT1]{fontenc}
\allowdisplaybreaks[1]

\begin{document}
\title{Unsourced Random Access in MIMO Quasi-Static Rayleigh Fading Channels: Finite Blocklength and Scaling Law Analyses\\}

\author{\IEEEauthorblockN{Junyuan~Gao,
Yongpeng~Wu, 
Giuseppe~Caire, 
Wei~Yang, 
H.~Vincent~Poor,  
and~Wenjun~Zhang
}
\thanks{This article was presented in part at the 2024 IEEE International Symposium on Information Theory (ISIT) \cite{Gao_ISIT}. (Corresponding author: Yongpeng Wu)}
\thanks{J. Gao, Y. Wu, and W. Zhang are with the Department of Electronic Engineering, Shanghai Jiao Tong University, Minhang 200240, China (e-mail: sunflower0515@alumni.sjtu.edu.cn, yongpeng.wu@sjtu.edu.cn, zhangwenjun@sjtu.edu.cn).}
\thanks{G. Caire is with the Communications and Information Theory Group, Technische Universit{\"a}t Berlin, Berlin 10587, Germany (e-mail: caire@tu-berlin.de).}
\thanks{W. Yang is with Qualcomm Technologies, Inc., San Diego, CA 92121, USA (e-mail: weiyang@qti.qualcomm.com).}
\thanks{H. V. Poor is with the Department of Electrical and Computer Engineering, Princeton University, Princeton, NJ 08544, USA (email: poor@princeton.edu).}
}

\maketitle

\begin{abstract}
  This paper considers the unsourced random access~(URA) problem with a random and unknown number of active users in multiple-input multiple-output~(MIMO) quasi-static Rayleigh fading channels. We derive non-asymptotic achievability bounds on the probability of incorrectly estimating the number of active users, and provide scaling laws on the gap between the estimated and true numbers of active users. We prove that the error probability reaches a plateau as the power~$P$ and blocklength~$n$ increase, whereas it decays exponentially with the number~$L$ of receive antennas and eventually vanishes. Then, we explore the fundamental limits of URA by deriving non-asymptotic achievability bounds and converse bounds (including two single-user converse bounds and one multi-user ensemble converse bound) on the minimum energy-per-bit required by each active user to transmit $J$~bits with blocklength~$n$ under misdetection and false-alarm constraints. Numerical results show that the extra required energy-per-bit due to the uncertainty in the number~${\rm{K}}_a$ of active users decreases as $L$ and $\mathbb{E}[{\rm{K}}_a]$ increase and the error requirement becomes milder. In the non-asymptotic regime, using codewords distributed on a sphere outperforms Gaussian random coding. Existing schemes are shown to exhibit a large gap to our bounds when the number of active users is large, calling for more advanced schemes that perform energy-efficiently in this case. In the asymptotic regime with $n\to\infty$, we establish scaling laws on the minimum required $P$~and~$L$ to reliably support ${\rm{K}}_a$ active users as functions of $n$, which highlight the potential of MIMO in enabling low-cost communication and indicate that it is possible for the minimum required $P$~and~$L$ to remain on the same order when the number of active users increases but stays below a threshold. 
\end{abstract}
\begin{IEEEkeywords}
  Energy efficiency, finite blocklength, MIMO unsourced random access, misdetection and false-alarm probabilities, random user activity.
\end{IEEEkeywords}

\section{Introduction} \label{section1}
\subsection{Background and Related Works}\label{section1_1}
  Driven by widespread Internet of Things~(IoT) applications such as smart traffic and factory automation, ultra-massive machine type communication~(umMTC) is expected to be a primary use case in future wireless networks. A key challenge in umMTC scenarios is to enable efficient and reliable random access for large numbers of devices, among which only a small fraction are active and transmit a relatively small amount of information to the base-station~(BS)~\cite{wuyp}. In traditional grant-based random access protocols, active users wait for the grant of dedicated transmission resources from the BS, leading to prolonged latency and reduced spectral efficiency. Consequently, several grant-free random access schemes have been proposed where active users are allowed to simultaneously access wireless networks without waiting for a grant~\cite{ref:Zhu1,Xxy,Lty}. Given that each user becomes active in an intermittent pattern and accesses the network without a grant, the number of active users is typically random and unknown to the receiver~\cite{noKa}.

  \begin{table*}[!t]
  \scriptsize
    \caption{Comparison of our work and existing results on random access} \label{table:compare}
    \renewcommand{\arraystretch}{1.1}
    \centering
    \begin{threeparttable}          
    \begin{tabular}{|c|c|c|c|c|c|c|c|}
      \hline
      result & codebook & channel model & \makecell[c]{variable-length/\\fixed-length code} & \makecell[c]{the number of\\ active users\tnote{1}} & \makecell[c]{error\\criterion} & \makecell[c]{non-asymptotic\\bound} & asymptotic scaling\tnote{2} 
      \\  \hline
      \cite{On_joint}&individual&\makecell[c]{general P2P channel}&fixed&\makecell[c]{random unknown}&PUPE&\makecell[c]{achievability\\converse}& -- 
      \\  \hline
      \cite{GuoDN} & individual & Gaussian & fixed & \makecell[c]{random unknown} & joint & -- & $K_{a,n} = \mathcal{O}(n)$ 
      \\  \hline
      \cite{Ravi}&individual*\tnote{3}&Gaussian&fixed&\makecell[c]{fixed known/\\random unknown}&both&--& \makecell[c]{ $\mathbb{E}\!\left[{\rm{K}}_{a,n}\right] \ln K_n \in $ \\$\{ o(n) , \Theta(n) , \omega(n) \}$\tnote{4}}
      \\  \hline
      \cite{finite_payloads_fading}&individual&\makecell[c]{single-receive-antenna\\fading}&fixed&\makecell[c]{fixed known}&PUPE*&--&$K_{a,n} = \mathcal{O}(n)$ 
      \\  \hline 
      \cite{TIT}&individual&MIMO fading&fixed&\makecell[c]{fixed known*}&PUPE&\makecell[c]{achievability\\converse}&$K_{a,n} = \mathcal{O}(n^2)$
      \\  \hline
      \cite{A_perspective_on}&common*&Gaussian&fixed&\makecell[c]{fixed known}&PUPE&achievability&$K_{a,n} = \mathcal{O}(n)$
      \\  \hline
      \cite{indivicommon} &common*&\makecell[c]{Gaussian, \\adder-erasure\tnote{5}}&variable&\makecell[c]{fixed unknown}&both&achievability&\makecell[c]{
      $K_{a,n} = \mathcal{O}(1)$} 
      \\  \hline 
      \cite{sphere}& common  & Gaussian & variable &\makecell[c]{fixed known/\\unknown}&joint & achievability & \makecell[c]{ 
       $K_{a,n} = \mathcal{O}(1)$} 
      \\  \hline
      \cite{noKa} &common&Gaussian&fixed&\makecell[c]{random unknown}&PUPE&\makecell[c]{achievability\\converse}&--
      \\  \hline
      \cite{RAC_fading}&common&\makecell[c]{single-receive-antenna\\fading}&fixed&\makecell[c]{fixed known}&PUPE&\makecell[c]{achievability\\converse}&$K_{a,n} = \mathcal{O}(n)$ 
      \\  \hline 
      \cite{letter} &common&MIMO fading&fixed&\makecell[c]{fixed known}&PUPE&\makecell[c]{achievability\\converse}&--
      \\  \hline 
      \cite{Caire1}&common*&MIMO fading&fixed&\makecell[c]{fixed known}&joint&--&  $\!\!K_{a,n}\! = \! \mathcal{O}\!\left(\!\frac{n^2}{\ln^2\!\left(\!\frac{M_n}{n^2}\!\right)}\!\right)\!\!$
      \\  \hline  
      \makecell[c]{this\\paper} &common&MIMO fading&fixed&\makecell[c]{random unknown*}&PUPE*&\makecell[c]{achievability\\converse}& $ \mathbb{E}\!\left[{\rm{K}}_{a,n}\right] \!=\! \mathcal{O} \!\left( \frac{n^2}{ \ln^2 \!n }  \right)$
      \\  \hline
    \end{tabular}
    \begin{tablenotes}    
        \footnotesize               
        \item[1] In this column, we use ``random'' to indicate that the number of active users is a random variable, and use ``fixed'' to indicate that it is deterministic. We use ``known'' and ``unknown'' to signify whether the number of active users is known or not. For the references listed in this table, the distribution of the number of active users is assumed to be known when it is treated as a random~variable. 
        \item[2] {The blocklength $n\to\infty$ in the asymptotic regime. We use $K_{a,n}$ with the subscript ``$n$'' to indicate that the number of active users scales with $n$ as $n\to\infty$; and similarly for other parameters. 
            }
        \item[3] {We use ``*'' to indicate that this regime is mainly considered in the corresponding paper. For example, most of the results in \cite{Ravi} are based on the assumption that each user has an individual codebook, although there are a few discussions related to~URA.}  
        \item[4] {Assuming ${\rm{K}}_{a,n}$ follows a binomial distribution, it was proved in~\cite{Ravi} that interference-free communication is feasible in the case of $\mathbb{E}\!\left[{\rm{K}}_{a,n}\right] \ln K_n = o(n)$ and no positive rate per unit-energy is feasible in the case of $\mathbb{E}\!\left[{\rm{K}}_{a,n}\right] \ln K_n = \omega(n)$.}
        \item[5] {The established results also hold for other channel models satisfying (2), (3), and (16)-(21) in \cite{indivicommon}.}        
    \end{tablenotes}             
    \end{threeparttable}     
  \end{table*}

  A brief comparison between information-theoretic results on random access is summarized in Table~\ref{table:compare}. In a grant-free scenario, a commonly adopted approach is to assign each potential user a unique pilot for user identification. Each active user transmits its pilot followed by a message. The classical information-theoretic treatment of such a scheme assumes each user has an individual codebook. Several attempts to explore fundamental limits of random access with individual codebooks can be found in~\cite{GuoDN,Ravi,finite_payloads_fading,TIT}. However, in scenarios with large numbers of potential users, such as umMTC, designing and allocating a unique pilot or codebook to each user may impose a significant burden on the system. It is also overly wasteful since many users may remain silent for a long period of time. To this end, unsourced random access~(URA) was proposed in~\cite{A_perspective_on}, where the BS only needs to recover a list of transmitted messages up to a permutation, irrespective of user identities. From an information-theoretic perspective, URA is captured by allowing users to share a common codebook, thereby avoiding the burden of designing and allocating a unique codebook for each user. Thus, the URA paradigm has attracted significant attention, and it is of great interest to analyze it from an information-theoretic viewpoint in order to understand fundamental limits of URA and provide guidelines for system~design.

  To draw guidelines for the design of URA architectures, it is relevant to consider finite-blocklength transmission due to stringent latency requirements. Several attempts to account for finite-blocklength effects when characterizing fundamental limits of URA can be found in the literature. Specifically, assuming the number of active users is fixed and known in advance and each active user transmits a fixed amount of bits to the BS over Gaussian channels, Polyanskiy~\cite{A_perspective_on} derived a non-asymptotic upper bound on the per-user probability of error~(PUPE). This bound was improved in~\cite{Improved_Ma}. The authors in~\cite{RAC_fading} further considered single-receive-antenna fading channels, and provided non-asymptotic achievability and converse bounds on the error probability. Numerical results highlighted the existence of a critical user density, below which the optimal URA architecture is able to achieve near-perfect cancellation of multi-user interference~(MUI). This finding was supported by the theoretical analysis in~\cite{Ravi}. Motivated by the fact that equipping the BS with large numbers of antennas provides substantial spectral efficiency while guaranteeing high reliability, the authors in~\cite{letter} derived non-asymptotic bounds on the minimum energy-per-bit required for reliable URA over multiple-input multiple-output~(MIMO) quasi-static Rayleigh fading channels. In the asymptotic regime with blocklength $n\to\infty$ and the number of active users $K_{a,n}$ and codebook size $M_n$ scaling with $n$, it was proved in~\cite{Caire1} that when the BS is equipped with a sufficiently large number of antennas and $M_n = \Theta\left(K_{a,n}^{\alpha}\right)$ for some constant $\alpha > 1$, the error probability vanishes in the case of $K_{a,n} = \mathcal{O} \left( \frac{n^2}{ \ln^2 n }  \right)$.

  The above theoretical results have been established under the assumption of knowing the number of active users in advance. This assumption implies that the receiver has prior knowledge of the sparsity level when treating data detection as a sparse support recovery problem. However, this assumption is overoptimistic in practically relevant random access channels. This is because users typically have intermittent or bursty communication patterns and access the network without a grant, thereby leading to variations in the number of active users over time. To this end, Yavas~et~al.~\cite{indivicommon} explored a URA scenario with a fixed and unknown number of active users, and derived an achievability bound on the error probability applying variable-length codes and feedback. This bound was improved in~\cite{sphere} by using random codewords uniformly distributed on a sphere and a maximum likelihood~(ML) decoder. In the case of $K_a = \mathcal{O}(1)$, the proposed Gaussian random-access code was proved to achieve the same third-order performance as the best known result for Gaussian multiple-access with known user activity~\cite{sphere}. Assuming the number of active users is random and unknown, a random-access code accounting for both misdetection and false-alarm probabilities was defined in~\cite{noKa}. On the basis of this code, achievability and converse bounds on the two probabilities were derived for Gaussian channels, where the converse bound is an ensemble converse bound and is specific to a scheme where the BS first estimates the number of active users and then chooses the number of returned messages in an interval around the estimated value. For point-to-point~(P2P) channels, non-asymptotic bounds on the maximum coding rate were established in~\cite{On_joint}. Moreover, the authors in~\cite{estimate_Ka} analyzed estimation performance of the number of active users in single-receive-antenna fading channels, where active users are required to transmit a common sequence before message. However, allocating additional channel uses for the transmission of a common sequence 
  may result in lower spectral efficiency.

  Despite these existing results, the fundamental limits of URA with a random and unknown number of active users are still not well understood. First, when each active user randomly transmits a codeword from a common codebook over fading channels, the estimation performance of the number of active users remains unclear, and how the lack of knowledge of the sparsity level affects data detection performance remains uncertain. Second, most existing results related to URA with an unknown sparsity level are confined to the setting with a single-antenna receiver. In the case where the BS is equipped with multiple antennas, it is unclear how the minimum required energy-per-bit behaves when each active user transmits a fixed number of bits using finite channel uses. Third, in the asymptotic regime with a random and unknown number of active users, the scaling behavior of the statistics of the number of active users, blocklength, power, and number of BS antennas is unclear. This work aims to address these issues in both non-asymptotic and asymptotic regimes applying information-theoretic tools.

\subsection{Main Contributions}\label{section1_2}

  In this work, we explore the fundamental limits of URA in MIMO quasi-static Rayleigh fading channels with random and unknown numbers of active users. To cope with the uncertainty in the sparsity level, we estimate the number of active users and analyze the estimation performance in both non-asymptotic and asymptotic regimes. Then, we delve into the data detection problem. We derive non-asymptotic achievability and converse bounds to explore the fundamental tradeoff among the payload size $J$, blocklength $n$, transmitting power $P$, number ${\rm K}_a$ of active users, number $L$ of BS antennas, and per-user probabilities of misdetection and false-alarm. The achievability and converse bounds under the criterion of joint error probability are also discussed. The derived finite-blocklength results play a crucial role in revealing the fundamental limits of URA with stringent latency and energy constraints, and provide benchmarks to evaluate practical schemes. In the asymptotic regime, we establish scaling laws to capture the underlying relationship among system parameters. Moreover, we present extensive simulation results and draw insightful conclusions from them. The contributions of this paper are as follows.

  \emph{Estimation of the Number of Active Users:} Consider a URA scenario in MIMO fading channels with a fixed and unknown number $K_a$ of active users. Each of them selects a message from the set $[M]$ and the received signal $\mathbf{Y}$ is used to estimate $K_a$ followed by decoding. We employ an energy-based estimator, which was used in \cite{sphere,noKa} to estimate $K_a$ in Gaussian channels. Define the error probability of estimating $K_a$ exactly as $K'_a \in [K] \backslash \{ K_a \}$ as $\mathbb{P}\left[ K_a \to K'_a \right] \triangleq \mathbb{P}\left[ {\rm K}'_a = K'_a | {\rm K}_a = K_a \right]$. In Theorem~\ref{Theorem_Ka}, we derive a non-asymptotic upper bound on this error probability. The advantage of using codewords uniformly distributed on a sphere over using a Gaussian codebook for the estimation of $K_a$ is revealed. Notably, the derived bound decays exponentially with $L$, but reaches a plateau as $P$ and $n$ grow. This is because the normalized squared Frobenius norm of $\mathbf{Y}$ concentrates around its mean as $L\to\infty$. However, channel coefficients do not concentrate as $P$ and $n$ grow. Thus, the decoder remains uncertain about them. Since different channel coefficients correspond to different $K_a$, the decoder also remains uncertain about $K_a$. Corollaries~\ref{Theorem_Ka_asymP} and \ref{Theorem_Ka_asymn} provide asymptotic bounds on $\mathbb{P} \left[ K_a  \to K'_{a} \right]$ in the cases of $P\to\infty$ and $n\to\infty$, respectively. Simulation results show that the error probability approaches to the asymptotic bounds at moderate values of $P$ and $n$.

  To provide further insights, we establish scaling laws on the gap between the estimated and true numbers of active users in Theorem~\ref{Theorem_scalinglaw_Ka_estimation}. We prove that if $\left| K'_a-K_a \right| = \omega \left( \max \left\{ \frac{K_a}{\sqrt{L}} , \frac{1}{P\sqrt{nL}} \right\} \right)$, then $\mathbb{P} \left[ K_a  \to K'_{a} \right] \to 0$ as $L\to\infty$. Their gap gradually reduces as $L$ increases, but reaches a plateau of $\omega \left( \frac{K_a}{\sqrt{L}} \right)$ as $n,P\to\infty$, indicating the limitation of increasing $P$ and $n$ in improving estimation performance. Moreover, we provide an interval estimation of $K_a$ with a confidence level close to $1$, and present conditions under which $K_a$ can be estimated with vanishing error~probability.

  \emph{Data Detection:} Consider a URA scenario with a random and unknown number ${\rm{K}}_a$ of active users. Theorem~\ref{Theorem_achievability} provides non-asymptotic achievability bounds on misdetection and false-alarm probabilities when each active user transmits $J$ bits with blocklength $n$ and power $P$ to a BS. Corollary~\ref{Theorem_achievability_energyperbit} characterizes how the minimum required energy-per-bit behaves under given error constraints. A challenge in scenarios with unknown ${\rm{K}}_a$ is how to determine the decoded list size. To this end, we employ the two-stage scheme proposed in \cite{noKa}. We first estimate the number of active users and then decode with the list size near the estimated value. A key difference is that we adopt the maximum \emph{a posteriori} probability~(MAP) criterion to decode instead of the ML criterion used in \cite{noKa}. Moreover, we consider MIMO fading channels instead of Gaussian~channels considered in~\cite{noKa}. In this case, we use Fano's bounding technique~\cite{1961} and a properly selected region around the linear combination of transmitted signals~\cite{TIT} to derive the achievability bounds.

  We derive three converse bounds for URA in MIMO quasi-static Rayleigh fading channels. Theorems~\ref{Theorem_converse_single} and \ref{Theorem_converse_single_noKa} are obtained by casting a URA code as a single-user code with Theorem~\ref{Theorem_converse_single_noKa} accounting for the uncertainty in user activity but only applicable when ${\rm K}_a$ follows a binomial distribution. Compared to the converse bound in~\cite{noKa}, our single-user bounds are not limited to the codebook ensemble and decoding scheme. Since single-user bounds can be loose when ${\rm K}_a$ is large, we address the challenge of massive access in Theorem~\ref{Theorem_converse_noCSI_Gaussian_cKa}, where we assume ${\rm K}_a$ is random but known to make the analysis tractable. Theorem~\ref{Theorem_converse_noCSI_Gaussian_cKa} is not limited to the decoding scheme, but is still an ensemble converse bound. These three parts complement each other and collectively characterize the lower bound on the minimum required energy-per-bit.

  We can further draw the following conclusions from simulation results included in this paper. First, in the non-asymptotic regime, the use of codewords distributed on a sphere outperforms Gaussian random coding for URA in MIMO fading channels, which is in line with the conclusions for P2P channels~\cite{Shannon} and Gaussian multiple access channels~\cite{sphere}. Second, the absence of knowledge regarding the number of active users incurs a minor penalty in energy efficiency when $L$ and $\mathbb{E}\left[ {\rm K}_a \right]$ are large and error constraints are mild. However, as $L$ and $\mathbb{E}\left[ {\rm K}_a \right]$ decrease and error requirements become stricter, the extra required energy-per-bit becomes more significant. Third, it is shown that large antenna arrays have substantial potential in improving spectral efficiency and enabling low-cost and highly reliable communication. Finally, the practical schemes proposed in~\cite{Fasura,Caire2,Duman2} exhibit a large gap to our theoretical bounds when the number of active users is large, calling for more advanced schemes that perform energy-efficiently in this case.

  Consider an asymptotic regime where the blocklength $n\to\infty$ and the number $L_n$ of BS antennas, number of active users, codebook size $M_n$, and power $P_n$ scale with $n$. We establish scaling laws between these parameters. First, we assume $K_{a,n}$ is fixed and known. In a sparse regime with $M_n$ much larger than $K_{a,n}$, it was proved in~\cite{Caire1} that the error probability vanishes if $K_{a,n} = \mathcal{O} \left( \frac{n^2}{ \ln^2 n }  \right)$. We prove that in this regime, the minimum required $P_n$ and $L_n$ satisfy $P_n^2L_n = \Theta\left( \frac{\ln n}{n^2} \right)$ under mild conditions given in Theorem~\ref{Theorem_scalinglaw_noCSI_Kaknown}. This implies that as $L_n$ increases, the minimum required $P_n$ decreases. Moreover, it is possible for the minimum required $P_n$ and $L_n$ to remain on the same order when $K_{a,n}$ increases from $\Theta(1)$ to $ \Theta \left( \frac{n^2}{ \ln^2 n }  \right)$. Then, we consider URA scenarios with random and unknown ${\rm K}_{a,n}$. Theorem~\ref{Theorem_scalinglaw_noCSI} shows that under some conditions on statistical properties of ${\rm K}_{a,n}$, one can reliably support ${\rm K}_{a,n}$ active user with mean $\mathbb{E}[{\rm K}_{a,n}] = \mathcal{O} \left( \frac{n^2}{ \ln^2 n }  \right)$ in the same regime of $P_n$ and $L_n$ as in the case with fixed and known~$K_{a,n}$.

  The organization of this paper is as follows. Our system model is introduced in Section~\ref{section2}. The non-asymptotic and asymptotic theoretical results on the estimation performance of the number of active users are provided in Section~\ref{Section:Ka}. Section~\ref{Section:Data} conducts information-theoretic analysis for data detection in both non-asymptotic and asymptotic regimes under the criterion of per-user probabilities of misdetection and false-alarm. In Section~\ref{Section:joint_error}, we briefly introduce the results under the constraint on joint error probability. In Section~\ref{Section:simulation}, we provide numerical results. Section~\ref{Section:conclusion} concludes the paper. Most proofs appear in appendices.

  \emph{Notation:} Throughout this paper, uppercase and lowercase boldface letters denote matrices and vectors, respectively. The notation $\left[\mathbf{x} \right]_{m}$ denotes the $m$-th element of $\mathbf{x}$ and $\left[\mathbf{X} \right]_{m,n}$ denotes the $\left( m,n \right)$-th element of $\mathbf{X}$. We use $\left(\cdot \right)^{T}$, $\left(\cdot \right)^{H}$, $\left|\mathbf{X}\right|$, $\left\|\mathbf{x} \right\|_{p}$, and $\left\|\mathbf{X} \right\|_{F}$ to denote transpose, conjugate transpose, determinant, ${\ell}_p$-norm, and Frobenius norm, respectively. We use $\operatorname{diag} \left\{ \mathbf{x} \right\}$ to denote a diagonal matrix with $\mathbf{x}$ comprising its diagonal elements and use $\operatorname{diag} \left\{ \mathbf{A}, \mathbf{B} \right\}$ to denote a block diagonal matrix. Given any complex variable, vector, or matrix, we use $\Re(\cdot)$ and $\Im(\cdot)$ to return its real and imaginary parts, respectively. We use $\cdot\backslash\cdot$ to denote set subtraction and $\left| \mathcal{A} \right|$ to denote the cardinality of a set $\mathcal{A}$. We use $k!$ to denote the factorial function. Let $x^{+} = \max\{x, 0\}$. For an integer $k > 0$, let $[k] = \left\{1,\ldots,k \right\}$; and for integers $k_2 \geq k_1\geq 0$, let $[k_1:k_2] = \left\{k_1,\ldots,k_2\right\}$. We use $1[\cdot]$ to denote the indicator function. For $0 \leq  p \leq 1$, we denote $h_2(p) = -p\log_2(p)-(1-p)\log_2(1-p)$ with $0\log_2 0$ defined to be $0$. We use $\mathcal{CN}(\cdot ,\cdot)$ and $\chi^2(\cdot)$ to denote the circularly symmetric complex Gaussian distribution and central chi-square distribution, respectively. We use $\gamma\left(x, y\right) = \int_{0}^{y} z^{x-1} e^{-z} dz $, $\Gamma\left(x, y\right) = \int_{y}^{\infty} z^{x-1} e^{-z} dz $, and $\Gamma\left(x\right) = \int_{0}^{\infty} z^{x-1} e^{-z} dz$ to denote the lower incomplete gamma function, upper incomplete gamma function, and gamma function, respectively. The complement of event $\mathcal{G}$ is denoted as $\mathcal{G}^c$. We use $\mathbb{P}[\cdot]$ and $\mathbb{E}[\cdot]$ to indicate the probability and expectation under the measure where the codewords satisfy a maximal power constraint and messages from distinct users may collide. After the change of measure, their counterparts are represented as $\mathbb{P}[\cdot]_{\rm new}$ and $\mathbb{E}[\cdot]_{\rm new}$. Working with the new measure simplifies the analysis. Let $f(x)$ and $g(x)$ be positive. The notation $f(x) = o \left( g(x)\right)$ means $\lim_{x\to\infty} f(x)/g(x) = 0$, $f(x) = \mathcal{O}  \left( g(x)\right)$ means $\limsup_{x\to\infty} f(x)/g(x) < \infty$, $f(x) = \Theta  \left( g(x)\right)$ means $f(x) = \mathcal{O}  \left( g(x)\right)$ and $g(x) = \mathcal{O}   \left( f(x)\right)$, $f(x) = \omega  \left( g(x)\right)$ means $g(x) = o  \left( f(x)\right)$, and $f(x) = \Omega  \left( g(x)\right)$ means $g(x) = \mathcal{O}  \left( f(x)\right)$.


\section{System Model} \label{section2}

  We consider the uplink of a wireless network with a total of $K$ potential single-antenna users, among which ${\rm{K}}_a$ users are active. Assume that ${\rm{K}}_a$ is random and unknown to the receiver, but its distribution is known in advance~\cite{noKa,GuoDN,Ravi,Yuwei_active}.\footnote{A discussion on how to extend our results based on the assumption of knowing the distribution of ${\rm{K}}_a$ to the scenario with imperfect knowledge of this distribution will be provided in Section~\ref{Section:Non-Asymptotic-Results-achievability}.} The probability of having exactly $K_a$ active users is denoted as $P_{{\rm{K}}_a}(K_a)$.\footnote{We use the non-italic symbol ${\rm K}_a$ to denote the random number of active users, and use the italic symbol $K_a$ to denote the deterministic counterpart. This specific notation differentiation is not extended to other quantities.} In a special case of $P_{{\rm{K}}_a}(K_a) = 1, \exists K_a \in \{0,1,\ldots,K\}$, the number of active users becomes fixed and known to the receiver. Both the setting with a random and unknown number of active users and the special case with a fixed and known number of active users are considered in this work. The set of potential users and that of active users are denoted as $\mathcal{K}$ and $\mathcal{K}_a$, respectively. All active users share $n$ complex channel uses, and each of them transmits $J = \log_2 M \geq 1$ bits to the BS. We assume that all users share a common codebook. Let the matrix $\mathbf{X} = \left[\mathbf{x}_{1},\ldots, \mathbf{x}_{M}\right] \in \mathbb{C}^{n\times M}$ denote the concatenation of all codewords.

  We consider a MIMO quasi-static Rayleigh fading channel model, where the BS is equipped with $L$ antennas and channel fading coefficients remain fixed during the transmission of an entire codeword. We assume neither the BS nor users know the instantaneous channel state information~(CSI), i.e. the realization of the channel fading coefficient, but they both know its distribution in advance. The received signal of the $l$-th antenna at the BS is given by
  \begin{equation} \label{eq_yl}
    \mathbf{y}_l = \sum_{k\in{\mathcal{K}}_a}{h}_{k,l} \mathbf{x}_{W_k} +\mathbf{z}_l \in \mathbb{C}^{n},
  \end{equation}
  where the fading coefficients $\{ {h}_{k,l}: k\in\mathcal{K}_a, l\in[L]\}$ are independent and identically distributed (i.i.d.) according to $\mathcal{CN}(0,1)$; the additive noises $\{ \mathbf{z}_l: l\in[L]\}$ are i.i.d. according to $\mathcal{CN}(\mathbf{0}, \mathbf{I}_n)$; $W_{k}$ denotes the message transmitted by active user~$k$, chosen uniformly at random from the set $[M]$; and $\mathbf{x}_{W_k}$ denotes the corresponding codeword transmitted by active user~$k$. The received signal $\mathbf{Y} = \left[ \mathbf{y}_1, \ldots, \mathbf{y}_L \right] $ among all antennas is given by
  \begin{equation} \label{eq_y}
    \mathbf{Y} =\mathbf{X}\boldsymbol{\Phi}\mathbf{H}+\mathbf{Z} \in \mathbb{C}^{n\times L}  ,
  \end{equation}
  where the matrix $\mathbf{H} = \left[\mathbf{h}_1, \ldots, \mathbf{h}_L \right] \in \mathbb{C}^{K\times L}$ with $\mathbf{h}_l = \left[ {h}_{1,l}, \ldots,{h}_{K,l} \right]^T$; $\mathbf{Z} = \left[\mathbf{z}_1, \ldots, \mathbf{z}_L \right] \in \mathbb{C}^{n\times L}$; and the matrix $\boldsymbol{\Phi} \in \{0,1\}^{M \times K}$  contains at most one ``1'' in each column since each user transmits one message if it is active or zero messages if it is inactive. The distributions of $\boldsymbol{\Phi}$ and $\mathbf{H}$ are known in advance.

  Based on the per-user probabilities of misdetection and false-alarm given in~\cite{Caire1} and the definition of a random-access code for the Gaussian multiple access channel in~\cite{noKa}, we provide the notion of a URA code for the scenario with a random and unknown number ${\rm K}_a$ of active users as follows:
  \begin{defi}\label{defi1}
    An $(n, M, \epsilon_{\mathrm{MD}}, \epsilon_{\mathrm{FA}}, P)$ URA code consists of
    \begin{enumerate}
      \item A random variable $\mathsf{U}$ defined on a set $\mathcal{U}$ that is revealed to both the transmitters and the receiver before the start of the transmission.\footnote{To derive the achievability bound in this work, we construct a codebook ensemble for which both \eqref{eq:MD} and \eqref{eq:FA} hold on average. However, this does not imply that there exists a single code in this ensemble that achieves both \eqref{eq:MD} and \eqref{eq:FA} at the same time, as explained in~\cite{noKa,Yury_feedback,indivicommon}. By introducing the random variable $\mathsf{U}$, this issue can be addressed using a randomized coding strategy which involves time-sharing among at most three deterministic codes (i.e., $|\mathcal{U}| \leq 3$) in this ensemble~\cite[Theorem 19]{Yury_feedback}. As in \cite{noKa}, we omit $\mathsf{U}$ in the encoding and decoding functions in the following for brevity. For further information on how this common randomness is utilized in the encoding and decoding functions, please refer to \cite{Yury_feedback,indivicommon}.} 
      \item 
          An encoder $\emph{f}_{\text{en}}: \mathcal{U} \times [M] \mapsto \mathbb{C}^{n}$ that maps the random variable $\mathsf{U}$ and the message $W_k$ to a codeword $\mathbf{x}_{W_k}$. The codewords in the codebook satisfy the maximal power constraint
          \begin{equation}\label{eq:power_constraint}
            \left\|\mathbf{x}_{m}\right\|_{2}^{2} \leq nP,\;\;\;\; \forall m\in[M].
          \end{equation}
      \item
          A decoder $\emph{g}_{\text{de}}: \mathcal{U} \times \mathbb{C}^{n\times L} \mapsto \binom{[M]}{|\hat{\mathcal{W}}|}$ that satisfies the constraints on the per-user probability of misdetection and the per-user probability of false-alarm as follows: 
          \begin{equation}\label{eq:MD}
            P_{\mathrm{MD}} = \mathbb{E}  \left[ 1 \left[ \mathrm{K}_a \geq 1 \right]   \cdot \frac{1}{\mathrm{K}_a}   \sum_{k\in {\mathcal{K}_a}}  1 \left[ W_{k} \notin \hat{\mathcal{W}}  \right]\right] \leq \epsilon_{\mathrm{MD}},
          \end{equation} 
          \begin{equation}\label{eq:FA}
            P_{\mathrm{FA}} = \mathbb{E}   \left[ 1 \left[ |\hat{\mathcal{W}}| \geq 1 \right]
            \cdot  \frac{1}{ |\hat{\mathcal{W}}| } \sum_{ \hat{w} \in \hat{\mathcal{W}} } 1 \left[ \hat{w} \notin \mathcal{W} \right]\right] \leq \epsilon_{\mathrm{FA}},
          \end{equation} 
          where $\mathcal{W} $ denotes the set of transmitted messages, $\hat{\mathcal{W}}$ denotes the set of decoded messages of size $ | \hat{\mathcal{W}}  | \leq K$, and the expectations are with respect to the random user activity, channel, and possibly random encoding and decoding functions. 
    \end{enumerate}
  \end{defi}
   
  Let $E_b = \frac{nP}{J}$ denote the energy-per-bit. The minimum energy-per-bit required for reliable URA is denoted as $E^{*}_{b}(n, M, \epsilon_{\mathrm{MD}}, \epsilon_{\mathrm{FA}}) = \inf \left\{E_b: \exists (n, M, \epsilon_{\mathrm{MD}}, \epsilon_{\mathrm{FA}}, P) \text{ code} \; \right\}$.

\section{Estimation of the Number of Active Users}\label{Section:Ka}

  To cope with the uncertainty in the number of active users, a feasible solution is to first estimate the number of active users based on the received signal $\mathbf{Y}$ given in~\eqref{eq_y} and then decode with the list size chosen from a predetermined interval around the estimated number of active users. In this section, we analyze the estimation performance of the number of active users in both non-asymptotic and asymptotic regimes. 
 
\subsection{Non-Asymptotic Bound}\label{Section:Ka1}

  Assume that there are ${\rm K}_a = K_a$ active users where $K_a$ is fixed and unknown in advance. As introduced in Section~\ref{section2}, the transmitted messages of active users are sampled uniformly at random from $[M]$. Based on the received signal $\mathbf{Y}$ given in~\eqref{eq_y}, the receiver outputs an estimate of the number $K_a$ of active users. The error probability of estimating $K_a$ exactly as $K'_a \in [K] \backslash \{ K_a \}$ is defined as
  \begin{equation} \label{eq:def_PKaKaprime}
      \mathbb{P}\left[ K_a \to K'_a \right] \triangleq \mathbb{P}\left[ {\rm K}'_a = K'_a | {\rm K}_a = K_a \right], 
  \end{equation} 
  where ${\rm K}'_a$ is the random variable denoting the estimated number of active users. This arrow notation is standard and commonly used in pairwise error probability. In Theorem~\ref{Theorem_Ka}, we provide a non-asymptotic upper bound on this error probability. To obtain Theorem~\ref{Theorem_Ka}, we construct a new codebook $\mathbf{C} = \left[ \mathbf{c}_1, \ldots, \mathbf{c}_M \right] \in \mathbb{C}^{n\times M}$ satisfying $\mathbb{E} [ \left\| \mathbf{c}_m \right\|_2^2  ] = nP'$ with $P'<P$ for $m \in [M]$. We employ an energy-based detector to estimate $K_a$, which was used in \cite{sphere,noKa} for Gaussian channels. This detector is effective in the considered MIMO quasi-static Rayleigh fading channels since the normalized squared Frobenius norm of $\mathbf{Y}$ concentrates around its mean when both $L$ and $n$ become large following from the law of large numbers. In a special case of using spherically distributed codewords, only $L$ is required to be large as will be detailed later in this section. The non-asymptotic bound in Theorem~\ref{Theorem_Ka} serves as the basis for the analysis of misdetection and false-alarm probabilities for the scenario with a random and unknown number of active users as given in Section~\ref{Section:Data}. 
\begin{Theorem} \label{Theorem_Ka} 
  Assume that there are exactly $K_a$ active users among $K$ potential users and the value of $K_a$ is fixed but unknown in advance. 
  The error probability of estimating $K_a$ exactly as $K'_a \in [K] \backslash \{ K_a \}$ is bounded as 
    \begin{equation}\label{eq_noCSI_noKa_pKa_Kahat}
      \mathbb{P} \left[ K_a  \to K'_{a} \right]
      \leq p_{K_a\to K'_a,{\rm achi}}
      = \min_{0<P'\leq P}  \left\{  p_{K_a, K'_a, P' }  + p_{0,K_a} + \frac{\binom{K_a}{2}}{M} \right\} ,
    \end{equation}
    where
    \begin{equation}\label{eq_noCSI_noKa_pKa_Kahat_p0}
       p_{0,K_a} = K_a \mathbb{P}\left[ \left\|\mathbf{c}_m \right\|_2^2 > nP \right],
    \end{equation}
    \begin{equation}\label{eq_noCSI_noKa_pKa_Kahat0}
       p_{K_a, K'_a, P'}
       =  \min 
       \left\{ p_{K_a, K'_a, P',1}  ,  p_{K_a, K'_a, P',2}  \right\} ,
    \end{equation}
    \begin{equation}
      p_{K_a, K'_a, P',1}
      = \mathbb{E} \left[ \min_{\rho\geq0} \exp \left\{  \rho nL \left( 1+   \left( K'_a  +  \frac{1}{2}  \right)  P^{\prime} \right)
      -L \ln  \left|  \mathbf{I}_n + \rho\mathbf{F} \right| \right\} \right] , \label{eq_noCSI_noKa_pKa_Kahat1}
    \end{equation}
    \begin{equation}
      p_{K_a, K'_a, P',2}
      = \mathbb{E} \left[
      \min_{ 0\leq \rho < \frac{1}{1 + \lambda'_{1}} }
      \exp \left\{ -\rho nL \left( 1 + \left( K'_a  -  \frac{1}{2}  \right)  P^{\prime} \right)
      - L \ln  \left|  \mathbf{I}_n - \rho\mathbf{F} \right|  \right\} \right] ,  \label{eq_noCSI_noKa_pKa_Kahat2}
    \end{equation} 
      \begin{equation}\label{eq_noCSI_noKa_pKa_Kahat_F}
        \mathbf{F} = \mathbf{I}_n + \mathbf{C}_{K_a} \mathbf{C}^H_{K_a} .
      \end{equation}
    Here, we assume without loss of generality (w.l.o.g.) that active users transmit the first $K_a$ codewords, i.e. $\mathbf{C}_{K_a} = \left[ \mathbf{c}_1, \ldots, \mathbf{c}_{K_a} \right] \in \mathbb{C}^{n\times K_a}$; $\lambda'_1$ denotes the maximum eigenvalue of $\mathbf{C}_{K_a} \mathbf{C}^H_{K_a}$; and the expectations in~\eqref{eq_noCSI_noKa_pKa_Kahat1} and \eqref{eq_noCSI_noKa_pKa_Kahat2} are with respect to $\mathbf{C}_{K_a}$. 
    \begin{IEEEproof}[Proof sketch] 
      We construct a codebook $\mathbf{C} = \left[ \mathbf{c}_1, \ldots, \mathbf{c}_M \right] \in \mathbb{C}^{n\times M}$ satisfying $\mathbb{E} [ \left\| \mathbf{c}_m \right\|_2^2  ] = nP'$ for $m \in [M]$. If user $k$ is active, it transmits $\mathbf{x}_{W_k} = \mathbf{c}_{W_k} 1 \left[ \left\|\mathbf{c}_{W_k}\right\|_{2}^{2} \leq n P \right]$ with the transmitted message $W_k \in [M]$ chosen uniformly at random. 
      Under the measure $\mathbb{P}[\cdot]$ in \eqref{eq:def_PKaKaprime}, the codeword associated with message $W_k$ is $\mathbf{x}_{W_k}$, which satisfies the power constraint in \eqref{eq:power_constraint}. We change the measure $\mathbb{P}[\cdot]$ to a new measure $\mathbb{P}[\cdot]_{\rm new}$ under which the codeword associated with $W_k$ is $\mathbf{c}_{W_k}$, and there is no message collision between users, i.e., $W_i \neq W_j$ for $i \neq j$. This change of measure step adds an error probability term that is bounded by $p_{0,K_a}+ \frac{\binom{K_a}{2}}{M}$ accounting the total variation distance between the two measures. An energy-based detector is used to estimate the number $K_a$ of active users. Specifically, $K_a$ is estimated as 
      \begin{equation}\label{eq:proof_noCSI_noKa_Kaestimate}
        K'_{a} = \arg \min_{\tilde{K}_a \in\left[0: K\right] }\; m(\mathbf{Y}, \tilde{K}_a),
      \end{equation}
      where the energy-based metric $m(\mathbf{Y}, \tilde{K}_a)$ is given by
      \begin{equation}\label{eq:proof_noCSI_noKa_Kaestimate_m}
        m(\mathbf{Y}, \tilde{K}_a) =\left| \frac{1}{nL}\left\| \mathbf{Y} \right\|_{F}^{2} -  \left( 1 + \tilde{K}_a P^{\prime} \right) \right|. 
      \end{equation} 
      Then, the probability of estimating $K_a$ exactly as $K'_a \in [K] \backslash \{ K_a \}$ is upper-bounded as
    \begin{equation} \label{eq:proof_noCSI_noKa_Ka_Kahat_p0}
      \mathbb{P} \left[ K_a  \to K'_{a} \right]
      \leq \min_{0<P'\leq P}    \left\{ \mathbb{P} \left[ K_a  \to K'_{a} \right]_{\text {new}} + p_{0,K_a} + \frac{\binom{K_a}{2}}{M} \right\} .
    \end{equation}
      Here, the probability $\mathbb{P}  \left[ K_a   \to  K'_{a} \right]_{\text {new}}$ of estimating $K_a$ exactly as $K'_a \in [K] \backslash \{ K_a \}$ under the new measure is upper-bounded by $p_{K_a, K'_a, P'}$ in~\eqref{eq_noCSI_noKa_pKa_Kahat0}, and the minimization is performed over $P'\in (0,P]$ considering that a smaller $P'$ is beneficial for reducing the variation distance $p_{0,K_a}$ but at the cost of increasing the estimation error probability $\mathbb{P}  \left[ K_a   \to  K'_{a} \right]_{\text {new}}$. A detailed proof is given in Appendix~\ref{Appendix_proof_Ka}. 
    \end{IEEEproof}
\end{Theorem}

  Theorem \ref{Theorem_Ka} can be adapted to both cases of applying i.i.d. Gaussian codewords and using codewords uniformly distributed on a sphere, as shown in the following corollary. 
  \begin{Corollary} \label{Theorem_Ka_C}
  In the case of using codewords drawn i.i.d. according to a Gaussian distribution, i.e. assuming that the columns of $\mathbf{C}$ satisfy $\mathbf{c}_{m} \stackrel{\mathrm{i.i.d.}}{\sim} \mathcal{CN} \left(0,P'\mathbf{I}_{n}\right)$ for $m \in [M]$ with $P'<P$, the error probability of estimating $K_a$ exactly as $K'_a \in [K] \backslash \{ K_a \}$ is bounded as in~\eqref{eq_noCSI_noKa_pKa_Kahat} with $p_{0,K_a}$ given by  
    \begin{equation}\label{eq_noCSI_noKa_pKa_Kahat_p0_Gaussian}
       p_{0,K_a} = \frac{K_a\Gamma \left(n, \frac{nP}{P'}\right)}{\Gamma\left(n\right)}. 
    \end{equation} 
  In the case of using codewords drawn uniformly i.i.d. from a sphere of radius $\sqrt{nP'}$, the error probability of estimating $K_a$ exactly as $K'_a \in [K] \backslash \{ K_a \}$ is bounded as in~\eqref{eq_noCSI_noKa_pKa_Kahat} by allowing $P'=P$ and $p_{0,K_a} = 0$. 
\end{Corollary}

  To obtain Corollary~\ref{Theorem_Ka_C}, we construct a common codebook $\mathbf{C}$ with codewords i.i.d. $\mathcal{CN} \left(0,P'\mathbf{I}_{n}\right)$ distributed or drawn uniformly from a sphere of radius $\sqrt{nP'}$. The power constraint violation probability is bounded by $p_{0,K_a}$ as introduced in Corollary~\ref{Theorem_Ka_C}. When the Gaussian random coding scheme is adopted, $P'$ needs to be smaller than $P$ to keep $p_{0,K_a}$ small. However, when using codewords uniformly distributed on a sphere, $P'$ can be as large as $P$ and we have $p_{0,K_a} = 0$. Thus, in the non-asymptotic regime, utilizing codewords distributed on a sphere is more advantageous since $P'$ can be larger, which is in line with the conclusions for P2P channels~\cite{Shannon} and Gaussian multiple access channels~\cite{sphere}.

  Moreover, the energy-based detector has a good performance when the normalized squared Frobenius norm of $\mathbf{Y}$ concentrates around its mean. When utilizing a Gaussian codebook, both $n$ and $L$ are required to be large to ensure a good concentration property. In contrast, when the spherically distributed codewords are adopted, only $L$ is required to be large as shown in Theorem~\ref{Theorem_Ka_asymL_concentration}. 
  \begin{Theorem} \label{Theorem_Ka_asymL_concentration}
  Assume that codewords are drawn uniformly i.i.d. from a sphere of radius $\sqrt{nP}$.
  For any constant $\delta>0$, the two-sided tail probability is bounded as
    \begin{equation} \label{eq_noCSI_noKa_pKa_Kahat_asym}
       \mathbb{P}\left[ \left| \frac{1}{nL} \left\|\mathbf{Y}\right\|_F^2 - \left( 1 + K_a P \right) \right| \geq \delta \right]
       \leq 2 \exp\!\left\{ -L  \min\!\left\{ \frac{  \left( n \delta \right)^2}{ 8( n + (nK_aP)^2 ) },  \frac{  n \delta  }{ 4(1+nK_aP) }  \right\}  \right\} . 
    \end{equation}
  \begin{IEEEproof}
    See Appendix~\ref{Appendix_proof_Ka_asymL_concentration}.
  \end{IEEEproof}
\end{Theorem} 

  Theorem~\ref{Theorem_Ka_asymL_concentration} shows that the two-sided tail probability in \eqref{eq_noCSI_noKa_pKa_Kahat_asym} decreases exponentially with $L$. Thus, $\frac{1}{nL} \left\|\mathbf{Y}\right\|_F^2$ concentrates around its mean as the number $L$ of receive antennas becomes large, even with a small blocklength $n$. This indicates that the spherically distributed codebook has better concentration property than the Gaussian codebook.

  Both the energy-based and ML-based detectors were used in~\cite{noKa} to estimate the number of active users. In this work, we only utilize the energy-based detector. This is because for Gaussian channels considered in~\cite{noKa}, the likelihood function $p(\mathbf{Y} | K_a)$ is a Gaussian probability density function, whereas for fading channels, the likelihood function is nontrivial due to the product of the random channel matrix and the codewords. Even when applying the energy-based detector, the coupling of the random channel matrix and codewords significantly increases the difficulty of bounding the error probability compared to Gaussian channels. Consequently, we resort to some techniques including the Chernoff bound and moment generating function of quadratic forms to obtain Theorem~\ref{Theorem_Ka}.

  Compared with~\cite{estimate_Ka}, where active users transmit a common training sequence before data for the estimation of $K_a$, we assume each active user independently selects a message from the set $[M]$ and the received signal $\mathbf{Y}$ is used to estimate $K_a$ followed by outputting a list of transmitted messages based on the estimates. In this way, there is no need for active users to sacrifice extra channel uses to transmit a common sequence, thereby achieving higher spectral efficiency.

\subsection{Asymptotic Analysis}\label{Section:Ka2}
  
  Applying similar ideas to those in the proofs of Theorem~\ref{Theorem_Ka} and Theorem~\ref{Theorem_Ka_asymL_concentration}, we can obtain that when the spherically distributed codewords are adopted, the derived upper bound on the error probability $\mathbb{P} \left[ K_a  \to K'_{a} \right]$ in~\eqref{eq_noCSI_noKa_pKa_Kahat} decays exponentially with the number $L$ of BS antennas, and eventually converges to $0$. However, for fixed $L$, this upper bound decreases initially but eventually reaches a plateau as the transmitting power $P$ and blocklength $n$ increase. This is because increasing $P$ only alleviates the impact of noise, rather than the randomness coming from the codebook and channel coefficients, resulting that the normalized squared Frobenius norm of $\mathbf{Y}$ cannot concentrate around its mean with sufficiently high probability. Moreover, increasing the blocklength $n$ mitigates the randomness of noise and codebook, but cannot completely eliminate the influence of random fluctuations in fading coefficients over quasi-static fading channels. That is, channel coefficients do not concentrate as $P$ and $n$ grow and different realizations of channel fading coefficients can lead to different estimates of the number of active users. Thus, neither increasing the blocklength nor boosting the transmitting power can bring $p_{K_a\to K'_a,{\rm achi}}$ in~\eqref{eq_noCSI_noKa_pKa_Kahat} down to $0$. From a communication perspective, increasing $P$ is ineffective in reducing MUI, and the channels of some users may remain in outage even if $n\to\infty$, hindering a consistent decrease in the error probability as $n,P\to\infty$. The following Corollary~\ref{Theorem_Ka_asymP} and Corollary~\ref{Theorem_Ka_asymn} show the limit of the upper bound on $\mathbb{P} \left[ K_a  \to K'_{a} \right]$ given in the right-hand side~(RHS) of~\eqref{eq_noCSI_noKa_pKa_Kahat} assuming $P\to\infty$ and $n\to\infty$, respectively.   
  \begin{Corollary} \label{Theorem_Ka_asymP}
  For fixed $n$ and $L$, in the case of $P',P\to\infty$, the error probability of estimating $K_a$ exactly as $K'_a \in [K] \backslash \{ K_a \}$ is bounded as
    \begin{equation}  
       \lim_{P\to\infty} \mathbb{P} \left[ K_a  \to K'_{a} \right]  
       \leq \min
       \left\{  p_{K_a, K'_{a}, 1} , p_{K_a, K'_{a}, 2} \right\} + \frac{\binom{K_a}{2}}{M} , \label{eq_noCSI_noKa_pKa_Kahat_asymP}
    \end{equation}
    where
    \begin{equation}
      p_{K_a,K'_a,1}  = \mathbb{E}
       \left[ \min_{\tilde{\rho}\geq0} \exp \left\{ \tilde{\rho} nL \left( K'_a + \frac{1}{2} \right)
      -L \ln  \left|  \mathbf{I}_n + \tilde{\rho} \tilde{\mathbf{C}}_{K_a} \tilde{\mathbf{C}}^H_{K_a}  \right|  \right\}  \right]  ,\label{eq_noCSI_noKa_pKa_Kahat1_asym}
    \end{equation}
    \begin{equation}
      p_{K_a,K'_a,2}
      =  \mathbb{E} 
      \left[
      \min_{ 0\leq \tilde{\rho} < \frac{1}{ \tilde{\lambda}'_{1} } }
      \exp \left\{ -\tilde{\rho} nL \left( K'_a - \frac{1}{2} \right)
      - L \ln \left|  \mathbf{I}_n - \tilde{\rho} \tilde{\mathbf{C}}_{K_a} \tilde{\mathbf{C}}^H_{K_a} \right|  \right\} \right]  .  \label{eq_noCSI_noKa_pKa_Kahat2_asym}
    \end{equation}
    Here, $\tilde{\mathbf{C}}_{K_a} = \frac{1}{P'} \mathbf{C}_{K_a}$ with $\mathbf{C}_{K_a}$ given in Theorem~\ref{Theorem_Ka}, i.e., the columns of $\tilde{\mathbf{C}}_{K_a} \in \mathbb{C}^{n\times M}$ satisfy $\mathbb{E} [ \left\| \tilde{\mathbf{c}}_m \right\|_2^2  ]=n$ for $m \in [M]$; and $\tilde{\lambda}'_1$ denotes the maximum eigenvalue of the matrix $\tilde{\mathbf{C}}_{K_a} \tilde{\mathbf{C}}^H_{K_a}$.
  \begin{IEEEproof}
    As in Theorem~\ref{Theorem_Ka}, we construct a common codebook $\mathbf{C} = \left[ \mathbf{c}_1, \ldots, \mathbf{c}_M \right] \in \mathbb{C}^{n\times M}$ satisfying $\mathbb{E} [ \left\| \mathbf{c}_m \right\|_2^2  ] = nP'$ for $m \in [M]$. As mentioned in the proof of Theorem~\ref{Theorem_Ka}, the total variation distance is upper-bounded by $p_{0,K_a}+ \frac{\binom{K_a}{2}}{M}$. Applying the Markov inequality, we have
    \begin{equation}
      p_{0,K_a}
      \leq \frac{K_a\mathbb{E}\left[ \left\|\mathbf{c}_m \right\|_2^2 \right]}{nP}
      = \frac{K_aP'}{P}  .
    \end{equation}
    In the case of $P'\to\infty$ and $P\to\infty$ while satisfying $\frac{P'}{P}\to 0$, we have $p_{0,K_a}=\frac{K_aP'}{P}\to 0$\footnote{In the case of $P\to\infty$, we assume $P'\to\infty$ with $\frac{P'}{P}\to 0$ to obtain a tight upper bound on the error probability $\mathbb{P} \left[ K_a  \to K'_{a} \right]$. This does not mean that it is optimal to choose $\frac{P'}{P}\to 0$ in the power-limited regime or the condition $\frac{P'}{P}\to 0$ must be satisfied to minimize the energy-per-bit under a fixed reliability constraint. Especially, in the power-limited regime with small $P$, the optimal $\frac{P'}{P}$ cannot approach $0$ because an extremely small $P'$ makes it impossible to estimate the number of active users and perform decoding.}. Moreover, when $P'\to\infty$, the term $ \left(  K'_a+  \frac{1}{2} \right) P'$ in \eqref{eq_noCSI_noKa_pKa_Kahat1} and $ \left(  K'_a - \frac{1}{2} \right) P'$ in \eqref{eq_noCSI_noKa_pKa_Kahat2} are much larger than $1$, and the matrix $\mathbf{C}_{K_a} \mathbf{C}^H_{K_a}$ in \eqref{eq_noCSI_noKa_pKa_Kahat_F} is dominant compared with $\mathbf{I}_n$. Denote $\tilde{\mathbf{C}}_{K_a} = \frac{1}{P'} \mathbf{C}_{K_a}$ and $\tilde{\rho} = \rho P'$. Then, we can obtain Corollary~\ref{Theorem_Ka_asymP} from Theorem~\ref{Theorem_Ka}.
  \end{IEEEproof}
\end{Corollary}
\begin{Corollary} \label{Theorem_Ka_asymn}
  In both cases of applying i.i.d. Gaussian codewords and using codewords uniformly distributed on a sphere, for fixed $P$ and $L$, as $n\to\infty$, the error probability of estimating $K_a$ exactly as $K'_a \in [K] \backslash \{ K_a \}$ is bounded as 
    \begin{equation} \label{eq_noCSI_noKa_pKa_Kahat_asymn}
       \lim_{n\to\infty} \mathbb{P} \left[ K_a  \to K'_{a} \right]  
       \leq  p_{K_a, K'_{a},P'} + \frac{\binom{K_a}{2}}{M},
    \end{equation}
    where
    \begin{equation}
      p_{K_a, K'_{a},P'}
      =
      \begin{cases}
      \exp \left\{ K_a L \left( 1 - \frac{2K'_a+1}{2K_a} + \ln \frac{ 2K'_a+1 }{2K_a} \right)  \right\} ,
      &  \text { if }  K'_a  < K_a\\
      \exp \left\{ K_a L \left( 1 - \frac{2K'_a-1}{2K_a} + \ln \frac{ 2K'_a-1 }{ 2K_a} \right)  \right\} ,
      &  \text { if }  K'_a  > K_a 
      \end{cases} . \label{eq_noCSI_noKa_pKa_Kahat1_asymn}
    \end{equation}
  \begin{IEEEproof}
    See Appendix~\ref{Appendix_proof_Ka_asymn}.
  \end{IEEEproof}
\end{Corollary}

  Corollary~\ref{Theorem_Ka_asymP} and Corollary~\ref{Theorem_Ka_asymn} provide error floors of the achievability bound on $\mathbb{P} \left[ K_a  \to K'_{a} \right]$ given in Theorem~\ref{Theorem_Ka} by allowing $P\to\infty$ and $n\to\infty$, respectively. To provide further insights, in Theorem~\ref{Theorem_scalinglaw_Ka_estimation}, we present scaling laws in terms of the distance between the estimated and true numbers of active users. 
  \begin{Theorem} \label{Theorem_scalinglaw_Ka_estimation}
    Assume that there are exactly $K_a$ active users among $K$ potential users. All users share a common codebook with each column drawn uniformly i.i.d. from a sphere of radius $\sqrt{nP}$. Let $L, K \to \infty$ satisfying $\frac{\ln K}{L} = o(1)$. For every $n,P,K_a$, and constant $\alpha>32$, we have
    \begin{itemize}
      \item For every $K'_a\in[0:K]$ satisfying the condition $\left| K'_a-K_a \right|   = \omega \left( \max \left\{ \frac{K_a}{\sqrt{L}} , \frac{1}{P\sqrt{nL}} \right\} \right)$, the error probability $\mathbb{P} \left[ K_a \to K'_{a} \right]$ converges to $0$.
      \item The probability of the event that the distance between the estimated and the true numbers of active users is no less than $\sqrt{ \frac{ \alpha \ln  K \left( n + (nK_aP)^2 \right) }{L n^2P^2} }$ vanishes, i.e., we have $\mathbb{P} \Big [ \left| {\rm K}'_a  - K_a \right|   \geq  \left.   \sqrt{ \alpha \ln K  \left(  \frac{K_a^2}{L}  +  \frac{ 1 }{ nL P^2} \right) } , {\rm K}'_a  \in [0:K]  \right| {\rm K}_a  = K_a  \bigg] = \mathcal{O}\left( \frac{1}{K} \right) \to  0$. 
      \item In the case of $K_a \leq \sqrt{\frac{L}{\alpha\ln K} - \frac{1}{nP^2} }$, the estimation error of $K_a$ vanishes, i.e., we have $\mathbb{P}[ \left. {\rm K}'_a  \neq K_a \right| {\rm K}_a  = K_a ] = \mathcal{O}\left( \frac{1}{K} \right) \to 0$. Especially, in a special case of $\sqrt{\frac{nL}{\ln K}}P \to \infty$, the number $K_a = \mathcal{O} \left(\sqrt{\frac{L}{\ln K}}\right)$ of active users can be estimated with vanishing error~probability. 
    \end{itemize} 
    \begin{IEEEproof}
    See Appendix \ref{Proof_scalinglaw_Ka_estimation_new}.
    \end{IEEEproof}
  \end{Theorem} 

  In Theorem~\ref{Theorem_scalinglaw_Ka_estimation}, we evaluate the estimation performance of the number of active users in the asymptotic regime. Theorem~\ref{Theorem_scalinglaw_Ka_estimation} holds for every positive $n, P,$ and $K_a$ under the condition that $L, K \to \infty$ satisfying $\frac{\ln K}{L} = o(1)$. It is demonstrated that $\forall K'_a\in [0:K]$ satisfying the condition $\left| K'_a-K_a \right| = \omega \left( \max \left\{ \frac{K_a}{\sqrt{L}} , \frac{1}{P\sqrt{nL}} \right\} \right)$, the error probability of estimating $K_a$ exactly as $K'_a \in [K] \backslash \{ K_a \}$ tends to $0$ as $L \to \infty$. This indicates that the distance $\left| K'_a-K_a \right|$ satisfying the condition $\mathbb{P} \left[ K_a \to K'_{a} \right] \to 0$ is dominated by $\frac{1}{\sqrt{nL}P}$ in the case of $\frac{1}{\sqrt{n}P} \geq K_a$ and dominated by $\frac{K_a}{\sqrt{L}}$ otherwise. Thus, this distance can be gradually reduced by increasing the number of antennas at the BS. In contrast, increasing the blocklength $n$ and transmitting power $P$ is beneficial for decreasing this distance initially, but it reaches a plateau of $\omega\left(\frac{K_a}{\sqrt{L}}\right)$ eventually, which shows the limitation of increasing the blocklength and boosting the transmitting power for improving the estimation performance as discussed before.

  Moreover, Theorem~\ref{Theorem_scalinglaw_Ka_estimation} provides an interval estimation of $K_a$. Specifically, it is shown that when there are exactly $K_a$ active users, the estimated value falls within an interval whose distance from the true value $K_a$ is less than $\sqrt{ \alpha \ln K  \left(  \frac{K_a^2}{L}  +  \frac{ 1 }{ nL P^2} \right) }$ with probability going to $1$. In a special case of $K_a \leq \sqrt{\frac{L}{\alpha\ln K} - \frac{1}{nP^2} }$, the distance between the estimated and true numbers of active users can be less than $1$, i.e., $K_a$ can be estimated correctly with overwhelming high~probability.

\section{Data Detection}\label{Section:Data}

  In this section, we conduct information-theoretic analysis for data detection in the context of URA with a random and unknown number of active users. The non-asymptotic achievability bounds and converse bounds are given in Section~\ref{Section:Non-Asymptotic-Results-achievability} and Section~\ref{Section:Non-Asymptotic-Results-converse}, respectively. Building on the derived non-asymptotic achievability and converse results, we establish asymptotic scaling laws in Section~\ref{Section:Asymptotic-Results}.

\subsection{Non-Asymptotic Achievability Bound} \label{Section:Non-Asymptotic-Results-achievability}

  Assuming that the number ${\rm K}_a$ of active users is random and unknown but its distribution is known in advance, Theorem~\ref{Theorem_achievability} presents achievability bounds on misdetection and false-alarm probabilities when each active user transmits $J$ information bits with blocklength $n$ and power $P$ to the BS over MIMO quasi-static Rayleigh fading channels. To obtain Theorem~\ref{Theorem_achievability}, we use a random coding scheme to generate a common codebook $\mathbf{C} = \left[ \mathbf{c}_{1}, \ldots, \mathbf{c}_{M} \right] \in \mathbb{C}^{n\times M}$ with each codeword drawn i.i.d. according to $\mathcal{CN} \left(0,P'\mathbf{I}_{n}\right)$ or drawn uniformly i.i.d. from a sphere of radius $\sqrt{nP'}$, as introduced in Corollary~\ref{Theorem_Ka_C}. To simplify the analysis, we only upper-bound the error probability for $\mathrm{K}_a \in [K_l : K_u]$ with $0 \leq K_l \leq K_u\leq K$, and set the upper-bound on the error probability as $1$ in the case of $\mathrm{K}_a \notin [K_l : K_u]$. When the number of active users is known in advance, a common assumption is that the receiver outputs a list of messages of size equal to the number of active users~\cite{A_perspective_on}. A difficulty in the scenario with a random and unknown number of active users lies in how to determine the size of the decoded list. To this end, we employ the two-stage scheme proposed in \cite{noKa}. Specifically, when there are ${\rm K}_a = K_a$ active users, we first estimate the number of active users based on an energy-based estimator as introduced in Section~\ref{Section:Ka1}. Considering that $K'_a$ is not necessarily equal to the true value $K_a$, the decoded list size $\hat{K}_a$ is not set to be $K'_a$. Instead, we borrow the idea in~\cite{noKa} and select $\hat{K}_a$ from the set $[K'_{a,l} : K'_{a,u}]$, where $K'_{a,l} = \max\left\{ K_l , K'_{a} -r' \right\}$, $K'_{a,u} = \min\left\{ K_u , K'_{a} +r' \right\}$, and $r'$ denotes a nonnegative integer referred to as the decoding radius. Compared with \cite{noKa}, where the ML criterion was used, we adopt the MAP criterion to determine $\hat{K}_a$ and decode. By incorporating a prior distribution of the set of transmitted codewords, the MAP criterion contributes to tighter achievability bounds shown in Theorem~\ref{Theorem_achievability}.
  \begin{Theorem} \label{Theorem_achievability}
    Consider URA in MIMO quasi-static Rayleigh fading channels where the number of active users is random and unknown but its distribution is known in advance. There exists an $(n,M,\epsilon_{\rm MD},\epsilon_{\rm FA},P)$ URA code such that for each $P'\in [0,P] , K_l\in[0:K] ,  K_u\in[K_l:K]$, and $r' \in \mathbb{N} $, the misdetection and false-alarm probabilities is upper-bounded as
      \begin{equation}
        P_{\mathrm{MD}} \leq 
        p_0 + \!\!
        \sum_{K_a = \max\{K_l,1\}}^{K_u}  \!\!\!  P_{{\rm{K}}_a}(K_a) \!\!
        \sum_{K'_a=K_l}^{K_u}
        \sum_{t \in \mathcal{T}_{K'_a} }  \!\!\frac{t \!+\! (K_a \!-\! K'_{a,u} )^{+}  }{K_a}  \!
        \min \!\left\{ \! 1,  \! \sum_{t' \in  \bar{\mathcal{T}}_{ K'_a ,t}}  \!\!    p_{K'_a,t,t'}, p_{K_a, K'_a,P'}   \!\right\} \! ,\label{eq_noCSI_noKa_epsilonMD}
      \end{equation}
      \begin{align} \label{eq_noCSI_noKa_epsilonFA}
        P_{\mathrm{FA}}  \leq
        p_0 + \!\!
        \sum_{K_a=K_l}^{K_u}   \!\! P_{{\rm{K}}_a}(K_a)\!
        \sum_{K'_a=K_l}^{K_u} \sum_{t \in \mathcal{T}_{K'_a} } \sum_{t' \in \mathcal{T}_{K'_a,t} } \!\!\frac{t'  +   ( K'_{a,l} - K_a  )^{+}}{ \hat{K}_a  }
        \min \!\left\{ 1, p_{K'_a,t,t'}, p_{K_a, K'_a,P'} \right\} \! ,
      \end{align}
    where
    \begin{equation} \label{eq_noCSI_p0}
      p_0  
       = \sum_{K_a =0}^{K}  P_{{\rm{K}}_a} (K_a)  p_{0,K_a} 
       +     \sum_{K_a \notin [K_l:K_u]}     P_{{\rm{K}}_a} (K_a) 
       + \sum_{K_a =0}^{K}  P_{{\rm{K}}_a} (K_a)  \frac{\binom{K_a}{2}}{M}  ,
    \end{equation}
      \begin{equation}\label{eq_noCSI_noKa_estimate_Tset1}
        \mathcal{T}_{K'_a }  =   \left[ 0   :
        \min \left\{K_a,K'_{a,u},M - K'_{a,l} - (K_a - K'_{a,u})^{+}\right\} \right]  ,
      \end{equation}
      \begin{equation}\label{eq_noCSI_noKa_estimate_Tset2bar}
        \bar{\mathcal{T}}_{K'_a,t}  =
        \left[ \left( (K_a  -  K'_{a,u} )^{+}   -  (K_a - K'_{a,l} )^{+}  + t \right)^{+}
         :  T_{upper} \right]  ,
      \end{equation}
      \begin{equation}\label{eq_noCSI_noKa_estimate_Tset2}
        \mathcal{T}_{K'_a,t}  =
        \left[ \left( (K_a  -  K'_{a,u} )^{+}   -  (K'_{a,l}  -  K_a)^{+}  +   \max\{K'_{a,l},1\}
         -  K_a +  t \right)^{+} : T_{upper} \right]  ,
      \end{equation}
      \begin{equation}
        T_{upper} = \min  \left\{ (K'_{a,u}   -  K_a)^{+}   -  (K'_{a,l}  -  K_a)^{+}  +  t ,
        M-\max \left\{ K_a , K'_{a,l} \right\} \right\} ,
      \end{equation}
      \begin{equation}\label{eq_noCSI_noKa_estimate_Kahat}
        \hat{K}_a = K_a - t - (K_a-K'_{a,u} )^{+} + t' + ( K'_{a,l}  - K_a )^{+} ,
      \end{equation}
      \begin{equation}\label{eq_noCSI_noKa_estimate_Kal}
        K'_{a,l}  = \max\left\{ K_l , K'_{a} -r' \right\} ,
      \end{equation}
      \begin{equation}\label{eq_noCSI_noKa_estimate_Kau}
        K'_{a,u}  = \min\left\{ K_u , K'_{a} +r' \right\} ,
      \end{equation}
    \begin{align} \label{eq_noCSI_noKa_estimate_ptt}
      p_{K'_a,t,t'}  =  \min_{ 0 \leq \omega \leq 1, 0 \leq \nu }
      & \left\{ q_{1,K'_a,t,t'}\left(\omega,\nu\right)
      + 1 \left[ t+(K_a - K'_{a,u})^{+} > 0 \right]
      q_{2,K'_a,t }\left(\omega,\nu\right) \right. \notag \\
      &  \; \left. +\; 1 \left[ t+(K_a - K'_{a,u})^{+}  = 0 \right] q_{2,K'_a,t,0 }\left(\omega,\nu\right) \right\}   ,
    \end{align}
    \begin{align}
      q_{1 ,K'_a ,t ,t'}  \left(\omega,\nu\right)
      = C_{K'_a ,t ,t'}
      & \mathbb{E}
      \left[
      \min_{ {u\geq 0,r\geq 0, \lambda_{\min} (\mathbf{B}) > 0}}
      \exp  \left\{  L rn\nu + b_{u,r} \right.
      \right. \notag \\
      & \;\;\;\; \;\; \left.  +  L   \left( u \ln  \left|\mathbf{F}''\right|
      - r \ln  \left|\mathbf{F}\right|
      - u \ln  \left| {\mathbf{F}'} \right|
      + r\omega \ln  \left| \mathbf{F}_{1} \right|
      - \ln  \left| \mathbf{B} \right| \right)
      \right\}  \bigg] , \label{eq_noCSI_noKa_estimation_q1t}
    \end{align}
      \begin{equation}\label{eq_noCSI_noKa_estimation_Ctt}
        C_{K'_a,t,t'} = \binom{K_a}{ t+(K_a - K'_{a,u})^{+} } \binom{M-K_a}{t'+(K'_{a,l} - K_a)^{+}}  ,
      \end{equation}
      \begin{equation}\label{eq_noCSI_noKa_pKa_Kahat_F_new}
        \mathbf{F} = \mathbf{I}_n + \mathbf{C}_{ \mathcal{W} } \mathbf{C}^H_{ \mathcal{W} } , 
      \end{equation}
      \begin{equation}\label{eq_noCSI_noKa_estimation_F1}
        \mathbf{F}_1  = \mathbf{I}_n + \mathbf{C}_{ \mathcal{W} \backslash \mathcal{W}_1 } \mathbf{C}^H_{ \mathcal{W} \backslash \mathcal{W}_1 } , 
      \end{equation}
      \begin{equation}\label{eq_noCSI_noKa_estimation_Fprime}
        \mathbf{F}'  = \mathbf{I}_n + \mathbf{C}_{ \mathcal{W} \backslash \mathcal{W}_1 \cup \mathcal{W}_2 } \mathbf{C}^H_{ \mathcal{W} \backslash \mathcal{W}_1 \cup \mathcal{W}_2 } ,
      \end{equation}
      \begin{equation}\label{eq_noCSI_noKa_estimation_Fprime2}
        \mathbf{F}''  = \mathbf{I}_n + \mathbf{C}_{ \mathcal{W} \backslash \mathcal{W}_{1,1} \cup \mathcal{W}_{2,1} } 
        \mathbf{C}^H_{ \mathcal{W} \backslash \mathcal{W}_{1,1} \cup \mathcal{W}_{2,1} } ,
      \end{equation}
      \begin{equation}\label{eq_noCSI_noKa_estimation_B}
        \mathbf{B} = (1+r) \mathbf{I}_n - u \left( \mathbf{F}'' \right)^{-1} \mathbf{F}
        + u \left( \mathbf{F}' \right)^{-1} \mathbf{F}
        - r\omega \mathbf{F}_{ 1}^{-1} \mathbf{F} ,
      \end{equation}
    \begin{equation}\label{eq_noCSI_noKa_estimation_bur}
      b_{u,r} = -u b'' + r b + u b' -r\omega b_1,
    \end{equation}
    \begin{equation}\label{eq_noCSI_noKa_estimation_b}
      b = \ln \left( P_{ {\rm{K}}_a} \left(K_a\right) \right) -
      \ln  \binom{M}{ K_a }  ,
    \end{equation}
    \begin{equation}
      b_1 = \ln \left( P_{{\rm{K}}_a} \left( K_a - t - (K_a-K'_{a,u})^{+} \right) \right)
      - \ln  \binom{M}{ K_a - t - (K_a-K'_{a,u})^{+} } ,\label{eq_noCSI_noKa_estimation_b1}
    \end{equation}
    \begin{equation}\label{eq_noCSI_noKa_estimation_bprime}
      b' = \ln \left( P_{{\rm{K}}_a} \left({\hat{K}}_a\right) \right) - \ln  \binom{M}{ \hat{K}_a },
    \end{equation}
    \begin{equation}
      b''  =  \ln \! \left( P_{{\rm{K}}_a}\!  \left( K_a  -  (K_a - K'_{a,u} )^{+} \!+ ( K'_{a,l}   -  K_a )^{+}  \right) \right) - \ln \! \binom{M}{ K_a  -  (K_a - K'_{a,u} )^{+} \!+ ( K'_{a,l}   -  K_a )^{+}  } ,\label{eq_noCSI_noKa_estimation_bprime2}
    \end{equation}
    \begin{align}
      q_{2,K'_a,t} \left(\omega,\nu\right) =  & \binom{K_a}{t + (K_a - K'_{a,u})^{+} }
      \min_{ \delta\geq0 }
      \Bigg\{ \frac{\Gamma \left( nL, nL \left( 1 +\delta \right)\right)}{\Gamma\left( nL \right)} \notag\\
      &  \left.  +  \mathbb{E}
      \left[   \frac{\gamma \left(  Lm,  \frac{ nL(1+\delta)(1-\omega) - \omega\left( L \ln\left|\mathbf{F}_1 \right| - b_1 \right) + L  \ln\left|\mathbf{F}\right| - b - nL\nu }
      {\omega \prod_{i=1}^{m}   \lambda_i^{1/m} }   \right)}{\Gamma\left( Lm \right)}  \right]
       \right\}   , \label{eq_noCSI_noKa_estimation_q2t}
    \end{align}
    \begin{equation}\label{eq_noCSI_noKa_estimation_q2t0}
      q_{2,K'_a,t,0} = \mathbb{E}
      \left[  \frac{\Gamma\left( nL, \frac{nL\nu}{1-\omega} - L \ln \left| \mathbf{F} \right| + b \right)}{\Gamma\left( nL \right)}  \right] .
    \end{equation} 
  Here, $p_{K_a, K'_a,P'}$ in~\eqref{eq_noCSI_noKa_epsilonMD} and \eqref{eq_noCSI_noKa_epsilonFA} is given in~\eqref{eq_noCSI_noKa_pKa_Kahat0} and $p_{0,K_a}$ in \eqref{eq_noCSI_p0} is defined in Corollary~\ref{Theorem_Ka_C}. In \eqref{eq_noCSI_noKa_pKa_Kahat_F_new}-\eqref{eq_noCSI_noKa_estimation_Fprime2}, for any subset $S \subset [M]$, the matrix $\mathbf{C}_S\in \mathbb{C}^{n\times |S|}$ denotes the concatenation of codewords in $\left\{ \mathbf{c}_i : i\in S \right\}$. Exploiting symmetry, we assume w.l.o.g. that the set $\mathcal{W}=[K_a]$ includes the transmitted messages of active users; the set $\mathcal{W}_1$ includes $t+(K_a-K'_{a,u})^{+}$ misdetected messages satisfying $\mathcal{W}_1  = \mathcal{W}_{1,1} \cup \mathcal{W}_{1,2}$, where $\mathcal{W}_{1,1} = [(K_a-K'_{a,u})^{+}]$ includes misdetections due to insufficient decoded list size and $\mathcal{W}_{1,2} = [(K_a-K'_{a,u})^{+}+1:(K_a-K'_{a,u})^{+}+t]$ includes additional $t$ misdetections occurring during decoding; we assume that the set $\mathcal{W}_{2}$ includes $t'+(K'_{a,l}-K_a)^{+}$ falsely alarmed messages satisfying $\mathcal{W}_{2}  = \mathcal{W}_{2,1} \cup \mathcal{W}_{2,2}$, where $\mathcal{W}_{2,1} = [K_a+1 : K_a+(K'_{a,l}-K_a)^{+}]$ includes falsely alarmed messages coming from excessive decoded list size and $t'$ messages in $\mathcal{W}_{2,2} = [K_a+(K'_{a,l}-K_a)^{+}+1: K_a+(K'_{a,l}-K_a)^{+}+t']$ are additionally falsely alarmed while decoding. In~\eqref{eq_noCSI_noKa_estimation_q2t}, $\lambda_1,\ldots, \lambda_m$ denote the non-zero eigenvalues of $\mathbf{F}_1^{-1} \mathbf{C}_{\mathcal{W}_1} \mathbf{C}^H_{\mathcal{W}_1} $ with $m=\min\left\{ n, t+(K_a-K'_{a,u})^{+} \right\}$. The expressions of the lower incomplete gamma function $\gamma\left(\cdot, \cdot\right)$, upper incomplete gamma function $\Gamma\left(\cdot, \cdot\right)$, and gamma function $\Gamma\left(\cdot\right)$ in~\eqref{eq_noCSI_noKa_estimation_q2t} and \eqref{eq_noCSI_noKa_estimation_q2t0} are provided in the introduction.  
  \begin{IEEEproof}[Proof sketch] 
    As introduced above, we use a random coding scheme to generate a common codebook $\mathbf{C}$. If user $k$ is active, it transmits $\mathbf{x}_{W_k} = \mathbf{c}_{W_k} 1 \left\{ \left\|\mathbf{c}_{W_k}\right\|_{2}^{2} \leq n P \right\}$, where the message $W_k$ is chosen uniformly at random from the set $[M]$. To upper-bound the error probability, we change the measure over which $\mathbb{E}$ is taken in \eqref{eq:MD} and \eqref{eq:FA} to the new one under which: 1)~there is no message collision, i.e. $W_i \neq W_j$ for $i\neq j$; 2)~the active user~$k$ transmits $\mathbf{x}_{W_k} = \mathbf{c}_{W_k}$ regardless of the power constraint; and 3)~the number of active users satisfies $\mathrm{K}_a \leq [K_l : K_u]$ with $0 \leq K_l \leq K_u\leq K$. The per-user probability of misdetection in~\eqref{eq:MD} is upper-bounded as
    \begin{align}
      P_{\mathrm{MD}} 
      & \leq \mathbb{P} [{\rm power\;violation}] 
      + \mathbb{P} [{\rm atypical\;} {\rm K}_a] 
      + \mathbb{P} [W_i = W_j {\rm\; for \;some\;} i\neq j]  \notag\\
      & \;\;\;\; + \mathbb{P} [{\rm MD \;under\; no\; power\; constraint\;} | {\rm \;typical\;} {\rm K}_a, {\rm no\; clash \;of \;messages}]. \label{eq:Theorem_achi_proofsketch0}
  \end{align} 
  The sum of the first three terms in the RHS of \eqref{eq:Theorem_achi_proofsketch0} is bounded by $p_0$ in~\eqref{eq_noCSI_p0}.

  We adopt a two-stage decoder to bound the last term on the RHS of \eqref{eq:Theorem_achi_proofsketch0}. Specifically, the decoder first obtains an estimate of the number of active users. Then, based on the MAP criterion, the decoder produces a set of decoded messages of size $\hat{K}_a \in [K'_{a,l}: K'_{a,u}]$, where $K'_{a,l}$ and $K'_{a,u}$ are determined by the estimate $K'_a$ of the number of active users and the decoding radius $r'$ as shown in~\eqref{eq_noCSI_noKa_estimate_Kal} and \eqref{eq_noCSI_noKa_estimate_Kau}. The output of the MAP decoder is given by 
  \begin{equation} \label{eq:decoderoutput_noCSI_W}
    \hat{\mathcal{W}} = \left\{\; \emph{f}_{\text{en}}^{-1}\left( \hat{\mathbf{c}} \right): \hat{\mathbf{c}} \in \hat{\mathcal{C}}  \right\},
  \end{equation}
  \begin{equation}\label{eq:decoderoutput_noCSI}
    \hat{\mathcal{C}}
    =\arg \min_{ \bar{\mathcal{C}} \subset \{ \mathbf{c}_1, \ldots, \mathbf{c}_M \}: \bar{K}_a = | \bar{\mathcal{C}} | \in [ K'_{a,l} : K'_{a,u} ] }
    g\;\! (  {\boldsymbol{\Gamma}} ) ,
  \end{equation}
  where both the sets $ \hat{\mathcal{W}}$ and $ \hat{\mathcal{C}}$ are of size $\hat{K}_a$, the matrix $ {\boldsymbol{\Gamma}} = \operatorname{diag} \left\{  {\boldsymbol{\gamma} } \right\}\in\{0,1\}^{M\times M}$ with $[ {\boldsymbol{\gamma} }]_i=1$ if $\mathbf{c}_i \in \bar{\mathcal{C}}$, and the MAP decoding metric $g (  {\boldsymbol{\Gamma}} )$ is given by 
  \begin{equation}
    g (  {\boldsymbol{\Gamma}} )
    = L \ln \big| \mathbf{I}_n + \mathbf{C} {\boldsymbol{\Gamma}}\mathbf{C}^H \big|
    +  \operatorname{tr}  \left(  \mathbf{Y}^{H}
    \big( \mathbf{I}_n  +  \mathbf{C} {\boldsymbol{\Gamma}}\mathbf{C}^H \big)^{-1}   \mathbf{Y}  \right)
     -  \ln  \left(  P_{{\rm{K}}_a} ( \bar{K}_a )  \right)
     +  \ln   \binom{M}{ \bar{K}_a }   .\label{eq:g_noCSI}
  \end{equation} 
  Here, the sum of the first two terms on the RHS of \eqref{eq:g_noCSI} is proportion to $-\ln P(Y|\bar{\mathcal{C}})$; $P_{{\rm{K}}_a} ( \bar{K}_a )$ denotes the probability of having exactly $\bar{K}_a$ active users; and $\frac{1}{\binom{M}{ \bar{K}_a }}$ denotes the probability of selecting any specific set of codewords of size $\bar{K}_a$ given that there are $\bar{K}_a$ active users.

  Denote the last term in the RHS of \eqref{eq:Theorem_achi_proofsketch0} as $p_1$. We have 
  \begin{align}
    p_1 
    & =  
    \mathbb{E}  \left[ 1 \left[ \mathrm{K}_a > 0 \right]
            \cdot \frac{1}{\mathrm{K}_a}  \sum_{k\in {\mathcal{K}_a}}  1  \left[ W_{k} \notin \hat{\mathcal{W}}  \right] \right]_{\text{new}} \label{eq:Theorem_achi_proofsketch1} \\
    & \leq   \sum_{K_a=\max\{K_l,1\}}^{K_u}    P_{{\rm{K}}_a}(K_a)
    \sum_{K'_a=K_l}^{K_u}
    \sum_{t \in \mathcal{T}_{K'_a} }
    \frac{t+(K_a - K'_{a,u})^{+}}{K_a}  \mathbb{P} \left[  \mathcal{F}_{t} \cap \left\{ K_a \to K'_{a} \right\} \right]_{\text{new}} ,  \label{eq:Theorem_achi_proofsketch2}
  \end{align}
  where $\mathcal{F}_{t} = \left\{ \sum_{k\in {\mathcal{K}_a}} 1 \left\{ W_k \notin \hat{\mathcal{W}} \right\} = t+(K_a-{K}'_{a,u})^{+} \right\}$ denotes the event that there are exactly $t+(K_a-K'_{a,u})^{+}$ misdetected codewords; the integer $t$ takes value in $\mathcal{T}_{K'_a}$ defined in~\eqref{eq_noCSI_noKa_estimate_Tset1} because the number of misdetected codewords is lower-bounded by $(K_a-K'_{a,u})^{+}$ and upper-bounded by the total number $K_a$ of transmitted messages and by $M-K'_{a,l}$ since at least $K'_{a,l}$ messages are returned; and $\left\{K_a \to K'_{a} \right\}$ denotes the event that $K_a$ is estimated exactly as $K'_a$. The probability $\mathbb{P} \left[  \mathcal{F}_{t} \cap \left\{ K_a \to K'_{a} \right\} \right]_{\text{new}}$ is bounded as
  \begin{equation}
    \mathbb{P} \left[  \mathcal{F}_{t} \cap \left\{ K_a \to K'_{a} \right\} \right]_{\text{new}}
    \leq \min \left\{ 1, \sum_{t' \in \bar{\mathcal{T}}_{K'_a,t} }   \mathbb{P}  \left[  \mathcal{F}_{t,t'}
    \left|  \hat{K}_{a}  \in  [K'_{a,l}:K'_{a,u} ] \right.\right]_{\text{new}}
    , \mathbb{P} \left[ K_a  \to  K'_{a}  \right]_{\text{new}} \right\} .  \label{eq:Theorem_achi_proofsketch3}
  \end{equation}
  Here, $\bar{\mathcal{T}}_{K'_a,t}$ is defined in \eqref{eq_noCSI_noKa_estimate_Tset2bar}; $\mathbb{P} \left[ K_a  \to  K'_{a}  \right]_{\text{new}}$ is bounded by $p_{K_a, K'_a,P'}$ given in~\eqref{eq_noCSI_noKa_pKa_Kahat0}; and $\mathbb{P}  \left[  \mathcal{F}_{t,t'} \left|  \hat{K}_{a}  \in  [K'_{a,l}:K'_{a,u} ] \right.\right]_{\text{new}}$ denotes the probability of the event that there are exactly $t+(K_a-K'_{a,u})^{+}$ misdetected messages and $t' + ( K'_{a,l} - K_a )^{+}$ falsely alarmed messages when we select the decoded list size $\hat{K}_{a}$ from the set $[K'_{a,l}:K'_{a,u} ]$ under the new measure, which is bounded by $p_{K'_a,t,t'}$ in~\eqref{eq_noCSI_noKa_estimate_ptt}. The decoding is performed based on the MAP criterion, and $p_{K'_a,t,t'}$ is obtained applying Fano's bounding technique~\cite{1961}, a properly selected region around the linear combination of the transmitted signals~\cite{TIT}, Chernoff bound~\cite{goodregion}, and moment generating function of quadratic forms~\cite{quadratic_form1}. Likewise, we can obtain an upper bound on the false-alarm probability as shown in \eqref{eq_noCSI_noKa_epsilonFA}. A detailed proof is given in Appendix~\ref{Appendix_proof_achievability}.  
    \end{IEEEproof}
  \end{Theorem}

  As mentioned above, Theorem \ref{Theorem_achievability} is obtained applying a two-stage scheme proposed in \cite{noKa} with a key difference that we adopt the MAP criterion introduced in \eqref{eq:decoderoutput_noCSI_W}-\eqref{eq:g_noCSI} to decode instead of the ML criterion used in \cite{noKa}. Compared to the ML criterion, the MAP criterion contributes to a tighter achievability bound by incorporating a prior distribution of the set of transmitted codewords. Moreover, the Gaussian channel model was considered  in~\cite{noKa}, where the main technique employed to upper-bound the error probability was Gallager's $\rho$-trick. In contrast, we consider the MIMO quasi-static Rayleigh fading channel model in this work. The incorporation of random channel fading coefficients increases the difficulty of upper-bounding the error probability, which hinders the application of Gallager's $\rho$-trick in this work. Thus, we utilize Fano's bounding technique~\cite{1961}, a properly selected region around the linear combination of the transmitted signals~\cite{TIT}, and moment generating function of quadratic forms~\cite{quadratic_form1} to derive upper bounds on the misdetection and false-alarm probabilities. Since different techniques are employed in this work compared with \cite{noKa}, Theorem~\ref{Theorem_achievability} cannot recover Theorem 1 in~\cite{noKa} in the case of $L = 1$ and no-fading.

  The following corollary provides an achievability bound on the minimum energy-per-bit required by each active user to transmit $J$ information bits using $n$ channel uses under fixed misdetection and false-alarm constraints in MIMO quasi-static Rayleigh fading channels with random and unknown ${\rm K}_a$.
  \begin{Corollary} \label{Theorem_achievability_energyperbit}
    For URA in MIMO quasi-static Rayleigh fading channels where the number of active users is random and unknown but its distribution is known in advance, there exists an $(n, M, \epsilon_{\mathrm{MD}}, \epsilon_{\mathrm{FA}}, P)$ URA code for which the minimum required energy-per-bit satisfies    
      \begin{equation}\label{eq_noCSI_noKa_EbN0}
        E^{*}_{b}(n,M,\epsilon_{\mathrm{MD}},\epsilon_{\mathrm{FA}})
        \leq E^{*}_{b,achi}
        = \inf \frac{n {P}}{J}.
      \end{equation}
    The $\inf$ is taken over all $P > 0$ satisfying that there exist $P'\in [0,P] , K_l\in[0:K] ,  K_u\in[K_l:K]$, and $r' \in \mathbb{N} $ such that
      \begin{equation}
        \epsilon_{\mathrm{MD}} \geq {\text{the RHS of }} \;\!
        \eqref{eq_noCSI_noKa_epsilonMD}, \label{eq_noCSI_noKa_epsilonMD_energyperbit}
      \end{equation}
      \begin{equation} \label{eq_noCSI_noKa_epsilonFA_energyperbit}
        \epsilon_{\mathrm{FA}}  \geq {\text{the RHS of }} \;\!
        \eqref{eq_noCSI_noKa_epsilonFA} . 
      \end{equation}
  \end{Corollary}

  Theorem~\ref{Theorem_achievability} and Corollary~\ref{Theorem_achievability_energyperbit} present achievability bounds on misdetection and false-alarm probabilities and the minimum required energy-per-bit for URA in MIMO quasi-static Rayleigh fading channels, respectively, which provide theoretical benchmarks for evaluating practical URA schemes. Theorem~\ref{Theorem_achievability} and Corollary~\ref{Theorem_achievability_energyperbit} are applicable when ${\rm K}_a$ is random and unknown in advance. Once it becomes fixed and known \emph{a priori}, Corollary~\ref{Theorem_achievability_energyperbit} reduces to Theorem~1 in~\cite{letter}.

  Different from the scenario where each user has an individual codebook considered in~\cite{TIT}, we focus on URA channels in which all users share a common codebook. Thus, in Theorem~\ref{Theorem_achievability} and Corollary~\ref{Theorem_achievability_energyperbit}, the message collision probability is taken into consideration. From \eqref{eq:MD} and \eqref{eq:FA} it is evident that the misdetection and false-alarm probabilities in URA scenarios are determined by the set difference between the set of transmitted messages and that of decoded messages. This is different from the scenario considered in~\cite{TIT}, where the focus is on determining whether the transmitted message matches its decoded counterpart for each user. Thus, the definition of error probability in URA is more relaxed than that in the case of individual codebooks. Moreover, when each user has an individual codebook, the aim of data detection is to recover the transmitted codewords from $K$ codebooks and at most one codeword can be returned from each codebook, which can be formulated as a block sparse support recovery problem with the byproduct of recovering user identities. In contrast, in URA scenarios, the receiver searches for the transmitted codewords from a common codebook comprised of $M$ codewords, which is essentially a sparse support recovery problem without block constraints. As we shall see in Section~\ref{Section:simulation}, compared to the scenario with individual codebooks, the minimum required energy-per-bit can be reduced when users share a common codebook due to the reduced search space for decoding and the relaxed definition of error probability.

  Similar to Theorem~\ref{Theorem_Ka}, Theorem~\ref{Theorem_achievability} and Corollary~\ref{Theorem_achievability_energyperbit} are applicable in both cases of using codewords drawn i.i.d. according to a Gaussian distribution and using codewords uniformly distributed on a sphere. As mentioned in Section~\ref{Section:Ka}, in the non-asymptotic regime, utilizing codewords distributed on a sphere is more advantageous, which will be confirmed by the simulation results in Section~\ref{Section:simulation}.

  Theorem~\ref{Theorem_achievability} and Corollary~\ref{Theorem_achievability_energyperbit} can be extended to the scenario where there is a mismatch between the assumed distribution of the number of active users and the actual one. Denote the inaccurate probability of having exactly $K_a$ active users as $\bar{P}_{{\rm{K}}_a} ( K_a )$. To derive the achievability bound, we treat the assumed distribution as perfect and substitute it into the decoding metric in~\eqref{eq:g_noCSI}. Then, the decoding metric becomes   
  \begin{equation}
    g (  {\boldsymbol{\Gamma}} )
    = L \ln \big| \mathbf{I}_n + \mathbf{C} {\boldsymbol{\Gamma}}\mathbf{C}^H \big|
    +  \operatorname{tr}  \left(  \mathbf{Y}^{H}
    \big( \mathbf{I}_n  +  \mathbf{C} {\boldsymbol{\Gamma}}\mathbf{C}^H \big)^{-1}   \mathbf{Y}  \right)
     -  \ln  \left(  \bar{P}_{{\rm{K}}_a} ( \bar{K}_a )  \right)
     +  \ln   \binom{M}{ \bar{K}_a }  . \label{eq:g_noCSI_MM}
  \end{equation}
  Along similar ideas as in the proof of Theorem~\ref{Theorem_achievability}, we can obtain the achievability bound for the scenario with imperfect knowledge of the distribution of ${\rm{K}}_a$ by changing $P_{{\rm{K}}_a} ( \cdot )$ in \eqref{eq_noCSI_noKa_estimation_b}-\eqref{eq_noCSI_noKa_estimation_bprime2} to $\bar{P}_{{\rm{K}}_a} ( \cdot )$. When $\ln \binom{M}{K_a }$ is dominant compared with $-  \ln  \left(  \bar{P}_{{\rm{K}}_a} ( K_a )  \right)$ and $-  \ln  \left(  P_{{\rm{K}}_a} ( K_a )  \right)$ for $K_a \in [K_l:K_u]$, the mismatch between the assumed and actual distributions of the number of active users does not significantly deteriorate the decoding performance. In this case, it is sufficient to consider only the prior probability of selecting any specific set of codewords of size $K_a$.

\subsection{Non-Asymptotic Converse Bound} \label{Section:Non-Asymptotic-Results-converse}

  In this section, we provide three converse bounds on the minimum required energy-per-bit for URA in MIMO quasi-static Rayleigh fading channels with random and unknown numbers of active users. Specifically, Theorem~\ref{Theorem_converse_single} presents a single-user type converse bound, which is obtained by casting a URA code as a single-user code with list decoding via assuming the activities of all users and the transmitted messages and channel coefficients of ${\rm K}_a-1$ active users are known \emph{a priori}. 
  \begin{Theorem} \label{Theorem_converse_single} 
   The minimum required energy-per-bit for the URA model described in Section~\ref{section2} can be lower-bounded as
    \begin{equation} \label{eq:P_tot_conv_singleUE_EbN0}
       E^{*}_{b}(n, M, \epsilon_{\mathrm{MD}}, \epsilon_{\mathrm{FA}})
      \geq \inf_{P} \frac{nP}{J}, 
    \end{equation} 
    under the constraints that $P>0$ and for any subset $S \subset \mathcal{K}$, there exist $\hat{K}_a \in [K]$ and $\epsilon_{K_a} \in [0,1]$ for $K_a \in S$ such that
      \begin{equation} \label{eq:P_tot_conv_singleUE1}
        J  - \log_2   \hat{K}_a   \leq
         - \log_2  { \mathbb{P} \left[ \chi^2(2L) \geq (1+(n + 1)P) \;  r_{K_a}
        \right] } , \forall K_a \in S , 
      \end{equation}
      \begin{equation}\label{eq:P_tot_conv_singleUE2}
        \mathbb{P} \left[ \chi^2(2L) \leq r_{K_a} \right] = \epsilon_{K_a} ,
      \end{equation}
      \begin{equation}\label{eq:P_tot_conv_singleUE3}
        \sum_{K_a \in S}  P_{{\rm K}_a}(K_a) \; \epsilon_{K_a} \leq \epsilon_{\mathrm{MD}},
      \end{equation}
      and
      \begin{equation} \label{P_tot_conv_noCSI_FA}
      \sum_{K_a \in S}  P_{{\rm K}_a}(K_a) \max \left\{ \frac{\hat{K}_a-K_a}{\hat{K}_a} , 0 \right\} \leq \epsilon_{\mathrm{FA}} ,
    \end{equation}
    are satisfied, where $\hat{K}_a$ and $\epsilon_{K_a}$ are functions of $K_a$ denoting the size of the decoded list and the misdetection error probability when there are $K_a$ active users, respectively.  
    \begin{IEEEproof}
      The converse result in~\cite[Theorem~2]{letter} is established under the assumption that the number of active users is fixed and known in advance. It is obtained by casting a URA code as a single-user code with list decoding of size no more than $K_a$ and by applying the single-user converse result in~\cite[Theorem~3]{noCSI_conv}. Theorem~\ref{Theorem_converse_single} is obtained by extending the converse result in~\cite[Theorem~2]{letter} to the scenario with a random and unknown number of active users applying the following changes: 1)~the optimal list size is uncertain beforehand; 2)~taking the expectation over ${\rm K}_a \in S \subset \mathcal{K}$ as in~\eqref{eq:P_tot_conv_singleUE3} to adapt to the case with random ${\rm K}_a$; and 3)~since at least $\max \{0,\hat{K}_a-K_a\}$ messages are false-alarmed when the size of the decoded list is $\hat{K}_a$, the constraint in~\eqref{P_tot_conv_noCSI_FA} should be satisfied.
    \end{IEEEproof}
  \end{Theorem}

  In the following, we provide another single-user type converse bound. Specifically, in Theorem~\ref{Theorem_converse_single_noKa}, we assume the activities of $K-1$ users, as well as the transmitted codewords and channel coefficients of active users among them, are known in advance. Theorem~\ref{Theorem_converse_single_noKa} takes the uncertainty in the activity of the remaining one user into account, but is only applicable when ${\rm K}_a$ follows a binomial distribution, though this is a common assumption as considered in~\cite{Yuwei_active,GuoDN,Ravi}. 
  \begin{Theorem} \label{Theorem_converse_single_noKa} 
    Assume that ${\rm K}_a\sim {\rm{Binom}}(K,p_a)$. The minimum required energy-per-bit for the URA model described in Section~\ref{section2} is lower-bounded as
    \begin{equation} \label{eq:P_tot_conv_singleUE_EbN0_noKa}
      E^{*}_{b}(n,M,\epsilon_{\rm MD},\epsilon_{\rm FA}) \geq \inf_P \frac{nP}{J}, 
    \end{equation}
    under the constraints that $P > 0$ and 
      \begin{equation} \label{eqR:beta_conv_AWGN_singleUE_1e}
        \frac{M}{K} \leq 
        \frac{ \epsilon_{1} }{ \mathbb{P}\left[ \chi^2(2L) \geq (1+(n+1)P)r
        \right] } . 
      \end{equation}
    Here, $\epsilon_{1}$ is given by    
    \begin{equation}\label{eqR:beta_conv_epsilon1}
        \epsilon_{1} = \min\left\{ 1, \frac{ \epsilon_{\rm{FA}} }{1-p_a} \right\} ,
      \end{equation}
    and $r$ is the solution of
      \begin{equation}\label{eqR:beta_conv_AWGN_singleUE_e_constraint1}
        \mathbb{P} \left[ \chi^2(2L) \leq r \right] = \epsilon_{2} , 
      \end{equation} 
      \begin{equation}\label{eqR:beta_conv_epsilon2}
        \epsilon_{2} = \min\left\{ 1, \frac{ \epsilon_{\rm{MD}} }{p_a} \right\} . 
      \end{equation} 
    \begin{IEEEproof}
      See Appendix~\ref{Appendix_proof_converse_single_noKa}.
    \end{IEEEproof}
  \end{Theorem}
  
  Both Theorem~\ref{Theorem_converse_single} and Theorem~\ref{Theorem_converse_single_noKa} are obtained by casting a URA code as a single-user code with list decoding. The main differences between them are as follows. First, the activity of the remaining user is assumed to be known beforehand in Theorem~\ref{Theorem_converse_single}, whereas it is unknown to the receiver in Theorem~\ref{Theorem_converse_single_noKa}. That is, Theorem~\ref{Theorem_converse_single_noKa} takes the uncertainty in user activity into consideration. It does not mean Theorem~\ref{Theorem_converse_single_noKa} is always tighter than Theorem~\ref{Theorem_converse_single}. This is because after casting a URA code as a single-user code, the mean of the number of active users under consideration is only $p_a$ under the assumption in Theorem~\ref{Theorem_converse_single_noKa}; in contrast, the remaining user is definitely active as long as ${\rm K}_a\geq 1$ under the assumption in Theorem~\ref{Theorem_converse_single}. Second, as aforementioned, Theorem~\ref{Theorem_converse_single_noKa} is only applicable when ${\rm K}_a$ follows a binomial distribution, but Theorem~\ref{Theorem_converse_single} holds for all kinds of distributions of ${\rm K}_a$.

  Theorem~\ref{Theorem_converse_single} and Theorem~\ref{Theorem_converse_single_noKa} are applicable to all codes. However, they can be loose in the large ${\rm K}_a$ regime since they are derived based on the knowledge of the transmitted messages of ${\rm K}_a-1$ active users and their instantaneous CSI. This knowledge is difficult to obtain when the number of active users is large since MUI is a significant bottleneck in this case. To this end, we consider the case without this knowledge in Theorem~\ref{Theorem_converse_noCSI_Gaussian_cKa}. To make the analysis tractable, we make a stronger assumption that the common codebook has i.i.d. entries with means $0$ and variances $P$, which reduces Theorem~\ref{Theorem_converse_noCSI_Gaussian_cKa} to a weaker ensemble converse bound. 
  \begin{Theorem}\label{Theorem_converse_noCSI_Gaussian_cKa} 
    Under the assumption that $\epsilon_{\mathrm{MD}} \leq \frac{M}{1+M}$ and the codebook has i.i.d. entries with means $0$ and variances $P$, the minimum required energy-per-bit for the URA model described in Section~\ref{section2} is lower-bounded as
    \begin{equation} \label{eq:P_tot_conv_EbN0_Gaussian}
      E^{*}_{b}(n, M, \epsilon_{\mathrm{MD}}, \epsilon_{\mathrm{FA}} )  \geq \inf_{P} \frac{nP}{J}, 
    \end{equation}
    under the constraints that $P>0$ and for any subset $S \subset \mathcal{K}$, there exists $\hat{K}_a \in [K]$ for $K_a \in S$ such that
    \begin{align}
    & \left( \mathbb{P}[ {\rm K}_a \in S] - \epsilon_{\rm MD} \right) J
    - h_2(\epsilon_{\rm MD})
    - \sum_{K_a \in S} P_{{\rm K}_a}(K_a) \log_2 \hat{K}_a  \notag\\
    & \leq \sum_{K_a \in S} P_{{\rm K}_a}(K_a) \left(
    \frac{ nL  \log_2 ( 1 + K_aP )}{K_a}
    - \frac{ L  \left( 1 - {\binom{K_a}{2}}/{M} \right) \mathbb{E}  \left[  \log_2 \left| \mathbf{I}_{n}  +  {\mathbf{X}}_{K_a}   {\mathbf{X}}_{K_a}^{ H}  \right|  \right]  }{K_a} \right)  , \label{P_tot_conv_noCSI}
    \end{align}
    and
    \begin{equation} \label{P_tot_conv_noCSI_FA_Fano}
      \sum_{K_a \in S}  P_{{\rm K}_a}(K_a) \max \left\{ \frac{\hat{K}_a-K_a}{\hat{K}_a} , 0 \right\} \leq \epsilon_{\mathrm{FA}} ,
    \end{equation}
    are satisfied, where $\hat{K}_a$ is a function of $K_a$ denoting the size of the decoded list when there are $K_a$ active users and ${\mathbf{X}}_{K_a} \in \mathbb{C}^{n\times K_{a}}$ has i.i.d. entries with means $0$ and variances $P$. 
  \begin{IEEEproof}
  See Appendix~\ref{Appendix_proof_converse_noCSI_Gaussian_noKa}.
  \end{IEEEproof}
  \end{Theorem}

  Theorem~\ref{Theorem_converse_single} and Theorem~\ref{Theorem_converse_noCSI_Gaussian_cKa} are obtained by extending the converse results in~\cite{letter}, where the number of active users is assumed to be fixed and known in advance and the decoded list size is equal to the number of active users, to the scenario with a random and known number of active users applying the following changes. First, since the optimal list size is uncertain, we use the false-alarm constraint to prevent the decoded list from being too large. Second, to adapt to the case with random ${\rm K}_a$, we take the expectation over ${\rm K}_a \in S \subset \mathcal{K}$. Choosing a subset $S$ instead of directly using $\mathcal{K}$ avoids the looseness coming from any user to affect the overall tightness of the converse bound, as well as being beneficial for reducing simulation complexity and fit for the case where the distribution of ${\rm K}_a$ is known only when ${\rm K}_a$ falls within a subset of $\mathcal{K}$.

  A converse bound for URA in Gaussian channels with a random and unknown number of active users was provided in~\cite[Theorem~3]{noKa}. This bound was derived by averaging over a random codebook ensemble and is only fit for the scheme where the BS first estimates the number of active users and then chooses the decoded list size from a given set around this estimate. To make it simpler, in this converse bound, the error coming from incorrect detection of the number of active users in the first stage is considered but it is assumed that there is no additional error in the decoding process. That is, this bound holds for a specific two-stage approach, and only the error occurring in the first stage is counted. In this work, we provide three converse bounds. They complement each other and collectively characterize the lower bound on the minimum required energy-per-bit for URA in MIMO quasi-static Rayleigh fading channels. Compared with~\cite[Theorem~3]{noKa}, all of our converse results deal with unknown instantaneous CSI, and we remove the restrictions on the codebook ensemble and specific decoding scheme in both Theorem~\ref{Theorem_converse_single} and Theorem~\ref{Theorem_converse_single_noKa}, and remove the restriction on the decoding scheme in Theorem~\ref{Theorem_converse_noCSI_Gaussian_cKa}.

  Since it is difficult to obtain a general converse bound when the number of active users is random and unknown, Theorem~\ref{Theorem_converse_single} and Theorem~\ref{Theorem_converse_noCSI_Gaussian_cKa} are derived assuming user activity is known to the receiver, although they are also converse in the case without this knowledge. Specifically, in Theorem~\ref{Theorem_converse_single}, we cast a URA code as a single-user code. This user is definitely active as long as ${\rm K}_a \geq 1$ as mentioned above. The multi-user converse bound in Theorem~\ref{Theorem_converse_noCSI_Gaussian_cKa} is established based on the knowledge of the number of active users and is an ensemble converse bound. This bound is not limited to a specific two-stage approach since there is no need to estimate the number of active users and we use the false-alarm constraint to control the decoded list size instead of using a specific decoding radius. It is challenging to derive a multi-user converse bound without exploiting the knowledge of the number of active users and the assumption on the codebook due to the following reasons. First, to obtain Theorem~\ref{Theorem_converse_noCSI_Gaussian_cKa}, we apply the standard Fano inequality for each active user assuming ${\rm K}_a$ is known in advance, where the Fano inequality provides a lower bound on the misdetection error probability. In the case without this knowledge, errors coming from activity detection and message misdetections for active users will be coupled, making it challenging to obtain the entropy and mutual information after applying the Fano inequality. Second, when instantaneous CSI is unknown to the receiver, the assumption of a specific codebook distribution makes the mutual information tractable. How to remove this restriction is still an open problem, even if the number of active users is known in advance.

  Different from Theorem~\ref{Theorem_converse_single} and Theorem~\ref{Theorem_converse_noCSI_Gaussian_cKa}, Theorem~\ref{Theorem_converse_single_noKa} takes the uncertainty in user activity into account but is only applicable in the scenario where ${\rm K}_a$ follows a binomial distribution. In this case, each user becomes active independently with probability $p_a$. Consequently, it is evident that the active probability of the remaining user conditioned on the activities of $K-1$ users is still $p_a$. However, if ${\rm K}_a$ follows other distributions, it is usually challenging to obtain this conditional probability.

  There are some differences between the converse bounds for the scenario where each user has an individual codebook~\cite{TIT} and those for URA given in this section. Specifically, a typical method to derive a converse bound is casting a multi-user access code as a single-user code. As in Theorem~\ref{Theorem_converse_single}, we assume the activities of all users and the transmitted messages and channel coefficients of $K_a-1$ active users are known \emph{a priori}. Denote the transmitted message of the remaining user as $W_1$ without loss of generality. In this way, the per-user probability of error for the scenario where each user has an individual codebook becomes $\mathbb{P} [ W_1 \neq \hat{W}_1 ]$, where $\hat{W}_1$ denotes the decoded message. In contrast, when all users share a common codebook, there may exist message collisions, and the number $B$ of different messages among the transmitted messages of $K_a-1$ active users satisfies $1 \leq B \leq K_a-1$. Therefore, the decoder aims to output another $K_a - B$ possible messages to recover $W_1$. Then, we loosen the list size from $K_a - B$ to $\hat{K}_a$ and the per-user probability of misdetection becomes $\mathbb{P} [ W_1 \notin \hat{\mathcal{W}} ]$. That is, the receiver is allowed to output more than one estimate, i.e., it performs list decoding in the URA scenario. It is evident that $\mathbb{P} [ W_1 \notin \hat{\mathcal{W}} ] \leq \mathbb{P} [ W_1 \neq \hat{W}_1 ]$. We can observe from the single-user converse bounds that compared to the scenario with individual codebooks, the error requirement is relaxed and the minimum required energy-per-bit is reduced when users share a common codebook.

  Moreover, the Fano inequality is commonly used when deriving a converse bound for massive access. As mentioned above, the misdetection and false-alarm probabilities are determined by the set difference between the set of transmitted messages and that of decoded messages when all users share a common codebook, whereas the aim is to determine whether the transmitted message matches its decoded counterpart for each user in the scenario with individual codebooks. This difference results in that the Fano inequality is applied differently when deriving multi-user converse bounds in the two scenarios, and the expressions for conditional entropies therein are distinct. As will be shown in Section~\ref{Section:simulation}, lower energy-per-bit is required in the URA paradigm compared to the individual codebook counterpart.

\subsection{Asymptotic Analysis} \label{Section:Asymptotic-Results}

  In this section, we consider an asymptotic regime where the blocklength $n\to\infty$ and the number of active users, the codebook size, the number of receive antennas, and the transmitting power scale with $n$. To indicate the relationship between these parameters and the blocklength $n$, we denote $K_a, {\rm K}_a, K, M, L$, and $P$ as $K_{a,n}, {\rm K}_{a,n}, K_n, M_n, L_n$, and $P_n$, respectively. We establish scaling laws in terms of these parameters building on the non-asymptotic results established in Section~\ref{Section:Non-Asymptotic-Results-achievability} and Section~\ref{Section:Non-Asymptotic-Results-converse}. In Theorem \ref{Theorem_scalinglaw_noCSI_Kaknown}, we provide conditions on the minimum $P_n$ and $L_n$ required to support a known number $K_{a,n}$ of active users, each reliably transmitting $\log_2M_n$ information bits to the BS under given error constraints. 
  \begin{Theorem} \label{Theorem_scalinglaw_noCSI_Kaknown}
    Assume that there are $K_{a,n}$ active users, which is known in advance, and the receiver is required to output a list of messages of size no more than $K_{a,n}$. In the case of $n\to \infty$, $P_n = \mathcal{O}\left( \frac{\ln n}{n^2} \right)$, and $M_n = \Theta \left( n^{c} \right)$ for any constant $c > 2$, to support $K_{a,n} = \mathcal{O} \left( \frac{n^2}{ \ln^2 n }  \right)$ active users each reliably transmitting $\log_2 M_n$ bits to the BS, the minimum required $P_n$ and $L_n$ satisfy $P^2_nL_n = \Theta\left( \frac{\ln n}{n^2} \right)$, regardless of whether the target error probabilities $\epsilon_{\rm {MD}}$ and $\epsilon_{\rm {FA}}$ are constants in $(0,1)$ or vanish.      
    \begin{IEEEproof}[Proof sketch]
  The data detection problem in the context of URA is essentially a sparse support recovery problem, in which the set $\mathcal{W}$ of transmitted messages is required to be recovered from the set $[M_n]$ but the size and multiplicity of $\mathcal{W}$ are unknown in advance since there may exist message collision. We require the receiver to output a list of messages of size no more than $K_{a,n}$ for simplicity. The achievability part is obtained by applying a common codebook with each codeword drawn uniformly i.i.d. from a sphere. Each active user independently selects a message from $[M_n]$. We change the measure to the new one under which the number of users who select the same message is no more than a constant $D$, and prove that the total variation distance between the true measure and the new one vanishes under the assumption of $M_n \geq K_{a,n}^{r_1}$ for a constant $r_1>1$. Under the new measure, the misdetection and false-alarm probabilities are upper-bounded via the application of the restricted isometry property result in \cite[Theorems~2 and~5]{Caire1}. Next, we prove that both the misdetection and false-alarm probabilities is bounded by a constant in $(0,1)$ or vanish in the asymptotic regime mentioned in Theorem~\ref{Theorem_scalinglaw_noCSI_Kaknown}. The converse part is obtained by performing asymptotic analysis based on the non-asymptotic single-user type converse bound in~\cite[Theorem 2]{letter}. The main tools used to establish the scaling law on the converse side are upper bounds for quantiles of the noncentral chi-square distribution~\cite{chi_bound}. A detailed proof is given in Appendix~\ref{Proof_scalinglaw_noCSI_Kaknown}. 
    \end{IEEEproof}
  \end{Theorem}

  It should be noted that we consider a sparse regime in Theorem~\ref{Theorem_scalinglaw_noCSI_Kaknown}, where the size of the common codebook is $M_n = \Theta \left( n^{c} \right)$ for any constant $c > 2$ and the number of active users scales as $K_{a,n} = \mathcal{O} \left( \frac{n^2}{ \ln^2 n }  \right)$. This regime is consistent with the logarithmic scaling regime introduced in~\cite{Caire1}, i.e. $M_n = \Theta\left(K_{a,n}^{\alpha}\right)$ for some constant $\alpha > 1$. In the considered regime, we have $\frac{K_{a,n}}{M_n}\to 0$. This is in line with practically relevant URA scenarios, where each active user typically transmits several hundred information bits. In this case, the codebook size is about $2^{100}$, which is far more than the number of active users.

  It was proved in~\cite{Caire1} that in the logarithmic regime, the error probability vanishes if the number of active users satisfies $K_{a,n} = \mathcal{O} \left( \frac{n^2}{ \ln^2 n }  \right)$. The contribution of Theorem~\ref{Theorem_scalinglaw_noCSI_Kaknown} lies in that we provide the scaling law on the number of receive antennas and transmitting power required to reliably support $K_{a,n} = \mathcal{O} \left( \frac{n^2}{ \ln^2 n }  \right)$ active users on both achievability and converse sides. From Theorem~\ref{Theorem_scalinglaw_noCSI_Kaknown}, we can observe that: 
  \begin{itemize}
    \item As the number of receive antennas increases, the minimum required transmitting power can be greatly reduced. This shows the significant potential of MIMO for enabling low-cost communication. Specifically, under the considered logarithmic regime, in the case of $L_n = \Theta\left( \frac{n^2}{\ln n} \right)$, the minimum energy-per-bit required to reliably support $K_{a,n} = \mathcal{O} \left( \frac{n^2}{\ln^2 n} \right)$ active users is on the order of $E_{b,n} = \Theta\left( \frac{1}{n} \right)$; and when the number of BS antennas increases to $L_n = \Theta\left( n^2 \right)$, the minimum required energy-per-bit can be reduced to $E_{b,n} = \Theta\left( \frac{1}{n\sqrt{\ln n}} \right)$.  
    \item In the considered logarithmic regime, it is possible to achieve a total spectral efficiency of $\frac{K_{a,n}\log_2 M_n}{n} = \mathcal{O} \left( \frac{n}{\ln n}\right)$ that grows without bound by employing larger blocks of dimension $n$, and minimize the error probability as small as desired. In contrast, in the regime where $M_n$ scales as $M_n = \delta n^2$ with some fixed $\delta\geq 1$, the number of reliably supported active users can be up to $K_{a,n} = \mathcal{O} \left( n^2 \right)$, and the corresponding sum spectral efficiency scales as $\frac{K_{a,n} \log_2 M_n}{n} = \mathcal{O} \left( n \ln n \right)$~\cite{Caire1}. As a result, when the codebook size decreases from $M_n = \Theta( n^c)$ with $c>2$ to $M_n = \Theta( n^2)$, the number of reliably supported active users and the sum spectral efficiency are increased by a logarithmic factor $\ln^2 n$ due to the reduced search space for sparse support recovery. Nevertheless, the more sparse regime considered in Theorem~\ref{Theorem_scalinglaw_noCSI_Kaknown} is more fit for practical URA scenarios as mentioned above.
    \item Theorem~\ref{Theorem_scalinglaw_noCSI_Kaknown} is proved from both achievability and converse sides. The achievability part is obtained by applying a multi-user type achievability bound, while the converse part is obtained based on the single-user type converse bound in~\cite[Theorem 2]{letter} (see also Lemma~\ref{Theorem_converse_single_Kaknown} in Appendix~\ref{Proof_scalinglaw_noCSI_Kaknown}). This converse bound is indeed very conservative because it is derived by casting a URA code as a single-user code. Nevertheless, it is enough to give the converse scaling law in the considered sparse regime. Since the single-user type converse scaling law matches the multi-user achievability part, we can draw the conclusion that when the number of active users increases from $\Theta(1)$ to the order of $K_{a,n} = \Theta \left( \frac{n^2}{ \ln^2 n }  \right)$, the minimum required $P_n$ and $L_n$ always satisfy $P_n^2L_n = \Theta\left( \frac{\ln n}{n^2} \right)$ under the conditions mentioned in Theorem~\ref{Theorem_scalinglaw_noCSI_Kaknown}. This means that it is possible for the minimum required $P_n$ and $L_n$ to remain on the same order when $K_{a,n}$ increases but stays below a threshold.
  \end{itemize}

  Next, we consider a URA scenario with a random and unknown number ${\rm K}_{a,n}$ of active users, and establish the scaling law in Theorem~\ref{Theorem_scalinglaw_noCSI}. 
  \begin{Theorem} \label{Theorem_scalinglaw_noCSI}
    Consider a URA scenario where there are ${\rm K}_{a,n}$ active users among $K_n$ potential users and ${\rm K}_{a,n}$ is random and unknown in advance.
    Assume that the variance of ${\rm K}_{a,n}$ satisfies $\frac{{\rm var}({\rm{K}}_{a,n})}{ \left( \frac{n^2}{\ln^2 n} \right)^{2} } = o(1)$ in the case of $K_n = \omega \left( \frac{n^2}{\ln^2 n} \right)$, and there is no requirement on ${\rm var}({\rm{K}}_{a,n})$ in the case of $K_n = \mathcal{O} \left( \frac{n^2}{\ln^2 n} \right)$. 
    When $n \to \infty$, $P_n = \mathcal{O}\left( \frac{\ln n}{n^2} \right)$, and $K_n^{r_1} \leq M_n = \Theta \left( n^{r_2} \right)$ with constants $r_1>1$ and $r_2> 2$, to support ${\rm{K}}_{a,n}$ active users each reliably transmitting $\log_2 M_n$ bits to the BS with the mean of ${\rm{K}}_{a,n}$ scaling as $ \mathbb{E} \left[{\rm{K}}_{a,n}\right] = \mathcal{O} \left( \frac{n^2}{ \ln^2 n }  \right)$, the minimum required $P_n$ and $L_n$ satisfy $P^2_nL_n = \Theta\left( \frac{\ln n}{n^2} \right)$, regardless of whether the target error probabilities $\epsilon_{\rm {MD}}$ and $\epsilon_{\rm {FA}}$ are constants in $(0,1)$ or vanish. 
    \begin{IEEEproof}[Proof sketch]
      Theorem~\ref{Theorem_scalinglaw_noCSI} considers a URA scenario with a random and unknown number of active users. To derive the achievability part, we first adopt a two-stage approach, where active users transmit a common sequence of length $n_0$ for the estimation of the number of active users in the first stage, and share a common codebook of length $n_1$ for data transmission in the second stage. Second, we change the measure to the new one under which: 1)~the number of users who select the same message is no more than a constant $D$; and 2)~the number of active users concentrates around its mean. The total variation distance is bounded applying Chebyshev's inequality and similar ideas used in Theorem~\ref{Theorem_scalinglaw_noCSI_Kaknown}. Third, under the new measure, we bound the error probability by the sum of $p_1$ and $p_2$. The term $p_1$ indicates the probability of the event that the estimated number of active users is larger than a threshold, which is bounded applying the asymptotic result in Theorem~\ref{Theorem_scalinglaw_Ka_estimation}. The term $p_2$ denotes the decoding error probability under the condition that the estimated number of active users is no more than a threshold, which is bounded along similar lines as in Theorem~\ref{Theorem_scalinglaw_noCSI_Kaknown} with the differences that the size of the decoded list varies in a range instead of being equal to $K_{a,n}$ and it is required to take the expectation over ${\rm K}_{a,n}$. The converse part is obtained by performing asymptotic analysis based on the non-asymptotic single-user type converse bound in Theorem~\ref{Theorem_converse_single}. A detailed proof is given in Appendix~\ref{Proof_scalinglaw_noCSI}.
    \end{IEEEproof}
  \end{Theorem}

  Under the assumptions in Theorem~\ref{Theorem_scalinglaw_noCSI}, the average spectral efficiency can be up to $\frac{\mathbb{E}[{\rm{K}}_{a,n}] \log_2 M_n}{n} = \mathcal{O} \left( \frac{n}{\ln n}\right)$. Similar to the case with a known number of active users, the minimum required energy-per-bit can be greatly reduced as $L_n$ increases in the case with random and unknown ${\rm{K}}_{a,n}$. Specifically, to reliably support ${\rm K}_{a,n}$ active users of mean $\mathbb{E} \left[{\rm{K}}_{a,n}\right] = \mathcal{O} \left( \frac{n^2}{ \ln^2 n }  \right)$, when $L_n$ increases from $\Theta\left( \frac{n^2}{\ln n} \right)$ to $\Theta\left( n^2 \right)$, the minimum required energy-per-bit can be reduced from $\Theta\left( \frac{1}{n} \right)$ to $\Theta\left( \frac{1}{n\sqrt{\ln n}} \right)$.

  Theorem~\ref{Theorem_scalinglaw_noCSI} indicates that when ${\rm K}_{a,n}$ is random and unknown and the number of potential users is $K_n = \mathcal{O} \left( \frac{n^2}{\ln^2 n} \right)$, in the same regime of $P_n$ and $L_n$ as introduced in the case with a known number of active users, one can reliably support ${\rm K}_{a,n}$ active users with mean $\mathbb{E} \left[{\rm{K}}_{a,n}\right] = \mathcal{O} \left( \frac{n^2}{ \ln^2 n }  \right)$. When $K_n$ becomes larger, i.e. $K_n = \omega \left( \frac{n^2}{\ln^2 n} \right)$, the mean of the number of active users can be also on the order of $\mathcal{O} \left( \frac{n^2}{ \ln^2 n }  \right)$ under the constraint $\frac{{\rm var}({\rm{K}}_{a,n})}{ \left( \frac{n^2}{\ln^2 n} \right)^{2} } = o(1)$. This is reasonable due to the following reasons. First, according to Chebyshev's inequality, the constraint $\frac{{\rm var}({\rm{K}}_{a,n})}{ \left( \frac{n^2}{\ln^2 n} \right)^{2} } = o(1)$ implies that ${\rm{K}}_{a,n}$ is on the order of $\mathcal{O} \left( \frac{n^2}{\ln^2 n} \right)$ with extremely high probability. Second, as shown in Theorem~\ref{Theorem_scalinglaw_Ka_estimation}, the number of active users is estimated to be in an interval around the true value with extremely high probability in the considered regime. Third, under the conditions given in Theorem~\ref{Theorem_scalinglaw_noCSI_Kaknown}, when the number of active users is known in advance, it is possible to reliably support $K_{a,n} = \mathcal{O} \left( \frac{n^2}{ \ln^2 n }  \right)$ active users.

\section{Extension to Joint Error Probability Constraint} \label{Section:joint_error}

  In this section, we consider the scenario with a fixed and known number $K_a$ of active users to compare the fundamental limits under the constraints on PUPE and classical joint error probability. In the following, we provide the notion of a URA code under the joint error probability constraint.  
    \begin{defi}\label{defi1_joint} 
    An $(n, M, \epsilon_{J}, P)$ URA code consists of 
    \begin{enumerate}
      \item
          An encoder $\emph{f}_{\text{en}}: [M] \mapsto \mathbb{C}^{n}$ that maps the message $W_k$ to a codeword $\mathbf{x}_{W_k}$ satisfying the maximal power constraint in \eqref{eq:power_constraint}. 
      \item
          A decoder $\emph{g}_{\text{de}}: \mathbb{C}^{n\times L} \mapsto \binom{[M]}{K_a}$ that satisfies the constraint on the joint error probability as follows:
          \begin{equation}\label{eq:joint_error}
            P_{e,J} = \mathbb{P}  \left[  \mathcal{W} \neq \hat{\mathcal{W}}  \right] \leq \epsilon_J,
          \end{equation} 
          where $\mathcal{W} $ denotes the set of distinct transmitted messages and $\hat{\mathcal{W}}$ denotes the set of decoded messages of size $K_a$. 
    \end{enumerate}
  \end{defi}

  The minimum energy-per-bit required for reliable URA under the joint error probability constraint is denoted as $E^{*}_{b,J}(n, M, \epsilon_{J} ) = \inf \left\{E_b: \exists (n, M, \epsilon_{J}, P) \text{ code} \; \right\}$.

  A non-asymptotic achievability bound on the minimum required energy-per-bit for the URA model under the joint error probability constraint introduced above is given in Theorem~\ref{Theorem_noCSI_achi_joint_fixed}.
  \begin{Theorem} \label{Theorem_noCSI_achi_joint_fixed}
    For URA in MIMO quasi-static Rayleigh fading channels with a fixed and known number of active users, there exists an $(n, M, \epsilon_{J}, P)$ URA code for which the minimum required energy-per-bit satisfies 
      \begin{equation} \label{eq_noCSI_EbN0_joint_fixed}
        E^{*}_{b,J}(n,M,\epsilon_J) \leq \inf \frac{n {P}}{J}.
      \end{equation}
    Here, the $\inf$ is taken over all $P > 0$ satisfying that
      \begin{equation} \label{eq_noCSI_PUPE_joint_fixed}
        \epsilon_J \geq \min_{0< P'< P} \left\{ p_0 + \sum_{t=1}^{K_a} p_{t} \right\},
      \end{equation}
    where 
    \begin{equation} \label{eq_noCSI_p0_joint_fixed}
      p_0 = \frac{\binom{K_a}{2}}{M} + p_{0,K_a},
    \end{equation} 
    \begin{equation} \label{eq_noCSI_pt_joint_fixed}
      p_t = \min_{ 0 \leq \omega \leq 1, 0 \leq \nu }
       \left\{ q_{1,t}\left(\omega,\nu\right)
       + q_{2,t}\left(\omega,\nu\right) \right\}  ,
    \end{equation}
    \begin{align}\label{eq_noCSI_q1t_joint_fixed}
      q_{1,t}  \left(\omega,\nu\right)
      = \binom{ K_a }{t}  \binom{M-K_a }{t}
      & \mathbb{E} \left[ 
      \min_{ {u\geq 0,r\geq 0, \lambda_{\min}\left(\mathbf{B}\right) > 0}}
      \exp \left\{  L rn\nu  \right\} \right. \notag \\
      & \!\! \cdot \exp \left\{  L  \left( (u-r) \ln \!\left|\mathbf{F}\right|
      -  u \ln \!\left| {\mathbf{F}' } \right|
      + r\omega  \ln \!\left| \mathbf{F}_{1} \right|
      - \ln \!\left| \mathbf{B} \right| \right)
      \right\} \!\bigg] ,
    \end{align}
    \begin{equation}
      q_{2,t}\!\left(\omega,\nu\right)
      =   \binom {K_a } {t}  \min_{ \delta\geq0 }
      \left\{ \frac{ \Gamma\left( nL, nL  \left( 1+\delta \right)\right)}{\Gamma\left( nL \right)} + \mathbb{E} \left[ \frac{\gamma\left( Lm,  c_{\delta} \right)}{\Gamma\left( Lm \right)} \right] \right\}
      , \label{eq_noCSI_q2t}
    \end{equation}
    \begin{equation} \label{eq_noCSI_cdelta_joint_fixed}
      c_{\delta} =  \frac{ L \left( n(1+\delta)(1-\omega) - \omega\ln\left|\mathbf{F}_1\right| + \ln\left|\mathbf{F}\right|  - n\nu \right) }
      {\omega \prod_{i=1}^{m}\lambda_i^{{1}/{m}}  },
    \end{equation}
    \begin{equation} \label{eq_noCSI_F_joint_fixed}
      \mathbf{F} = \mathbf{I}_n+ \mathbf{C}_{ \mathcal{W} } \mathbf{C}^H_{ \mathcal{W} } , 
    \end{equation}
    \begin{equation} \label{eq_noCSI_F1_joint_fixed}
       \mathbf{F}_1 = \mathbf{I}_n+ \mathbf{C}_{ \mathcal{W} \backslash \mathcal{W}_1 }\mathbf{C}^H_{ \mathcal{W} \backslash \mathcal{W}_1 },
    \end{equation}
    \begin{equation} \label{eq_noCSI_Fprime_joint_fixed}
       \mathbf{F}'  = \mathbf{I}_n+ \mathbf{C}_{ \mathcal{W} \backslash \mathcal{W}_1\cup \mathcal{W}_2 }  \mathbf{C}^H_{ \mathcal{W} \backslash \mathcal{W}_1\cup \mathcal{W}_2 },
    \end{equation}
    \begin{equation}  \label{eq_noCSI_B_joint_fixed}
      \mathbf{B} = (1-u+r) \mathbf{I}_n
      + u (\mathbf{F}')^{-1} \mathbf{F}
      - r \omega \mathbf{F}_{ 1}^{-1} \mathbf{F}.
    \end{equation} 
  Here, $p_{0,K_a}$ in \eqref{eq_noCSI_p0_joint_fixed} is defined in Corollary~\ref{Theorem_Ka_C}; $\lambda_1, \ldots, \lambda_n$ in \eqref{eq_noCSI_cdelta_joint_fixed} denote the eigenvalues of the matrix $\mathbf{F}_1^{-1} \mathbf{C}_{\mathcal{W}_1}\mathbf{C}^H_{\mathcal{W}_1}$ of rank $m=\min \left\{ n,t \right\}$ in decreasing order; for any subset $S \subset [M]$, the matrix $\mathbf{C}_S\in \mathbb{C}^{n\times |S|}$ denotes the concatenation of codewords in $\left\{ \mathbf{c}_i : i\in S \right\}$; and $\mathcal{W}$, $\mathcal{W}_1$, and $\mathcal{W}_2$ are given in Theorem~\ref{Theorem_achievability}.  
  \begin{IEEEproof}
     As introduced in Section~\ref{Section:Non-Asymptotic-Results-achievability}, we adopt the random coding scheme to generate a common codebook $\mathbf{C} = \left[ \mathbf{c}_{1}, \ldots, \mathbf{c}_{M} \right] \in \mathbb{C}^{n\times M}$ with each codeword  drawn i.i.d. according to $\mathcal{CN} \left(0,P'\mathbf{I}_{n}\right)$ or drawn uniformly i.i.d. from a sphere of radius $\sqrt{nP'}$. If user $k$ is active, it transmits $\mathbf{x}_{W_k} = \mathbf{c}_{W_k} 1 \left\{ \left\|\mathbf{c}_{W_k}\right\|_{2}^{2} \leq n P \right\}$, where the message $W_k$ is chosen uniformly at random from the set $[M]$. To upper-bound the error probability, we change the measure to the new one under which: 1)~there is no message collision, i.e. $W_i \neq W_j$ for $i\neq j$; and 2)~the active user~$k$ transmits $\mathbf{x}_{W_k} = \mathbf{c}_{W_k}$. The total variation distance between the true measure and the new one is bounded by $p_0$ in~\eqref{eq_noCSI_p0_joint_fixed}. Then, we can bound the joint error probability as
  \begin{equation} \label{eq_PUPE_upper_noCSI_joint_fixed}
    P_{e,J} \leq \sum_{t=1}^{K_a}  \mathbb{P} \left[ \mathcal{F}_t \right]_{\rm new} + p_0 , 
  \end{equation}
  where $\mathbb{P} \left[ \mathcal{F}_t \right]_{\rm new}$ denotes the probability of the event that there are exactly $t$ misdecoded messages under the new measure and is bounded following similar lines as in \cite[Theorem 1]{letter}. 
  \end{IEEEproof}
  \end{Theorem}
  
  The joint error probability can be upper-bounded as in \eqref{eq_PUPE_upper_noCSI_joint_fixed}; in contrast, the PUPE can be upper-bounded by $P_{e} \leq \sum_{t=1}^{K_a}  \frac{t}{K_a} \mathbb{P} \left[ \mathcal{F}_t \right]_{\rm new} + p_0$~\cite{letter}. As we can see, the scaling factor of the probability of existing $t$ misdecoded messages is $\frac{t}{K_a}$ for the PUPE but is equal to $1$ for the joint error probability. This indicates that the joint error probability is larger than the PUPE especially in a scenario where the number of wrongly detected messages is more likely to be far less than $K_a$.

  In Theorem \ref{Theorem_converse_noCSI_Gaussian_cKa_joint_fixed}, we provide a converse bound on the minimum required energy-per-bit subject to the joint error probability constraint. 
  \begin{Theorem}\label{Theorem_converse_noCSI_Gaussian_cKa_joint_fixed}  
    Assume that $ \epsilon_J \leq \frac{ \binom{M}{K_a}}{1+\binom{M}{K_a}} \frac{ M !  }{ M^{K_a} (M-K_a)! } $ and the codebook has i.i.d. entries with means $0$ and variances $P$. Applying the criterion of joint error probability, the minimum required energy-per-bit is lower-bounded as
    \begin{equation} \label{eq:P_tot_conv_EbN0_Gaussian_joint_fixed}
      E^{*}_{b,J}(n, M, \epsilon_{J} )  \geq \inf_{P} \frac{nP}{J}, 
    \end{equation} 
    under the constraints that $P>0$ and 
    \begin{align}
      & \left(1 - \frac{\epsilon_J M^{K_a} (M-K_a)! }{ M !  } \right)\log_2 \binom{M}{K_a} - h_2\left( \frac{\epsilon_J M^{K_a} (M-K_a)! }{ M !  } \right)   \notag\\
      & \leq nL  \log_2 ( 1 + K_aP )  - L   \mathbb{E}  \left[  \log_2 \left| \mathbf{I}_{n}  +  {\mathbf{X}}_{K_a}   {\mathbf{X}}_{K_a}^{ H}  \right|  \right]   , \label{P_tot_conv_noCSI_joint_fixed}
    \end{align} 
    where ${\mathbf{X}}_{K_a} \in \mathbb{C}^{n\times K_{a}}$ has i.i.d. entries with means $0$ and variances $P$. 
  \begin{IEEEproof} 
    See Appendix~\ref{Appendix_proof_converse_noCSI_Gaussian_noKa_joint_fixed}.  
  \end{IEEEproof}
  \end{Theorem}

  This converse bound is derived applying the Fano inequality. In a typical scenario where $M \gg K_a$, we have $\frac{M^{K_a} (M-K_a)! }{ M !  } \approx 1$ and $\binom{M}{K_a} \approx \left( \frac{M}{K_a} \right)^{K_a}$. In this case, the left-hand side~(LHS) of \eqref{P_tot_conv_noCSI_joint_fixed} is close to $ (1-\epsilon_J)K_a \left( J - \log_2 K_a \right) - h_2(\epsilon_J)$. In contrast, in the converse bound under the PUPE constraint, this term is reduced to $(1-\epsilon)K_a \left( J - \log_2 K_a \right) - K_a h_2(\epsilon)$ as shown in \cite[Theorem 3]{letter}. Moreover, the single-user type converse bound under the PUPE constraint given in \cite[Theorem 3]{letter} is also converse under the joint error probability constraint considering that the two error probabilities are the same in the scenario with only a single user.

\section{Numerical Results} \label{Section:simulation}


\subsection{Estimation of the Number of Active Users} \label{Section:simulation1}

\begin{figure}
	\centering
    \subfigure[]{\includegraphics[width=0.48\linewidth]{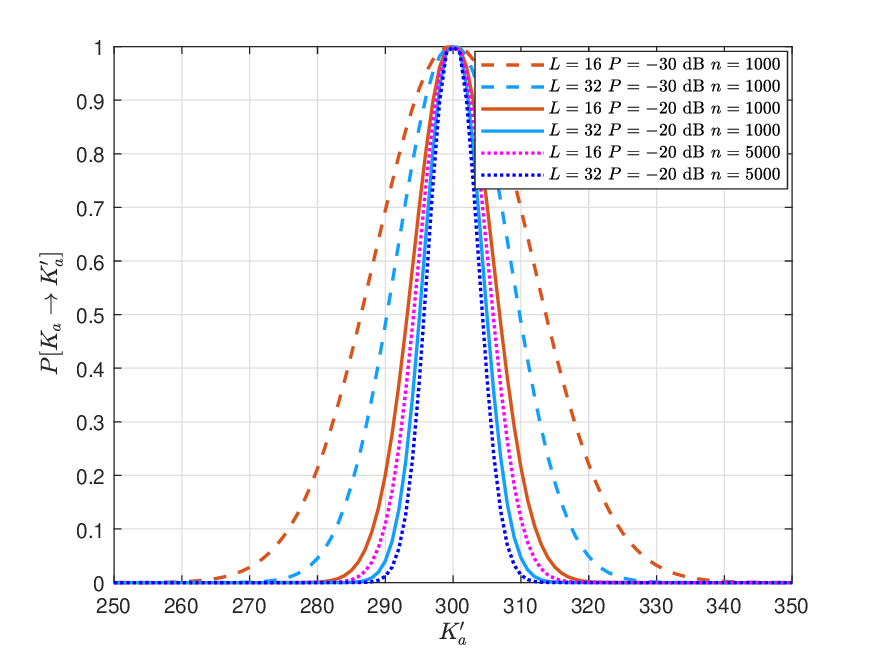}
		\label{fig:Ka_estimate}}
    \subfigure[]{\includegraphics[width=0.48\linewidth]{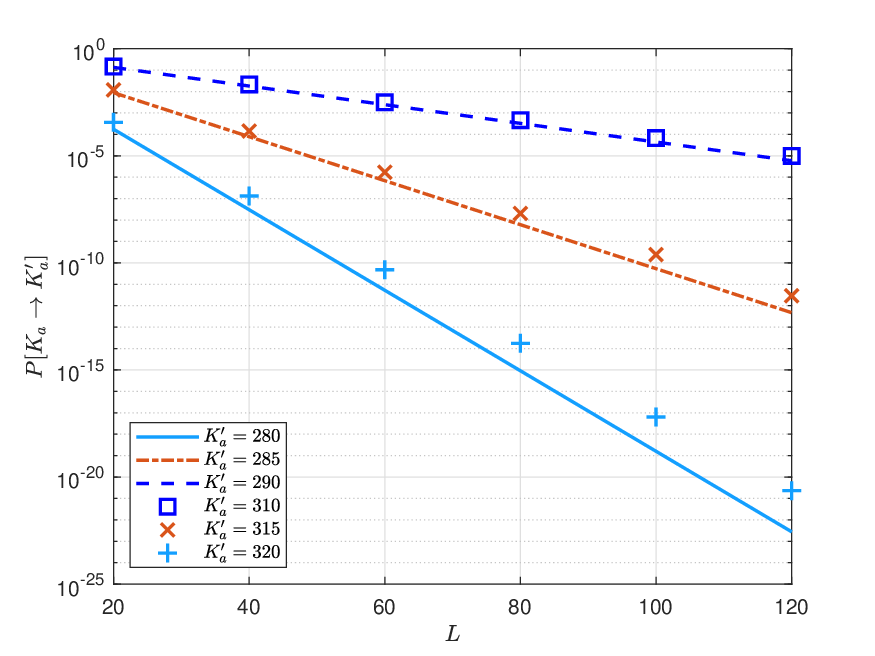}
		\label{fig:Ka_estimateL}}
    \subfigure[]{\includegraphics[width=0.48\linewidth]{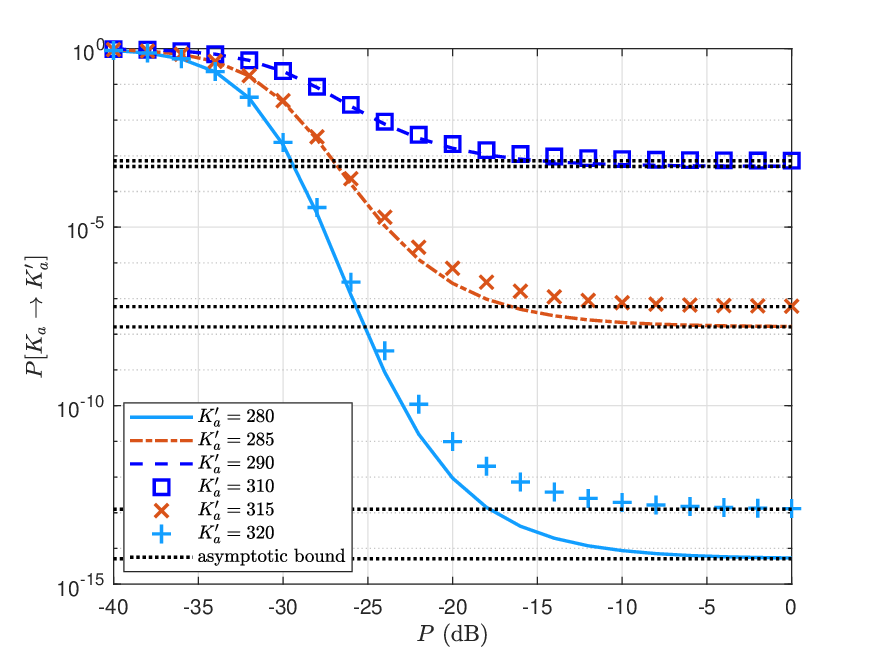}
		\label{fig:Ka_estimateP}}
	\subfigure[]{\includegraphics[width=0.48\linewidth]{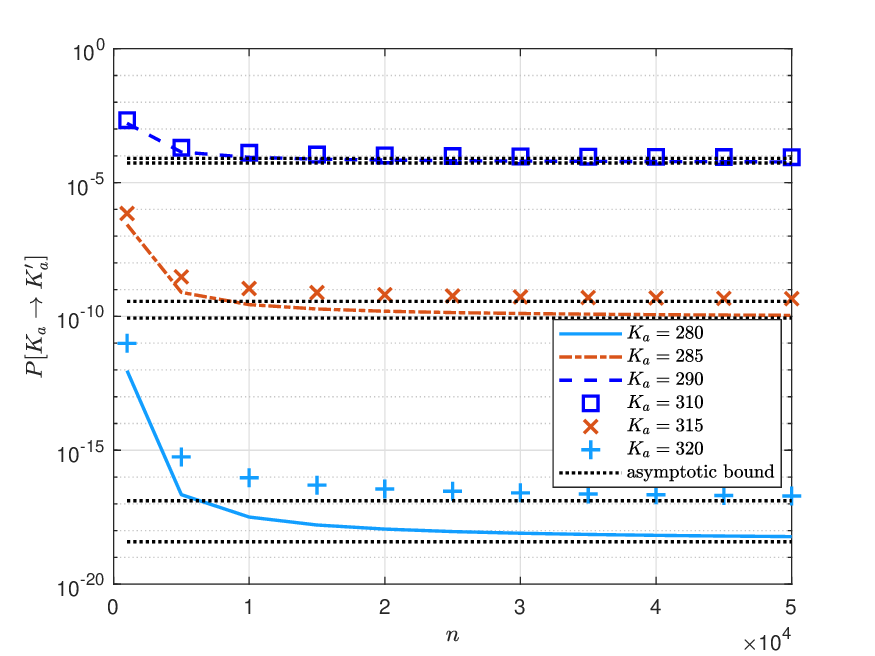}
		\label{fig:Ka_estimaten}}
	\caption{The achievability bound on $\mathbb{P}\left[ K_a \to K'_a \right]$ with $K_a=300$ and $K=600$:
(a)~$\mathbb{P}\left[ K_a \to K'_a \right]$ versus $K'_a$ with $L \in \{16,32\}$, $P \in \{-20,-30\}$~dB, and $n\in\{1000,5000\}$;
(b)~$\mathbb{P}\left[ K_a \to K'_a \right]$ versus $L$ for different values of $K'_a$ with $n=1000$ and $P=-20$~dB;
(c)~$\mathbb{P}\left[ K_a \to K'_a \right]$ versus $P$ for different values of $K'_a$ with $n=1000$ and~$L=64$;
(d)~$\mathbb{P}\left[ K_a \to K'_a \right]$ versus $n$ for different values of $K'_a$ with $L=64$ and $P=-20$~dB.
}\label{fig:Ka_estimate_all}
\end{figure}

  In Fig.~\ref{fig:Ka_estimate_all}, we assume there are exactly $K_a=300$ active users and present the upper bound~(in Theorem~\ref{Theorem_Ka}) on the error probability $\mathbb{P}\left[ K_a \to K'_a \right]$ of estimating $K_a$ exactly as $K'_a \in [K] \backslash \{ K_a \}$, where all users share a common codebook with codewords uniformly distributed on a sphere. In Fig.~\ref{fig:Ka_estimate}, we plot the upper bound on $\mathbb{P}\left[ K_a \to K'_a \right]$ versus $K'_a$. As we can see, as the number $L$ of receive antennas increases from $16$ to $32$, the transmitting power $P$ increases from $-30$~dB to $-20$~dB, or the blocklength increases from $1000$ to $5000$, the error probability decreases and the estimated value $K'_a$ becomes more concentrated around the true value $K_a=300$. From Fig.~\ref{fig:Ka_estimateL} we can observe that increasing the number $L$ of receive antennas leads to an exponential decrease in the error probability $\mathbb{P}\left[ K_a \to K'_a \right]$ for different values of $K'_a$, which is in line with the prediction in Section~\ref{Section:Ka2}. On the other hand, Fig.~\ref{fig:Ka_estimateP} shows that the upper bound on $\mathbb{P}\left[ K_a \to K'_a \right]$ initially drops at a certain transmitting power and then saturates to the error floor provided in Corollary~\ref{Theorem_Ka_asymP}. Likewise, Fig.~\ref{fig:Ka_estimaten} depicts that as the blocklength $n$ increases, the estimation error probability initially decreases but eventually approaches to the error floor provided in Corollary~\ref{Theorem_Ka_asymn}. The reasons for this behavior are explained in Section~\ref{Section:Ka2}. Moreover, it is shown that in the considered regime, the number $K_a$ of active users is more likely to be estimated as $K_a + \Delta K_a$ rather than $K_a -\Delta K_a$ for $\Delta K_a \in \{10,15,20\}$. One reason for this behavior may be the asymmetry of the probability density function of $\left\| \mathbf{Y} \right\|_F^2$.

\subsection{Data Detection Performance} \label{Section:simulation2}

  \begin{figure}
    \centering
    \includegraphics[width=0.7\linewidth]{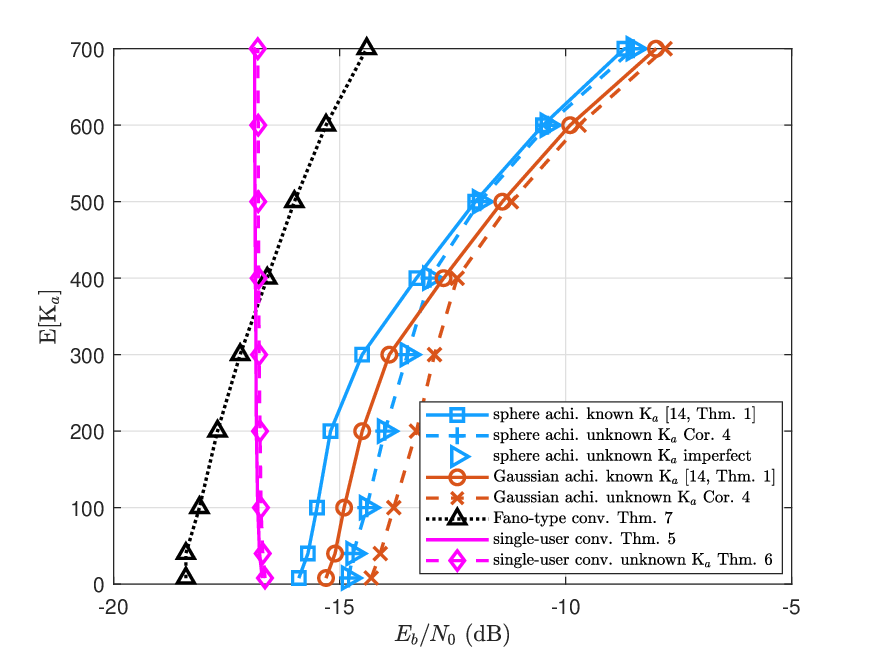}\\
    \caption{The mean of the number of active users versus $E_b/N_0$ with ${\rm K}_a \sim {\rm Binom}(K,0.5)$, $n=1000$, $J = 100$~bits, $L=128$, and $\epsilon_{\rm MD} = \epsilon_{\rm FA} = 0.001$.}
  \label{fig:randomKa_EbKa}
  \end{figure}
  In Fig.~\ref{fig:randomKa_EbKa}, we consider the scenario with $n\!=\!1000$, $J\!=\!100$~bits, $L \!=\! 128$, ${\rm K}_a \!\sim\! {\rm Binom}(K,0.5)$, and $\epsilon_{\rm MD} = \epsilon_{\rm FA} = 0.001$. We compare our achievability and converse bounds in terms of the minimum required $E_b/N_0$ for different values of $\mathbb{E}[{\rm K}_a]$, where $N_0$ denotes the noise power per complex degree of freedom and is equal to $1$ in our setting. We provide an explanation of how each curve is obtained in the following: 
  \begin{enumerate}
    \item The achievability bounds for the scenario where the number of active users is random and unknown but its distribution is known in advance are plotted based on Corollary~\ref{Theorem_achievability_energyperbit}. Two kinds of codebooks, namely the one drawn according to a Gaussian distribution and the one with codewords uniformly distributed on a sphere, are compared. We choose $K_l$ to be the largest value and $K_u$ the smallest value for which $\mathbb{P} \left[ {\rm K}_a \notin [K_l:K_u] \right] \leq 10^{-6}$ to simplify the simulation complexity. The decoding radius $r'$ is numerically selected from the set $\left\{1,2,\ldots,30\right\}$ to minimize the required $E_b/N_0$. The selected $r'$ varies for different values of $\mathbb{E}[{\rm K}_a]$ and error requirements. Generally, as explained in \cite{noKa}, the selected $r'$ is progressively increased to support a larger number of active users and meet stricter error requirements. 
        
    \item To characterize the impact of a mismatch between the assumed imperfect distribution of ${\rm K}_a$ and the actual one, we consider the scenario where the receiver assumes the probabilities of having different numbers of active users are the same. The achievability bound is obtained along similar lines as introduced in Section~\ref{Section:Non-Asymptotic-Results-achievability}.

    \item The achievability bounds for the case with a random and known number of active users are plotted applying Theorem~1 in~\cite{letter} followed by averaging over a binomial distribution of ${\rm K}_a$. Two kinds of codebooks are compared along similar lines as in Corollary~\ref{Theorem_achievability_energyperbit}.

    \item The single-user and Fano-type converse bounds correspond to Theorem~\ref{Theorem_converse_single} and Theorem~\ref{Theorem_converse_noCSI_Gaussian_cKa}, respectively, which are converse regardless of whether ${\rm K}_a$ is known or not since the optimal list size is uncertain in both cases. The Fano-type converse bound is plotted under the assumption of Gaussian codebook. The single-user converse bound for the case with random and unknown ${\rm K}_a$ is plotted based on Theorem~\ref{Theorem_converse_single_noKa}. The tail probabilities of chi-squared distribution in \eqref{eq:P_tot_conv_singleUE1} and \eqref{eqR:beta_conv_AWGN_singleUE_1e} can be lower-bounded applying \cite[Lemma 1]{chi_bound_L}. 
  \end{enumerate}
  Numerical experiments indicate that the single-user converse bound is dominant in the small $\mathbb{E}[{\rm K}_a]$ regime, while the Fano-type ensemble converse bound is dominant in the large $\mathbb{E}[{\rm K}_a]$ regime. This is because the single-user bound relies on knowledge of the transmitted messages of ${\rm K}_a-1$ active users and their instantaneous CSI, which becomes an overoptimistic assumption in the large $\mathbb{E}[{\rm K}_a]$ regime since MUI is a significant bottleneck in this case. The finite-blocklength achievability bounds show that using codewords distributed on a sphere outperforms Gaussian random coding by about $0.6$~dB in all $\mathbb{E}[{\rm K}_a]$ regimes regardless of whether ${\rm K}_a$ is known or not, which is consistent with the prediction in Section~\ref{Section:Non-Asymptotic-Results-achievability}. On the achievability side, the extra required $E_b/N_0$ due to the uncertainty in the exact number of active users is about $1.1$~dB when $\mathbb{E}[{\rm K}_a] \leq 300$. However, in the large $\mathbb{E}[{\rm K}_a]$ regime, this gap reduces to be less than $0.3$~dB considering that MUI, rather than the lack of knowledge of ${\rm K}_a$, becomes the bottleneck. The gap between our single-user converse bounds with and without known ${\rm K}_a$ is slight since they are derived by casting a URA code as a single-user code, resulting in that only one user's activity is unknown. This gap can be more significant in the scenario with stricter error requirements, which will be shown in Fig.~\ref{fig:PeEb}. Similar to the case with a fixed and known number of active users~\cite{letter}, we can observe that in the considered MIMO fading channels with random and unknown ${\rm K}_a$, only a small increase in $E_b/N_0$ is needed as $\mathbb{E}[{\rm K}_a]$ increases, provided that it remains below a certain threshold. Numerical results confirm the tightness of our bounds especially when $\mathbb{E}[{\rm K}_a]\leq 400$, with the gap between our achievability (using spherically distributed codewords) and converse bounds less than $4$~dB in this case. Moreover, the achievability bounds in the scenarios with and without perfect knowledge of the distribution of the number of active users are close. As explained in Section~\ref{Section:Non-Asymptotic-Results-achievability}, this is because in the considered regime, $\ln \binom{M}{K_a }$ is significantly larger than $-  \ln  \left(  \bar{P}_{{\rm{K}}_a} ( K_a )  \right)$ and $-  \ln  \left(  P_{{\rm{K}}_a} ( K_a )  \right)$ for $K_a \in [K_l:K_u]$.

  \begin{figure}
    \centering
    \includegraphics[width=0.7\linewidth]{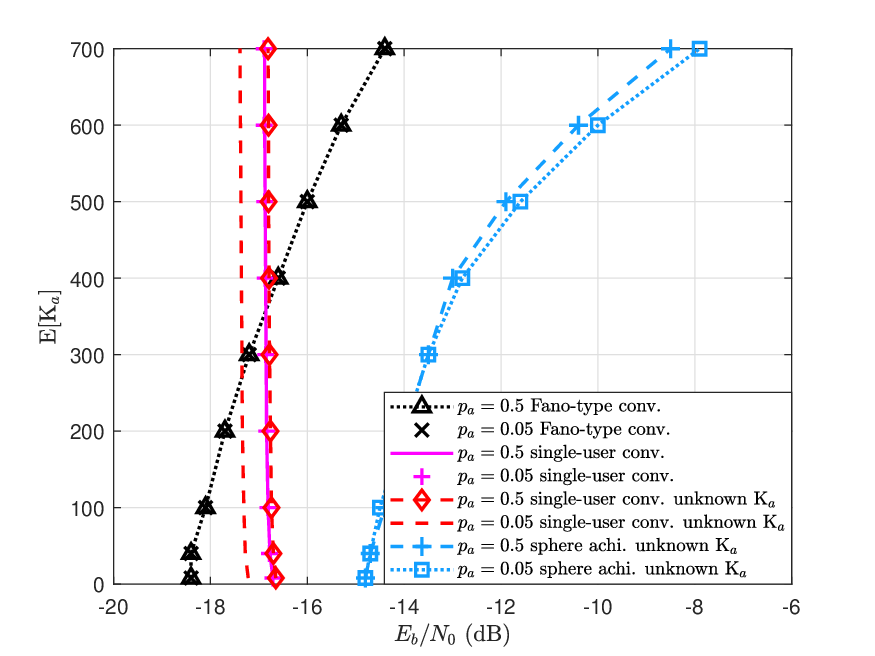}\\
    \caption{The mean of the number of active users versus $E_b/N_0$ in the setting where ${\rm K}_a$ follows the binomial distribution with active probability $p_a\in\{0.5,0.05\}$, $n=1000$, $J = 100$~bits, $L=128$, and $\epsilon_{\rm MD} = \epsilon_{\rm FA} = 0.001$.}
  \label{fig:randomKa_EbKa005}
  \end{figure}
  In Fig.~\ref{fig:randomKa_EbKa005}, we compare the derived non-asymptotic bounds with known active probability $p_a \in \{0.5 , 0.05\}$, which are plotted in a similar way to those in Fig.~\ref{fig:randomKa_EbKa}. For fixed $\mathbb{E}\left[ {\rm K}_a \right]$, the number $K$ of potential users is larger when $p_a = 0.05$ than when $p_a = 0.5$. On the converse side, the single-user bound with unknown ${\rm K}_a$ (Theorem~\ref{Theorem_converse_single_noKa}) is tighter than that with known ${\rm K}_a$ (Theorem~\ref{Theorem_converse_single}) in the case of $p_a = 0.5$, whereas Theorem~\ref{Theorem_converse_single} dominates in the case of $p_a = 0.05$. This is because the mean of the number of active users under consideration is only $p_a$ under the assumption in Theorem~\ref{Theorem_converse_single_noKa} but is $1$ as long as ${\rm K}_a \geq 1$ under the assumption in Theorem~\ref{Theorem_converse_single}, as explained in Section~\ref{Section:Non-Asymptotic-Results-converse}. Thus, Theorem~\ref{Theorem_converse_single_noKa} can be looser than Theorem~\ref{Theorem_converse_single} when $p_a$ is small, even though Theorem~\ref{Theorem_converse_single_noKa} takes the unknown user activity into consideration. For fixed $\mathbb{E}\left[ {\rm K}_a \right]$, the achievability results show that higher $E_b/N_0$ is required as $p_a$ decreases from $0.5$ to $0.05$. The reasons are as follows. First, for fixed $\mathbb{E}\left[ {\rm K}_a \right]$, the variance of ${\rm K}_a$, given by $Kp_a(1-p_a) = \mathbb{E}\left[ {\rm K}_a \right] (1-p_a)$, increases as $p_a$ decreases, thereby increasing the uncertainty in the number of active users. Second, as can be seen from the achievability result, the minimum required $E_b/N_0$ increases significantly with an increase in the number of active users when it is above a threshold. Compared to the case with $p_a = 0.5$, ${\rm K}_a$ is more likely to be larger in the case of $p_a = 0.05$ and higher $E_b/N_0$ is required to support a possibly larger number of active users in this case.

  \begin{figure}
    \centering
    \includegraphics[width=0.7\linewidth]{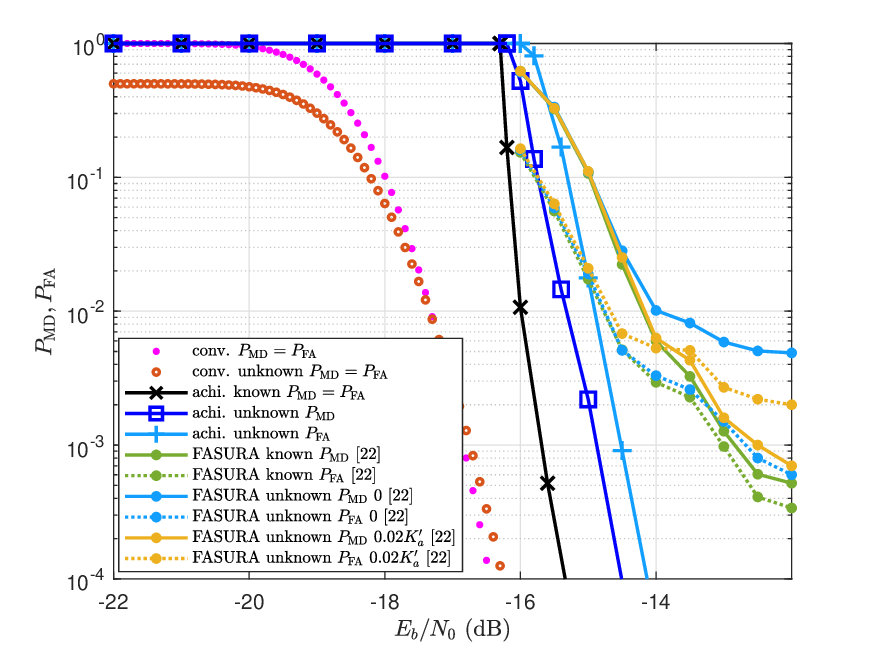}\\
    \caption{Per-user probabilities of misdetection and false-alarm versus $E_b/N_0$ with ${\rm K}_a \sim {\rm Binom}(200,0.5)$, $n=1000$, $J =100$~bits, and $L = 128$.}
  \label{fig:PeEb}
  \end{figure}

  In Fig.~\ref{fig:PeEb}, we compare non-asymptotic achievability and converse bounds on $P_{\rm MD}$ and $P_{\rm FA}$, as well as the performance of the FASURA scheme proposed in~\cite{Fasura}, as a function of $E_b/N_0$ for the setting with ${\rm K}_a\sim {\rm Binom}(200,0.5)$, $n=1000$, $J = 100$~bits, and $L=128$. The achievability bound for the scenario with random and unknown ${\rm K}_a$ is plotted based on Theorem~\ref{Theorem_achievability} with codewords distributed on a sphere, where the decoding radius is chosen from $[10:18]$. When ${\rm K}_a$ is random and known in advance, the achievability bound is plotted along similar lines as in Fig.~\ref{fig:randomKa_EbKa}. In this case, we have $P_{\rm MD}=P_{\rm FA}$ since the size of the decoded list is set to be ${\rm K}_a$ and the message collision probability is negligible. The single-user converse bounds correspond to Theorem~\ref{Theorem_converse_single} and Theorem~\ref{Theorem_converse_single_noKa}, which are plotted by allowing $P_{\rm MD}=P_{\rm FA}$. Although Theorem~\ref{Theorem_converse_single_noKa} takes the unknown user activity into consideration, it is not always tighter than Theorem~\ref{Theorem_converse_single} as explained above. From the achievability results we can observe that false-alarms are more likely to occur than misdetections in the considered regime. This may be due to an overestimation of the number of active users as explained in Section~\ref{Section:simulation1} and the use of relaxed upper bounds. For instance, as shown in \eqref{eq_noCSI_noKa_epsilonMD} and \eqref{eq_noCSI_noKa_epsilonFA}, the summation over $t'$ is within the minimum function for the misdetection probability but outside the minimum function for the false-alarm probability, resulting in that the upper bound on the false-alarm probability is looser than that of the misdetection probability. Our achievability and converse bounds are tight especially in the scenario with strict error requirements. It is shown that the extra required $E_b/N_0$ due to the uncertainty in user activity increases as the error requirement becomes stricter, which is in line with the observation for URA in Gaussian channels in~\cite{noKa}. This suggests that existing results established under the assumption of knowing the number of active users may provide an overoptimistic assessment of URA. The absence of this knowledge should be taken into account and our bounds serve as benchmarks for URA in this case.

  The FASURA scheme with NOPICE proposed in~\cite{Fasura} is evaluated in Fig.~\ref{fig:PeEb}. In this scheme, the message is divided into two parts with $16$~bits and $84$~bits, respectively. The first part features ``pilot'' of length $488$ and the second part is of length $512$ with spreading sequence of length $2$. ``NOPICE'' means apart from the pilot part, the second part is also used to enhance channel estimation performance. An iterative receiver is employed. When the number $K_a$ of active users is fixed and known in advance, the iteration process continues until the output list contains $K_a$ messages or there is no improvement between two consecutive rounds of iterations. To adapt this scheme to the case with a random and known number of active users, we average the error probability over the binomial distribution of ${\rm K}_a$. To adapt to the case with a random and unknown number of active users, we first obtain the estimate $K'_a$ of the number of active users in the case of ${\rm K}_a = K_a$ applying an energy-based detector, and then decode with the iteration process continuing until the output list contains $K'_a+\left[\Delta\right]_{N}$ messages or there is no improvement between two consecutive rounds. We set $\Delta \in \left\{ 0,0.02 K'_a \right\}$ in Fig.~\ref{fig:PeEb} and use $\left[\Delta\right]_{N}$ to denote the integer closest to $\Delta$. As we can see, the misdetection probability of this scheme can be larger than the false-alarm probability when ${\rm K}_a$ is known since the decoded list size might be smaller than the number of active users based on the stopping criterion mentioned above. In both cases with and without known ${\rm K}_a$, the gap between the FASURA scheme and our achievability bound increases as the error requirement becomes more stringent. When the number of active users is unknown in advance, by adjusting $\Delta$, we can adjust the size of the decoded list, thereby making a tradeoff between misdetection and false-alarm probabilities. Nevertheless, the performance of the FASURA scheme with known ${\rm K}_a$ is close to that with unknown ${\rm K}_a$ in the scenario with mild error constraints. However, as the target error probability reduces, a significantly higher $E_b/N_0$ is required when ${\rm K}_a$ is unknown compared to the case with known ${\rm K}_a$. This suggests that under mild error constraints, it is energy-efficient to adapt the practical scheme proposed for the case with known ${\rm K}_a$ to the case with unknown ${\rm K}_a$ by simply incorporating an estimation phase as mentioned above, but this adaption is inefficient in scenarios with stringent error requirements, which is in line with the observation in \cite{noKa}, calling for more advanced methods to address the challenge posed by the uncertainty in the number of active users.

\begin{figure}
	\centering
    \subfigure[]{\includegraphics[width=0.48\linewidth]{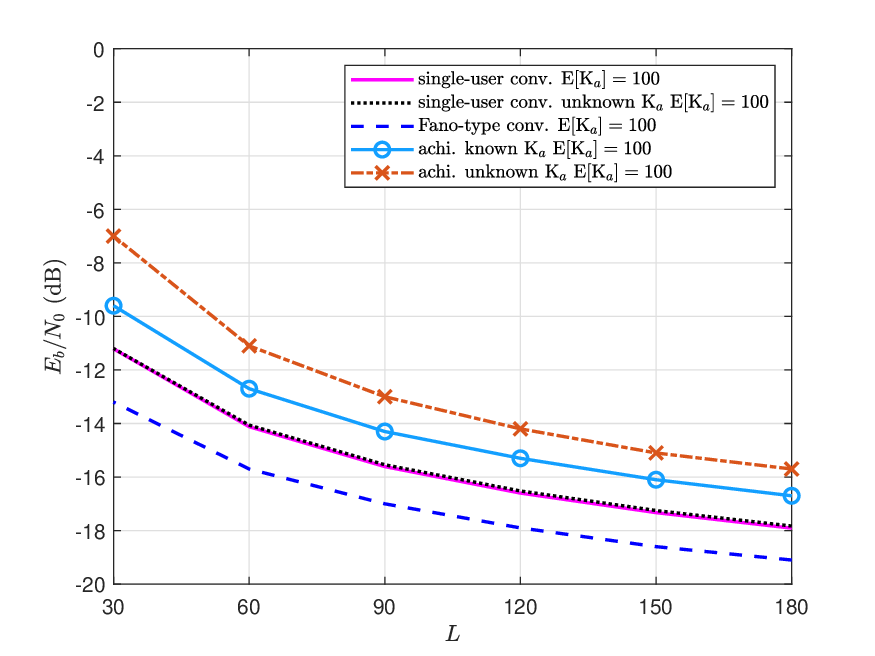}
		\label{fig:EbL1}}
    \subfigure[]{\includegraphics[width=0.48\linewidth]{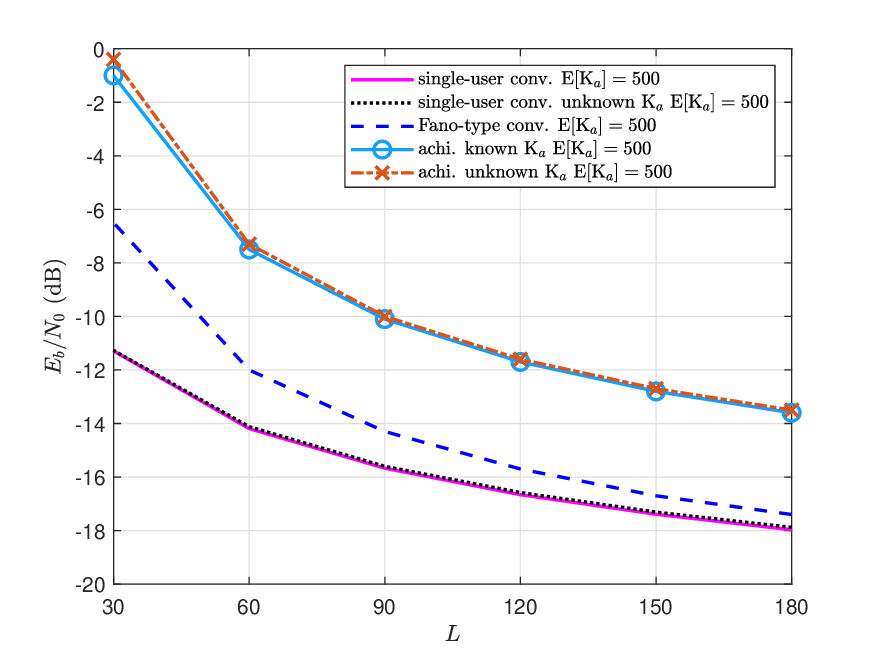}
		\label{fig:EbL2}}
	\caption{The required $E_b/N_0$ versus the number of BS antennas with $n=1000$, $J = 100$~bits, and $\epsilon_{\rm MD} = \epsilon_{\rm FA} = 0.001$:
(a)~${\rm K}_a\sim{\rm Binom}(200,0.5)$; (b)~${\rm K}_a\sim{\rm Binom}(1000,0.5)$.
}\label{fig:EbL}
\end{figure}

In Fig.~\ref{fig:EbL}, we compare the achievability and converse bounds in terms of the required $E_b/N_0$ for different values of $L$ in the setting with $n=1000$, $J = 100$~bits, $\epsilon_{\rm MD} = \epsilon_{\rm FA} = 0.001$, and $\mathbb{E}[{\rm K}_a] \in \{ 100, 500 \}$. The bounds are computed in similar ways to those in Fig.~\ref{fig:randomKa_EbKa}. We observe again that the single-user converse bound dominates when $\mathbb{E}[{\rm K}_a]=100$, while the Fano-type converse bound prevails when $\mathbb{E}[{\rm K}_a]=500$. The gap between our achievability bounds with and without known number of active users narrows as the number of BS antennas grows larger since estimation performance of the number of active users is improved at the same time. In contrast to the case with small $\mathbb{E}[{\rm K}_a]$, the extra required $E_b/N_0$ on the achievability side due to the uncertainty in the exact value of ${\rm K}_a$ is reduced in the large $\mathbb{E}[{\rm K}_a]$ regime, attributing to the fact that MUI instead of the lack of knowledge of ${\rm K}_a$ is the bottleneck in this case. This finding is in agreement with the result presented in Fig.~\ref{fig:randomKa_EbKa}. Moreover, we observe that in both cases of $\mathbb{E}[{\rm K}_a]=100$ and $\mathbb{E}[{\rm K}_a]=500$, the minimum required $E_b/N_0$ decreases as the number of BS antennas increases, highlighting the potential of large antenna arrays for enabling low-cost communication.

  \begin{figure}
    \centering
    \includegraphics[width=0.7\linewidth]{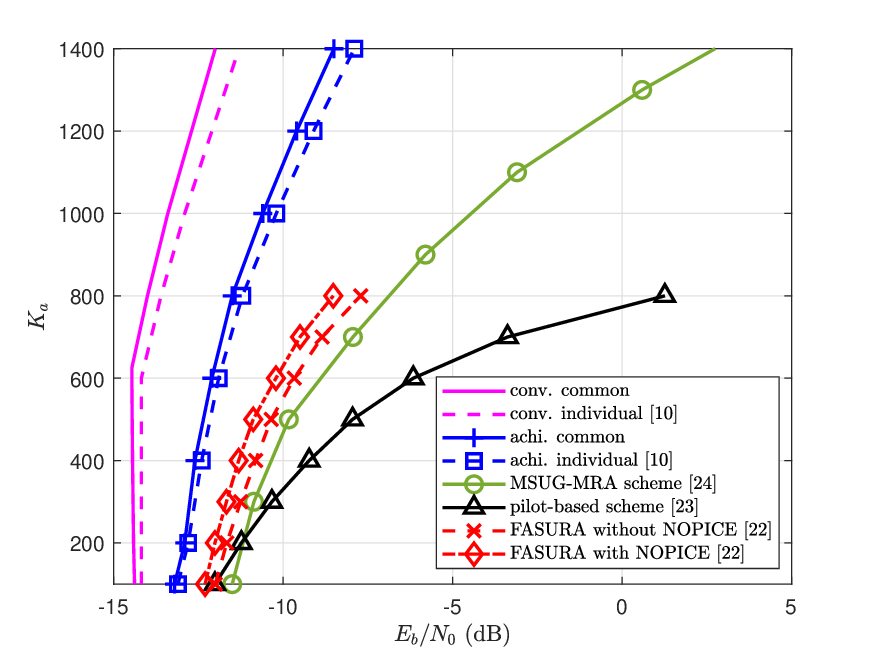}\\
    \caption{Comparison of existing schemes and theoretical bounds with individual codebooks and a common codebook under the assumption that $K_a$ is fixed and known in the case of $n=3200$, $J = 100$~bits, $L=50$, $\epsilon_{\rm MD} = \epsilon_{\rm FA} = 0.025$, and $K=K_a/0.4$ for the scenario with individual codebooks.}
  \label{fig:EbL50}
  \end{figure}
  In Fig.~\ref{fig:EbL50}, we consider the scenario with a fixed and known number $K_a$ of active users. We compare theoretical bounds with individual codebooks and a common codebook, as well as the schemes proposed in~\cite{Caire2,Fasura,Duman2}, in the setting with $n=3200$, $J=100$~bits, $L=50$, and $\epsilon_{\rm MD} = \epsilon_{\rm FA} = 0.025$. We provide an explanation of how each curve is obtained as follows:
  \begin{itemize}
    \item The achievability and converse bounds for the scenario where all users share a common codebook are plotted applying similar lines as in Fig.~\ref{fig:randomKa_EbKa}. The codebook with each codeword uniformly distributed on a sphere is adopted in the achievability bound.
    \item The achievability bound for the scenario where each user has an individual codebook is plotted by applying Corollary~7 in~\cite{TIT} and changing the Gaussian codebook therein to the one with each codeword uniformly distributed on a sphere. The converse bound with individual codebooks is plotted based on Theorem~9 in~\cite{TIT}. 
    \item The pilot-based scheme is evaluated as in~\cite[Fig.~7]{Caire2}. The data is split into two parts with $16$~bits and $84$~bits, respectively. The first part is coded as ``pilot'' of length $1152$ and the second one is coded by a polar code of length $2048$.
    \item The MSUG-MRA scheme is evaluated as in~\cite[Fig.~4]{Duman2}, where the blocklength $n$ is divided into $S$ slots and each active user randomly selects a single slot to transmit $J$ bits of information. The $J$-bit message is divided into $D$ orthogonal pilot parts each of length $B_p$~bits and one coded part of length $J-DB_p$~bits. There are $G$ groups, each being assigned unique interleaver and power pair. 
    \item The FASURA schemes with and without NOPICE are evaluated as in~\cite[Fig.~4]{Fasura}, where the message is divided into two parts with $16$~bits and $84$~bits, respectively. The first part features ``pilot'' of length $896$ and the second part is of length $2304$ with the spreading sequence of length $9$. 
  \end{itemize}
  Compared to the scenario with individual codebooks, errors in user identity recovery are not counted and the search space for decoding is reduced in the URA paradigm, leading to a decrease in the minimum required $E_b/N_0$ for URA. The saved $E_b/N_0$ becomes more obvious as $K_a$ increases. Moreover, it is shown that existing schemes are energy-efficient when $K_a$ is small, but exhibit a large gap to our bounds in the large $K_a$ regime. That is, as $K_a$ increases, these schemes suffer from more performance degradation and require higher $E_b/N_0$ compared with our achievability bound. One reason for this observation is that our achievability bound is derived in a non-coherent way, where the transmitted messages are detected by estimating statistical information. However, there is a pilot transmission phase before data transmission in the practical schemes proposed in~\cite{Caire2,Fasura,Duman2}. That is, these schemes attempt to accomplish a more difficult task, i.e., instantaneous CSI estimation. When $K_a$ is extremely large, it is difficult to obtain precise estimates using limited channel uses, thereby deteriorating the decoding performance. This calls for more advanced methods that may not require a specific pilot transmission stage to explicitly estimate instantaneous CSI but perform energy-efficiently in the case with a large number of active users.

  \begin{figure}
    \centering
    \includegraphics[width=0.7\linewidth]{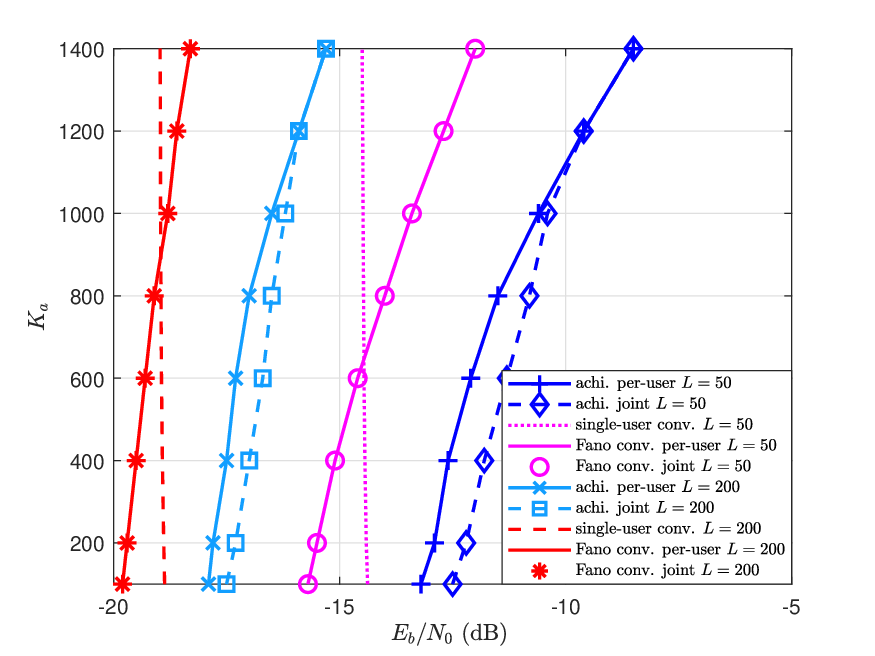}\\
    \caption{Comparison of theoretical bounds under the constraints on per-user probability of error and joint error probability assuming that $K_a$ is fixed and known in the case of $n=3200$, $J = 100$~bits, $L\in\{50,200\}$, $\epsilon_{\rm MD} = \epsilon_{\rm FA} = 0.025$ for per-user probability of error, and $\epsilon_{J} = 0.025$ for joint error probability.}
  \label{fig:EbL50L200}
  \end{figure} 
  In Fig.~\ref{fig:EbL50L200}, we consider the scenario with a fixed and known number $K_a$ of active users. We compare theoretical bounds for URA with $n=3200$, $J=100$~bits, and $L\in\{50,200\}$. Both the joint error probability constraint with target error probability $\epsilon_J = 0.025$ and the PUPE constraint with $\epsilon_{\rm MD} = \epsilon_{\rm FA} = 0.025$ are considered. The achievability bound, single-user converse bound, and Fano-type converse bound under the constraint on PUPE are plotted in a similar way to those in Fig.~\ref{fig:EbL50}. Under the joint error probability constraint, the achievability and Fano-type converse bounds are plotted based on Theorem~\ref{Theorem_noCSI_achi_joint_fixed} and Theorem~\ref{Theorem_converse_noCSI_Gaussian_cKa_joint_fixed}, respectively. The single-user converse bound subject to the PUPE constraint is also converse under the joint error probability constraint. We observe that increasing the number $L$ of BS antennas substantially decreases the minimum $E_b/N_0$ required to reliably support $K_a$ active users, and increases the threshold of the number of active users below which only a small increase in $E_b/N_0$ is needed as $K_a$ increases. Moreover, the Fano-type converse bounds under the PUPE constraint and the joint error probability constraint are close in the considered regime. On the achievability side, when $K_a$ is small, higher $E_b/N_0$ is required under the joint error probability constraint than under the PUPE constraint. This gap narrows as $K_a$ increases. This is because the number of wrongly detected messages is more likely to be close to $1$ when $K_a$ is small, whereas the probability of existing nearly $K_a$ misdetections dominates when $K_a$ is large, as observed in \cite{A_perspective_on}. Considering that PUPE introduces a scaling factor $\frac{t}{K_a}$ for the error term corresponding to $t$ out of $K_a$ messages decoded incorrectly, while the scaling factor remains $1$ for the joint error probability, the joint error probability can be obviously larger than the PUPE when the number of incorrectly decoded messages is more likely to be substantially less than $K_a$, leading to an increase in the minimum required $E_b/N_0$ under the joint error probability constraint than under the PUPE constraint in the case with small $K_a$.

  \begin{figure}
    \centering
    \includegraphics[width=0.7\linewidth]{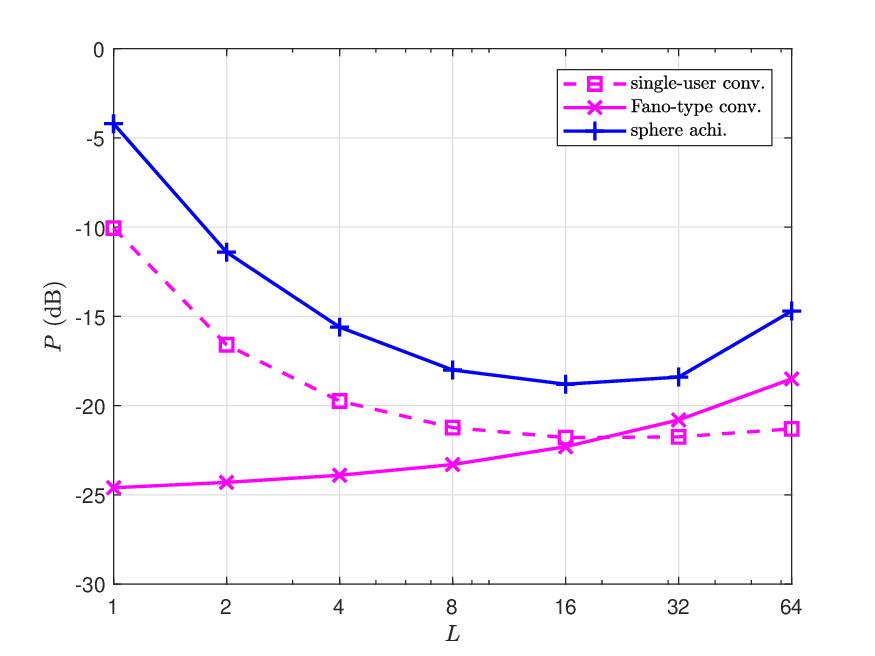}\\
    \caption{The minimum required transmitting power $P$ versus the number $L$ of BS antennas 
    in the setting with $K_a = 200$ (fixed and known), $J = 100$~bits, $n' = nL = 25600$, and $\epsilon_{\rm MD} = \epsilon_{\rm FA} = 0.025$.}
  \label{fig:nprime}
  \end{figure}
  
  In Fig.~\ref{fig:nprime}, we compare the minimum required transmitting power $P$ versus the number $L$ of BS antennas for fixed $n'=nL$. We assume $K_a = 200$, $J = 100$~bits, $n' = 25600$, and $\epsilon_{\rm MD} = \epsilon_{\rm FA} = 0.025$. As can be seen, the single-user converse bound dominates when $L$ is small (or equivalently, $n$ is large), while the Fano-type converse bound dominates otherwise. For a fixed $n'$, the dominating converse bound and the achievability bound indicate that as $L$ increases, the minimum required transmitting power first decreases and then increases. This is because when $L$ is small, adding the number of receive antennas is beneficial for avoiding the channels of some users to be in outage; however, when $L$ is large, the bottleneck becomes the limited blocklength for carrying information.

\section{Conclusion} \label{Section:conclusion}

In this paper, we have shed light on the fundamental limits of URA over MIMO quasi-static Rayleigh-fading channels, where the number of active users is random and unknown in advance. To characterize the estimation performance of the number of active users, we have derived non-asymptotic achievability bound on the error probability of estimating $K_a$ as $K'_a$. The principal finding is that it decays exponentially with $L$ and eventually converges to $0$. However, it reaches a plateau as $P$ and $n$ increase. The error floors for the cases of $P\to\infty$ and $n\to\infty$ have been provided, respectively. We have also presented scaling laws on the distance between the estimated and true numbers of active users.

Building on these results, we have further derived non-asymptotic achievability and converse bounds on the minimum energy-per-bit required by each active user to transmit $J$ bits using $n$ channel uses subject to constraints on per-user probabilities of misdetection and false-alarm with random and unknown ${\rm K}_a$. Our non-asymptotic bounds provide theoretical benchmarks to evaluate existing schemes, which are shown to be energy-inefficient when the number of active users is large, calling for more advanced schemes that perform energy-efficiently in this case. Compared with random access with individual codebooks, the minimum required energy-per-bit is reduced in URA. Using codewords distributed on a sphere outperforms Gaussian random coding in the non-asymptotic regime. Simulation results have highlighted the potential of MIMO in enabling low-cost and highly reliable communication and improving system spectral efficiency. Moreover, as $L$ and $\mathbb{E}\left[ {\rm K}_a \right]$ decrease and the error requirement becomes stricter, the energy efficiency penalty suffering from the lack of knowledge of the exact value of ${\rm K}_a$ becomes more significant.

Finally, we have considered the asymptotic regime where the number of active users, the codebook size, the number of BS antennas, and the transmitting power scale with $n$ as $n\to\infty$. We have established scaling laws in terms of these parameters in a sparse regime. Consider a URA scenario with a fixed and known number $K_{a,n}$ of active users. We have proved that under some conditions, to reliably serve $K_{a,n} = \mathcal{O} \left( \frac{n^2}{ \ln^2 n }  \right)$ active users, the minimum required $P_n$ and $L_n$ satisfy $P_n^2L_n = \Theta\left( \frac{\ln n}{n^2} \right)$. As can be seen, the minimum required $P_n$ is greatly reduced by increasing $L_n$. Moreover, it is possible for the minimum required $P_n$ and $L_n$ to remain on the same order when $K_{a,n}$ increases but stays below a threshold. In URA scenarios with random and unknown ${\rm K}_{a,n}$, we have proved that under some conditions on statistical properties of ${\rm K}_{a,n}$, one can reliably support ${\rm K}_{a,n}$ active users with mean $\mathbb{E} \left[{\rm K}_{a,n}\right] = \mathcal{O} \left( \frac{n^2}{ \ln^2 n }  \right)$ in the same regime of $P_n$ and $L_n$ as in the case with a fixed and known number of active users.


\appendices

\section{Proof of Theorem~\ref{Theorem_Ka}} \label{Appendix_proof_Ka}

  Assume that there are $K_a$ active users, which is fixed and unknown in advance. In this appendix, we derive a non-asymptotic upper bound on the error probability $\mathbb{P} \left[ K_a  \to K'_{a} \right]$ of estimating $K_a$ exactly as $K'_a \in [K] \backslash \{ K_a \}$. First, we can upper-bound the probability $\mathbb{P} \left[ K_a  \to K'_{a} \right]$ as in~\eqref{eq:proof_noCSI_noKa_Ka_Kahat_p0}, where the term $p_{0,K_a}$ is given in \eqref{eq_noCSI_noKa_pKa_Kahat_p0}. Next, we aim to derive an upper bound on the term $\mathbb{P} \left[ K_a  \to K'_{a} \right]_{\text {new}} $ in~\eqref{eq:proof_noCSI_noKa_Ka_Kahat_p0}. Applying the energy-based metric $m(\cdot, \cdot)$ given in~\eqref{eq:proof_noCSI_noKa_Kaestimate_m}, we have
  \begin{align}
    \mathbb{P} \left[ K_a \to K'_{a} \right]_{\text {new}}
    & \leq \mathbb{P} \left[ \cap_{K''_a \neq K'_a} \left\{  
      m(\mathbf{Y},K'_a) \leq m(\mathbf{Y},K''_a) \right\} \right]  \label{eq:proof_noCSI_noKa_Ka_Kahat0} \\
    & =  \mathbb{P} \left[ \left\{  
      m(\mathbf{Y},K'_a) \leq m(\mathbf{Y},K'_a+1) \right\} 
      \cap \left\{  
      m(\mathbf{Y},K'_a) \leq m(\mathbf{Y},K'_a - 1) \right\} 
      \right]  \label{eq:proof_noCSI_noKa_Ka_Kahat1} \\
      & \leq \min \left\{ \mathbb{P} \left[ m(\mathbf{Y},K'_a) \leq m(\mathbf{Y},K'_a+1)  \right] ,
      \mathbb{P} \left[ m(\mathbf{Y},K'_a) \leq m(\mathbf{Y},K'_a - 1)  
      \right]
      \right\}. \label{eq:proof_noCSI_noKa_Ka_Kahat}
  \end{align}

  Exploiting symmetry, we assume w.l.o.g. that the first $K_a$ codewords are transmitted by active users. The matrix $\mathbf{C}_{K_a} = \left[ \mathbf{c}_1, \ldots, \mathbf{c}_{K_a} \right] \in \mathbb{C}^{n\times K_a}$ includes the transmitted codewords. Denote $C_{K'_a,P'}  =  1 + \left( K'_a + \frac{1}{2} \right) P^{\prime}$. The probability $\mathbb{P} \left[ m(\mathbf{Y},K'_a) \leq m(\mathbf{Y},K'_a+1)  \right]$ in~\eqref{eq:proof_noCSI_noKa_Ka_Kahat} can be bounded as follows:
  \begin{align}
    & \mathbb{P} \left[ m(\mathbf{Y},K'_a) \leq m(\mathbf{Y},K'_a+1)  \right] \notag\\
    & = \mathbb{P} \left[ \left\| \mathbf{Y} \right\|_{F}^{2} \leq nL C_{K'_a, P'} \right] 
    \label{eq:proof_noCSI_noKa_Ka_Kahat_Chernoff_conditiona0} \\
    & = \mathbb{E} \left[  \mathbb{P} \left[ \left. \left\| \mathbf{Y} \right\|_{F}^{2} \leq nL C_{K'_a, P'} \right|  \mathbf{C}_{K_a}  \right]   \right] \label{eq:proof_noCSI_noKa_Ka_Kahat_Chernoff_conditional} \\
    & \leq  \mathbb{E}    \left[ \min_{\rho\geq0}
    \exp \left\{ \rho nLC_{ K'_a , P'} \right\}
     \mathbb{E}
      \left[ \left.  \exp  \left\{  - \rho  \left\|\mathbf{Y}\right\|_F^{2} \right\} \right|  \mathbf{C}_{K_a} \right] \right] , \label{eq:proof_noCSI_noKa_Ka_Kahat_Chernoff_11}  
  \end{align}
  where \eqref{eq:proof_noCSI_noKa_Ka_Kahat_Chernoff_11} follows by applying the Chernoff bound $\mathbb{P}\left[W>0\right] \leq \mathbb{E}\left[ \exp\left\{\rho W\right\} \right]$ for $\rho\geq 0$~\cite{elements_IT} to the conditional probability in~\eqref{eq:proof_noCSI_noKa_Ka_Kahat_Chernoff_conditional}. The conditional expectation in~\eqref{eq:proof_noCSI_noKa_Ka_Kahat_Chernoff_11} is given by
  \begin{align}
    \mathbb{E} \left[ \left.  \exp  \left\{  - \rho  \left\|\mathbf{Y}\right\|_F^{2} \right\} \right|  \mathbf{C}_{K_a} \right]  
    & = \left( \mathbb{E} \left[ \left.  \exp  \left\{  - \rho  \mathbf{y}_l^H \mathbf{y}_l \right\} \right|  \mathbf{C}_{K_a} \right]  \right)^L \label{eq:proof_noCSI_noKa_Ka_Kahat_Chernoff_Lemma1_1} \\
    & = \left|\mathbf{I}_{n} + \rho \mathbf{F} \right|^{-L} . \label{eq:proof_noCSI_noKa_Ka_Kahat_Chernoff_Lemma1_2}  
  \end{align}
  where \eqref{eq:proof_noCSI_noKa_Ka_Kahat_Chernoff_Lemma1_1} follows because conditioned on $\mathbf{C}_{K_a}$, we have $\mathbf{y}_{l} \stackrel{ \rm{i.i.d.} }{\sim} \mathcal{CN} \left( \mathbf{0}, \mathbf{F} \right)$ for $l\in[L]$ with $\mathbf{F}$ given in \eqref{eq_noCSI_noKa_pKa_Kahat_F}, and \eqref{eq:proof_noCSI_noKa_Ka_Kahat_Chernoff_Lemma1_2} follows by applying Lemma~\ref{expectation_bound} shown below to the conditional expectation in~\eqref{eq:proof_noCSI_noKa_Ka_Kahat_Chernoff_Lemma1_1}. 
  \begin{Lemma}[\cite{quadratic_form1,TIT}]\label{expectation_bound}
    Assume that ${\mathbf{x}} \in \mathbb{C}^{p\times 1}$ is distributed as $ {\mathbf{x}} \sim \mathcal{CN}\left( \mathbf{0},  {\boldsymbol{\Sigma}} \right)$. Let $\mathbf{B}\in\mathbb{C}^{p\times p}$ be a Hermitian matrix. For any $\gamma$, if the eigenvalues of $\mathbf{I}_{p} - \gamma {\boldsymbol{\Sigma}}\mathbf{B}$ are positive, we have
    \begin{equation}\label{eq:lemma_expectation}
      \mathbb{E}\left[ \exp \left\{ \gamma \mathbf{x}^{H}\mathbf{B}\mathbf{x} \right\} \right] = \left|\mathbf{I}_{p} - \gamma \boldsymbol{\Sigma}\mathbf{B}\right|^{-1} .
    \end{equation}
  \end{Lemma}
  
  Then, we have
  \begin{align}
    \mathbb{P} \left[ m(\mathbf{Y},K'_a) \leq m(\mathbf{Y},K'_a+1)  \right]  
    \leq  \mathbb{E}   \left[ \min_{\rho\geq0} \exp \left\{ \rho nLC_{K'_a, P'}
     -  L   \ln \left| \mathbf{I}_n    +   \rho\mathbf{F} \right| \right\} \right] \label{eq:proof_noCSI_noKa_Ka_Kahat11} .
  \end{align}
  Denote the RHS of \eqref{eq:proof_noCSI_noKa_Ka_Kahat11} as $p_{K_a, K'_a, P', 1}$, which is given in~\eqref{eq_noCSI_noKa_pKa_Kahat1}. Likewise, we can derive an upper bound on the probability $\mathbb{P} \left[ m(\mathbf{Y},K'_a) \leq m(\mathbf{Y},K'_a - 1)  \right]$, which is denoted as $p_{K_a, K'_a, P',2}$ as in~\eqref{eq_noCSI_noKa_pKa_Kahat2}.

\section{Proof of Theorem~\ref{Theorem_Ka_asymL_concentration}} \label{Appendix_proof_Ka_asymL_concentration}  
  
  We construct a codebook $\mathbf{C} = \left[ \mathbf{c}_1, \ldots, \mathbf{c}_M \right] \in \mathbb{C}^{n\times M}$ with each codeword drawn uniformly i.i.d. from a sphere of radius $\sqrt{nP}$. For $\forall l \in [L]$, we have
  \begin{align} 
    \mathbb{E} \left[ \frac{1}{n} \left\| \mathbf{y}_l \right\|_2^2\right] 
    & = \frac{1}{n}\mathbb{E} \left[\left\| \mathbf{C}_{K_a} \mathbf{h}_{K_a,l} +\mathbf{z}_l \right\|_2^2\right] \label{eq:proof_expectation1} \\ 
    & =  \frac{1}{n} \mathbb{E} \left[ \operatorname{tr} \left( \mathbf{C}_{K_a}^H \mathbf{C}_{K_a}  \mathbb{E} \left[ \mathbf{h}_{K_a,l}\mathbf{h}_{K_a,l}^H \right] \right)  \right] + \frac{1}{n} \mathbb{E} \left[ \mathbf{z}_l^H \mathbf{z}_l \right] \label{eq:proof_expectation2} \\
    & = \frac{1}{n} \operatorname{tr} \left( \mathbf{C}_{K_a}^H \mathbf{C}_{K_a} \right) + 1  \label{eq:proof_expectation3} \\
    & = 1 + K_a P ,\label{eq:proof_expectation4}
  \end{align}
  where $\mathbf{C}_{K_a} \in \mathbb{C}^{n\times K_a}$ includes codewords transmitted by $K_a$ active users, the elements of $\mathbf{h}_{K_a,l} \in \mathbb{C}^{K_a}$ follow the distribution of $\mathcal{CN}(0,1)$ independently, and $\mathbf{z}_l \in \mathbb{C}^{n}$ includes i.i.d. $\mathcal{CN}(0,1)$ elements.

  Denote $\mu =  1 + K_a P $. We have $\mathbb{E} \left[ \frac{1}{nL} \left\| \mathbf{Y} \right\|_F^2 \right] = \mu$. Then, for any constant $\delta > 0$, the two-sided probability can be bounded as  
  \begin{align}
    & \mathbb{P} \left[ \left| \frac{1}{nL}  \left\| \mathbf{Y}  \right\|_F^2 - \mu \right| \geq \delta \right]  \notag\\
    & = \mathbb{E} \left[ \mathbb{P} \left[ \left. \left|  \left\| \mathbf{Y}  \right\|_F^2 - nL\mu \right| \geq nL\delta \right| \mathbf{C}_{K_a} \right]  \right] 
    \label{eq:proof_noCSI_noKa_Ka_Kahat_scalinglaw_cond3_concentrate0} \\
    & =  \mathbb{E} \left[  \mathbb{P}   \left[ \left. \left| \sum_{l\in[L]}
    \begin{bmatrix} \Re^{H}  \left( \tilde{\mathbf{y}}_l \right) &      \Im^{H}  \left( \tilde{\mathbf{y}}_l \right) \end{bmatrix}
     \begin{bmatrix} \boldsymbol{\Lambda} & \\ &       \boldsymbol{\Lambda} \end{bmatrix}
      \begin{bmatrix} \Re \left( \tilde{\mathbf{y}}_l \right)\\ \Im \left( \tilde{\mathbf{y}}_l \right) \end{bmatrix}
     -  nL\mu \right|
    \geq  n L \delta \right| \mathbf{C}_{K_a} \right] \right] \label{eq:proof_noCSI_noKa_Ka_Kahat_scalinglaw_cond3_concentrate} \\
    & =  \mathbb{E} \left[  \mathbb{P}   \left[ \left. \left|  
    \sum_{i\in[n]} \left(  \frac{1}{2} \lambda_i   x_i  -  L \lambda_i  \right)  \right|
    \geq  n L \delta \right| \mathbf{C}_{K_a} \right] \right] \label{eq:proof_noCSI_noKa_Ka_Kahat_scalinglaw_cond3_concentrate_chi} \\
    & =  \mathbb{E} \!\left[  \mathbb{P}   \left[ \left.  \sum_{i\in[n]} \!\left(  \frac{1}{2} \lambda_i   x_i  \!-\!  L \lambda_i  \right)
    \geq  n L \delta \right| \mathbf{C}_{K_a} \right] \right]   
    + \mathbb{E} \!\left[  \mathbb{P}   \left[ \left. - \sum_{i\in[n]} \!\left(  \frac{1}{2} \lambda_i   x_i  \!-\!  L \lambda_i  \right)
    \geq  n L \delta \right| \mathbf{C}_{K_a} \right] \right] .
    \label{eq:proof_noCSI_noKa_Ka_Kahat_scalinglaw_cond3_concentrate2} 
  \end{align}
  Here, \eqref{eq:proof_noCSI_noKa_Ka_Kahat_scalinglaw_cond3_concentrate} holds because $\mathbf{y}_{l} \stackrel{ \rm{i.i.d.} }{\sim} \mathcal{CN} \left( \mathbf{0}, \mathbf{F} \right)$ conditioned on the transmitted codewords, where the matrix $\mathbf{F} = \mathbf{I}_n + \mathbf{C}_{K_a} \mathbf{C}_{K_a}^H = \mathbf{U} \boldsymbol{\Lambda} \mathbf{U}^H$ for a unitary matrix $\mathbf{U}$ and a diagonal matrix $\boldsymbol{\Lambda} = \operatorname{diag} \left\{ \lambda_1, \ldots, \lambda_n \right\}$ with eigenvalues $ \lambda_1 \geq \cdots \geq \lambda_n $, and the vectors $\tilde{\mathbf{y}}_{l} \stackrel{ \rm{i.i.d.} }{\sim} \mathcal{CN} \left( \mathbf{0}, \mathbf{I}_n \right)$, $\Re \left( \tilde{\mathbf{y}}_{l} \right) \stackrel{ \rm{i.i.d.} }{\sim} \mathcal{CN} \left( \mathbf{0}, \frac{1}{2} \mathbf{I}_n \right) $, and $\Im \left( \tilde{\mathbf{y}}_{l} \right) \stackrel{ \rm{i.i.d.} }{\sim} \mathcal{CN} \left( \mathbf{0}, \frac{1}{2} \mathbf{I}_n \right) $ for $l\in[L]$. The random variable $x_i$ in \eqref{eq:proof_noCSI_noKa_Ka_Kahat_scalinglaw_cond3_concentrate_chi} satisfies $x_i \stackrel{ \rm{i.i.d.} }{\sim} \chi^2(2L)$ for $i\in[n]$, and \eqref{eq:proof_noCSI_noKa_Ka_Kahat_scalinglaw_cond3_concentrate_chi} holds because $\sum_{i=1}^{n} \lambda_i = n(1+K_aP) = n\mu$ and for $k$ independent standard normal random variables, the sum of their squares is distributed according to the chi-squared distribution with $k$ degrees of freedom.

  The conditional probability in the first term of the RHS of~\eqref{eq:proof_noCSI_noKa_Ka_Kahat_scalinglaw_cond3_concentrate2} can be bounded as follows:
  \begin{align}
    & \mathbb{P}   \left[ \left.  \sum_{i\in[n]}  \left( \frac{1}{2} \lambda_i   x_i  -  L\lambda_i \right)
    \geq  n L \delta \right| \mathbf{C}_{K_a} \right]  \notag  \\ 
    & \leq \min_{\rho \in \left[ 0, \frac{1}{\bar{\alpha}} \right) } \exp\left\{ - \rho n L \delta\right\} 
    \prod_{i\in[n]}  \mathbb{E} \left[ \left. \exp\left\{  \frac{1}{2} \rho \lambda_i   x_i  -  \rho L\lambda_i   \right\} \right| \mathbf{C}_{K_a} \right] \label{eq:proof_noCSI_noKa_Ka_Kahat_scalinglaw_cond3_concentrate_chi_Chernoff}\\ 
    & = \min_{\rho \in \left[ 0, \frac{1}{\bar{\alpha}} \right) } \exp\left\{ - \rho n L \delta\right\} 
    \prod_{i\in[n]} \left( \frac{ \exp\left\{ - \frac{\rho  \lambda_i}{2} \right\} }{  \sqrt{ 1-\rho\lambda_i } } \right)^{2L}
    \label{eq:proof_noCSI_noKa_Ka_Kahat_scalinglaw_cond3_concentrate_chi_Chernoff_1}\\  
    & \leq \min_{\rho \in \left[ 0, \frac{1}{\bar{\alpha}} \right) } \exp\left\{ - \rho n L \delta  +  \rho^2 L  \sum_{i\in[n]} \lambda_i^2 \right\}  
    \label{eq:proof_noCSI_noKa_Ka_Kahat_scalinglaw_cond3_concentrate_chi_subexp} ,
  \end{align}
  where \eqref{eq:proof_noCSI_noKa_Ka_Kahat_scalinglaw_cond3_concentrate_chi_Chernoff} follows by applying the Chernoff bound $\mathbb{P}\left[W>0\right] \leq \mathbb{E}\left[ \exp\left\{\rho W\right\} \right]$ for $\rho\geq 0$~\cite{elements_IT} to the conditional probability; \eqref{eq:proof_noCSI_noKa_Ka_Kahat_scalinglaw_cond3_concentrate_chi_Chernoff_1} follows because for the chi-squared random variable $Z \sim \chi^2(k)$, the moment-generating function is given by $\mathbb{E}\left[ \exp\left\{ t Z \right\} \right] = (1-2t)^{-k/2}$ for $t<\frac{1}{2}$; \eqref{eq:proof_noCSI_noKa_Ka_Kahat_scalinglaw_cond3_concentrate_chi_subexp} follows because $\frac{\exp\{-t\}}{\sqrt{1-2t}} \leq \exp\{2t^2\}$ for all $|t|<\frac{1}{4}$; and the conditions $\rho\geq 0$, $\frac{1}{2} \rho \lambda_i<\frac{1}{2}$, and $|\frac{1}{2} \rho \lambda_i|<\frac{1}{4}$ for $i\in[n]$ lead to $\rho \in \left[ 0, \frac{1}{\bar{\alpha}} \right) $ with $\bar{\alpha} = 2 \lambda_1$.

  Let $g(\rho) = - \rho n L \delta  +  \frac{\rho^2 \bar{v}^2}{2}  $ with $\bar{v}^2 = 2L  \sum_{i\in[n]} \lambda_i^2$. The unconstrained minimum of $g(\rho)$ is at $\rho^{*} = \frac{ n L \delta  }{ \bar{v}^2 } $. In the case of $\frac{ n L \delta  }{ \bar{v}^2 }  \leq  \frac{1}{ \bar{\alpha} } $, we have $\min_{\rho\in \left[ 0,\frac{1}{\bar{\alpha}} \right) }  g(\rho)  =  g(\rho^{*})  =  - \frac{ (n L \delta)^2 }{ 2 \bar{v}^2 }$. In the case of $\frac{ n L \delta  }{ \bar{v}^2 }  >  \frac{1}{ \bar{\alpha} } $, we have $\min_{\rho\in \left[ 0,\frac{1}{\bar{\alpha}} \right) }  g(\rho)  =  g\left( \frac{1}{\bar{\alpha}} \right) = - \frac{ n L \delta }{ \bar{\alpha} }  +  \frac{ \bar{v}^2 }{ 2\bar{\alpha}^2  } \leq   - \frac{n L \delta}{ 2\bar{\alpha} }  $. Therefore, we have $\min_{\rho\in \left[ 0,\frac{1}{\bar{\alpha}} \right) }  g(\rho) \leq - \min \left\{ \frac{ (n L \delta)^2 }{ 2 \bar{v}^2 } ,  \frac{n L \delta}{ 2\bar{\alpha} }  \right\}$, which contributes to 
  \begin{align}
    & \mathbb{P}   \left[ \left.  \sum_{i\in[n]}  \left( \frac{1}{2} \lambda_i   x_i  -  L\lambda_i \right)
    \geq  n L \delta \right| \mathbf{C}_{K_a} \right]  \notag  \\  
    & \leq \exp \left\{  - \min \left\{   \frac{ L (n \delta)^2 }{ 4  \sum_{i\in[n]} \lambda_i^2 } ,  
    \frac{n L \delta}{ 4 \lambda_1 }  \right\}    \right\} \label{eq:proof_noCSI_noKa_Ka_Kahat_scalinglaw_cond5_concentrate_term1_0}  \\
    & \leq  \exp  \left\{  -  \min  \left\{  \frac{ L   \left( n \delta \right)^2}{ 8( n + (nK_aP)^2 ) },  
    \frac{  n L \delta  }{ 4(1+nK_aP) } \right\} \right\} . \label{eq:proof_noCSI_noKa_Ka_Kahat_scalinglaw_cond5_concentrate_term1} 
  \end{align}  
  Here, \eqref{eq:proof_noCSI_noKa_Ka_Kahat_scalinglaw_cond5_concentrate_term1}  follows from the inequalities $ \lambda_1 \leq 1 + n K_a P$ and 
  \begin{align}
  \sum_{i\in[n]} \lambda_i^2 
  & = \operatorname{tr}\left( \mathbf{I}_n + 2\mathbf{C}_{K_a} \mathbf{C}^H_{K_a} + \mathbf{C}_{K_a} \mathbf{C}^H_{K_a} \mathbf{C}_{K_a} \mathbf{C}^H_{K_a} \right) \label{eq:proof_noCSI_noKa_Ka_Kahat_scalinglaw_F0} \\
  & \leq n + 2\operatorname{tr}\left( \mathbf{C}_{K_a} \mathbf{C}^H_{K_a} \right) + \operatorname{tr}^2\left( \mathbf{C}_{K_a} \mathbf{C}^H_{K_a} \right)  \label{eq:proof_noCSI_noKa_Ka_Kahat_scalinglaw_F} \\
  & = n + 2nK_aP + (nK_aP)^2 \label{eq:proof_noCSI_noKa_Ka_Kahat_scalinglaw_F1} \\
  & \leq 2 \left( n + (nK_aP)^2 \right) , \label{eq:proof_noCSI_noKa_Ka_Kahat_scalinglaw_F2}
  \end{align}
  where \eqref{eq:proof_noCSI_noKa_Ka_Kahat_scalinglaw_F} follows because $\operatorname{tr}\left( \mathbf{AB}\right) \leq \operatorname{tr}\left( \mathbf{A}\right) \operatorname{tr}\left( \mathbf{B}\right)$ for positive semidefinite matrices $\mathbf{A}$ and $\mathbf{B}$ and \eqref{eq:proof_noCSI_noKa_Ka_Kahat_scalinglaw_F2} holds when $n\geq 1$.

  Likewise, the conditional probability in the second term of the RHS of~\eqref{eq:proof_noCSI_noKa_Ka_Kahat_scalinglaw_cond3_concentrate2} can be bounded by $\exp  \left\{  -  \min  \left\{  \frac{ L   \left( n \delta \right)^2}{ 8( n + (nK_aP)^2 ) }, \frac{  n L \delta  }{ 4(1+nK_aP) } \right\} \right\}$. Then, we have 
  \begin{equation}
    \mathbb{P} \left[ \left| \frac{1}{nL}  \left\| \mathbf{Y}  \right\|_F^2 - \mu \right| \geq \delta \right]  
    \leq  2  \exp  \left\{  -  \min  \left\{  \frac{ L   \left( n \delta \right)^2}{ 8( n + (nK_aP)^2 ) },  
    \frac{  n L \delta  }{ 4(1+nK_aP) } \right\} \right\} . \label{eq:proof_noCSI_noKa_Ka_Kahat_scalinglaw_cond5_concentrate} 
  \end{equation} 
  It concludes the proof of Theorem~\ref{Theorem_Ka_asymL_concentration}.

\section{Proof of Corollary~\ref{Theorem_Ka_asymn}} \label{Appendix_proof_Ka_asymn}
    
    Assume all users share a common codebook $\mathbf{C} = \left[ \mathbf{c}_1, \ldots, \mathbf{c}_M \right] \in \mathbb{C}^{n\times M}$ with each element drawn i.i.d. according to $\mathcal{CN}(0,P')$. As in Theorem~\ref{Theorem_Ka}, we perform a change of measure, which adds a total variation distance bounded by $p_{0,K_a} + \frac{\binom{K_a}{2}}{M}$. The power constraint violation probability $p_{0,K_a}$ can be further bounded as follows: 
    \begin{align}
      p_{0,K_a} 
      & = K_a \mathbb{P}\left[ \chi^2(2n)  \geq \frac{2nP}{P'} \right] \\
      & \leq \min_{0\leq r < \frac{1}{2}} \left\{ K_a \exp\left\{-r\frac{2nP}{P'}\right\} \mathbb{E}\left[ \exp\left\{ r \chi^2(2n) \right\} \right] \right\} \label{eq_noCSI_noKa_p0_asymn1} \\
      & = \min_{0\leq r < \frac{1}{2}} \left\{ K_a \exp\left\{-r\frac{2nP}{P'} - n \ln (1-2r) \right\} \right\} \label{eq_noCSI_noKa_p0_asymn2}  \\
      & = K_a \exp\left\{-n \left( \frac{P}{P'} - 1 - \ln \frac{P}{P'} \right) \right\},\label{eq_noCSI_noKa_p0_asymn3}
    \end{align}
    where \eqref{eq_noCSI_noKa_p0_asymn1} follows from the Chernoff bound $\mathbb{P}\left[W>0\right] \leq \mathbb{E}\left[ \exp\left\{rW\right\} \right]$ for $r\geq 0$~\cite{elements_IT}; \eqref{eq_noCSI_noKa_p0_asymn2} follows from the moment-generating function $\mathbb{E}\left[ \exp\left\{ t \chi^2(k) \right\} \right] = (1-2t)^{-k/2}$ for $t<\frac{1}{2}$; and \eqref{eq_noCSI_noKa_p0_asymn3} follows because $ r\frac{2nP}{P'} + n \ln (1-2r)$ is maximized in the case of $r = \frac{1}{2}\left( 1-\frac{P'}{P} \right)$. When $n \to \infty$ and $\frac{P'}{P}=c$ with the constant $c\in(0,1)$, we have $p_{0,K_a} \to 0$.

    In the following, we proceed to obtain $p_{K_a, K'_a, P',1}$ and $p_{K_a, K'_a, P',2}$ assuming all users share a common Gaussian codebook without power constraint and there is no message collision among $K_a$ active users. Let the matrix $\mathbf{C}_{K_a} \in \mathbb{C}^{n\times K_a}$ include the transmitted codewords of $K_a$ active users. Following from~\eqref{eq_noCSI_noKa_pKa_Kahat1}, we have
    \begin{align}
      & p_{K_a, K'_a, P',1} \notag\\
      & = \!\mathbb{E}_{\mathbf{C}_{K_a}} \!\! \left[ \min_{\rho\geq0} \exp \!\left\{  \rho nL \left( 1+ \left( K'_a + \frac{1}{2} \right)  P^{\prime} \right)
      -L \ln  \left|  \mathbf{I}_n + \rho \left( \mathbf{I}_n + \mathbf{C}_{K_a} \mathbf{C}_{K_a}^H \right) \right| \right\} \right] \\
      & = \!\mathbb{E}_{\mathbf{C}_{K_a}} \!\! \left[ \min_{\rho\geq0} \exp \!\left\{  \rho nL \!\left( 1 + \left( K'_a + \frac{1}{2} \right)  P^{\prime} \right)
      - L(n-K_a)\ln(1+\rho)  \right.\right. \notag\\
      & \;\;\;\;\;  \;\;\;\;\;  \;\;\;\;\; \;\;\;\;\;  \;\;\;\;\;  \;\;\;
       - L \ln \! \left| \mathbf{I}_{K_a} \!+\! \rho \left( \mathbf{I}_{K_a} \!+\! \mathbf{C}_{K_a}^H \mathbf{C}_{K_a} \right) \right| \bigg\} \bigg] \!.
    \end{align}

    In the case of $K'_a < K_a$, $n\to\infty$, and $\rho n\to c_1$ with constant $c_1>0$, we have
    \begin{align}
      & p_{K_a, K'_a, P',1} \notag\\ 
      & = \min_{\rho\geq0} \exp \!\left\{  \rho nL \!\left( 1+ \left( \!K'_a \!+\! \frac{1}{2} \right)  P^{\prime} \right)
      - L(n\!-\!K_a)\ln(1+\rho)
      - L K_a \ln  \left( 1+\rho(1\!+\! P' n) \right) \right\} \!+ o(1) \label{eq_noCSI_noKa_p1_asymn1}\\
      & = \min_{\rho\geq0} \exp \left\{  \rho nL \left( 1+ \left( K'_a + \frac{1}{2} \right)  P^{\prime} \right)
      - Ln\ln(1+\rho)
      - L K_a \ln  \left( 1+\rho P' n \right) \right\}  + o(1) \label{eq_noCSI_noKa_p1_asymn2}\\
      & = \min_{\rho\geq0} \exp \left\{  \rho nL \left( K'_a + \frac{1}{2} \right)  P^{\prime} - L K_a \ln  \left( 1+\rho P' n \right) \right\}  + o(1) \label{eq_noCSI_noKa_p1_asymn3}\\
      & = \min_{\tilde{\rho}\geq0} \exp \left\{  \tilde{\rho} L \left( K'_a + \frac{1}{2} \right) - L K_a \ln  \left( 1+\tilde{\rho} \right) \right\}  + o(1) \label{eq_noCSI_noKa_p1_asymn4}\\
      & = \exp \left\{ K_a L \left(  1 - \frac{2K'_a+1}{2K_a} + \ln \frac{2K'_a+1}{2K_a} \right)  \right\}  + o(1),\label{eq_noCSI_noKa_p1_asymn5}
    \end{align}
    where \eqref{eq_noCSI_noKa_p1_asymn1} follows because the eigenvalues of $\frac{1}{P'} \mathbf{C}_{K_a}^H \mathbf{C}_{K_a}$ converge to $n+o(n)$ as $n\to\infty$~\cite{eigenvalue_wishart}; \eqref{eq_noCSI_noKa_p1_asymn3} follows because $\rho n - n\ln(1+\rho) = \mathcal{O}(\rho^2 n) = o(1)$; \eqref{eq_noCSI_noKa_p1_asymn4} holds by setting $\tilde{\rho} = \rho P' n$; and \eqref{eq_noCSI_noKa_p1_asymn5} follows because $\tilde{\rho} L \left( K'_a + \frac{1}{2} \right) - L K_a \ln  \left( 1+\tilde{\rho} \right)$ is minimized when $\tilde{\rho} = \frac{K_a}{  K'_a + \frac{1}{2}  }-1$ in the case of $K'_a < K_a$. When $K'_a \geq K_a$, we denote $p_{K_a, K'_a, P',1} = 1$. Thus, in the case of $n\to \infty$, we have
    \begin{equation}
      p_{K_a, K'_{a},P',1}
      = \begin{cases}
      \exp \left\{ K_a L \left( 1 - \frac{2K'_a+1}{2K_a} + \ln \frac{ 2K'_a+1 }{2K_a} \right)  \right\} ,
      &  \text { if }  K'_a  < K_a\\
      1,
      &  \text { if }  K'_a \geq K_a
      \end{cases}  .
    \end{equation}
    Likewise, $p_{K_a, K'_{a},P',2}$ is given by
    \begin{equation}
      p_{K_a, K'_{a},P',2}
      = \begin{cases}
      \exp \left\{ K_a L \left( 1 - \frac{2K'_a-1}{2K_a} + \ln \frac{ 2K'_a-1 }{ 2K_a} \right)  \right\} ,
      &  \text { if }  K'_a  > K_a\\
      1,
      &  \text { if }  K'_a \leq K_a
      \end{cases}  .
    \end{equation}
    Then, we have
    \begin{align} 
       p_{K_a, K'_a, P'}
       & =  \min  \left\{ p_{K_a, K'_a, P',1}  ,  p_{K_a, K'_a, P',2}  \right\}  \\
       & = \begin{cases}
      \exp \left\{ K_a L \left( 1 - \frac{2K'_a+1}{2K_a} + \ln \frac{ 2K'_a+1 }{2K_a} \right)  \right\} ,
      &  \text { if }  K'_a  < K_a\\
      \exp \left\{ K_a L \left( 1 - \frac{2K'_a-1}{2K_a} + \ln \frac{ 2K'_a-1 }{ 2K_a} \right)  \right\} ,
      &  \text { if }  K'_a  > K_a 
      \end{cases} .
    \end{align} 
    
    When using codewords uniformly distributed on a sphere, we can bound $\mathbb{P}\left[ K_a \to K'_a\right]$ in the case of $n\to\infty$ in a similar way. It concludes the proof of Corollary~\ref{Theorem_Ka_asymn}.

\section{ Proof of Theorem \ref{Theorem_scalinglaw_Ka_estimation}  }
\label{Proof_scalinglaw_Ka_estimation_new}
  
  Assume that all users share a common codebook with each column drawn uniformly i.i.d. from a sphere of radius $\sqrt{nP}$. Let the matrix $\mathbf{C}_{K_a} \in \mathbb{C}^{n\times K_a}$ include the transmitted codewords of $K_a$ active users. The error probability of estimating $K_a$ exactly as $K'_a \in [K] \backslash \{ K_a \}$ can be bounded as
  \begin{align}
    \mathbb{P} \left[ K_a \to K'_{a} \right]   
    & \leq \mathbb{P} \left[  
      m(\mathbf{Y},K'_a) \leq m(\mathbf{Y},K_a) \right] \label{eq:proof_noCSI_noKa_Ka_Kahat_scalinglaw0} \\ 
    & = 1  \left[ K'_a  < K_a \right]
    \mathbb{P} \left[ \left\| \mathbf{Y} \right\|_{F}^{2} \leq n L \left( 1+ \frac{K'_a + K_a}{2}  P \right) \right] \notag\\
    & \;\;\;\;
    + 1  \left[ K'_a > K_a \right]
    \mathbb{P} \left[ \left\| \mathbf{Y} \right\|_{F}^{2} \geq n L \left( 1+ \frac{K'_a + K_a}{2} P \right) \right] \label{eq:proof_noCSI_noKa_Ka_Kahat_scalinglaw2} \\
    & = 1  \left[ K'_a   <  K_a \right]
    \mathbb{P}   \left[ \left\| \mathbf{Y} \right\|_{F}^{2} - nL( 1 + K_aP )  \leq n L P \frac{K'_a - K_a}{2} \right] \notag\\
    & \;\;\;\; +  1  \left[ K'_a  >  K_a \right]
    \mathbb{P}  \left[ \left\| \mathbf{Y} \right\|_{F}^{2}  - nL( 1 + K_aP )  \geq  n L P \frac{K'_a - K_a}{2} \right] \label{eq:proof_noCSI_noKa_Ka_Kahat_scalinglaw3} \\
    & \leq  \mathbb{E}_{\mathbf{C}_{K_a}}  \left[  \mathbb{P}  \left[ \left. \left| \left\| \mathbf{Y} \right\|_{F}^{2}  - nL( 1 +  K_aP ) \right| \geq n L P \frac{ \left| K'_a - K_a \right| }{2} \right| \mathbf{C}_{K_a} \right] \right] , \label{eq:proof_noCSI_noKa_Ka_Kahat_scalinglaw6}
  \end{align}
  where \eqref{eq:proof_noCSI_noKa_Ka_Kahat_scalinglaw0} follows from \eqref{eq:proof_noCSI_noKa_Ka_Kahat0}, and \eqref{eq:proof_noCSI_noKa_Ka_Kahat_scalinglaw6} follows by loosing one-sided tails in~\eqref{eq:proof_noCSI_noKa_Ka_Kahat_scalinglaw3} as two-sided tails in both cases of $K'_a  < K_a$ and~$K'_a > K_a$. The conditional probability in~\eqref{eq:proof_noCSI_noKa_Ka_Kahat_scalinglaw6} can be upper-bounded as
  \begin{align} 
    &\mathbb{P} \left[ \left. \left| \left\| \mathbf{Y} \right\|_{F}^{2}  - nL( 1 +  K_aP ) \right| \geq n L P \frac{ \left| K'_a - K_a \right| }{2} \right| \mathbf{C}_{K_a} \right] \notag \\
    & \leq  2  \exp  \left\{  -  \min  \left\{  \frac{ c_1     \left( n L P  \left| K'_a  - K_a \right|  \right)^2}{ L \left( n + (nK_aP)^2 \right) },  \frac{ c_2 n L P  \left| K'_a  - K_a \right|  }{ 1+nK_aP } \right\} \right\} \label{eq:proof_noCSI_noKa_Ka_Kahat_scalinglaw_cond5} \\
    & =
    \begin{cases}
      2 \exp \left\{- \frac{c_1 \left( n L P \left| K'_a-K_a \right|  \right)^2}{ L \left( n + (nK_aP)^2 \right) } \right\},
      &       \text { if }   \left| K'_a - K_a \right| \leq  K_{a,\text{thre}} \\
      2 \exp \left\{- \frac{ c_2 n L P \left| K'_a-K_a \right|  }{ 1+nK_aP } \right\},
      &       \text { if }   \left| K'_a - K_a \right| >  K_{a,\text{thre}}
    \end{cases}  \label{eq:proof_noCSI_noKa_Ka_Kahat_scalinglaw_cond6} \\
    & \leq
    \begin{cases}
       2  \exp  \left\{ - \frac{c_1  \left( n L P \left| K'_a -K_a \right|  \right)^2}{ L \left( n + (nK_aP)^2 \right) }  \right\} ,
      &       \text { if }   \left| K'_a - K_a \right| \leq  K_{a,\text{thre}} \\ 
       2 \exp  \left\{ - \frac{c_2^2 L}{2c_1}   \left(  1  +  \frac{ n-1 }{ (1+nK_aP)^2 }  \right)   \right\} ,
      &       \text { if }   \left| K'_a - K_a \right| >  K_{a,\text{thre}} 
    \end{cases}, \label{eq:proof_noCSI_noKa_Ka_Kahat_scalinglaw_cond7}
  \end{align}
  where \eqref{eq:proof_noCSI_noKa_Ka_Kahat_scalinglaw_cond5} follows from \eqref{eq:proof_noCSI_noKa_Ka_Kahat_scalinglaw_cond5_concentrate} with constants $c_1 = \frac{1}{32}$ and $c_2 = \frac{1}{8}$, $K_{a,\text{thre}}$ in~\eqref{eq:proof_noCSI_noKa_Ka_Kahat_scalinglaw_cond6} is given by
  \begin{equation}\label{eq:proof_noCSI_noKa_Ka_Kahat_scalinglaw_thre}
    K_{a,\text{thre}} = \frac{c_2 \left( n + (nK_aP)^2 \right)}{ c_1 nP (1+nK_aP) },
  \end{equation}
  and \eqref{eq:proof_noCSI_noKa_Ka_Kahat_scalinglaw_cond7} follows by substituting $\left| K'_a-K_a \right| > K_{a,\text{thre}} \geq \frac{c_2}{ 2c_1 nP }  \left( 1 +nK_aP   + \frac{n-1}{1+nK_aP}\right)$ into the term $ \frac{ c_2 n L P \left| K'_a-K_a \right|  }{ 1+nK_aP } $ in~\eqref{eq:proof_noCSI_noKa_Ka_Kahat_scalinglaw_cond6}.

  Combining \eqref{eq:proof_noCSI_noKa_Ka_Kahat_scalinglaw6} and  \eqref{eq:proof_noCSI_noKa_Ka_Kahat_scalinglaw_cond7}, we can obtain that the error probability $\mathbb{P} \left[ K_a \to K'_{a} \right]$ vanishes as $L \to \infty$ under the condition that
  \begin{equation}
    \left| K'_a-K_a \right|
    = \omega \left( \max \left\{ \frac{K_a}{\sqrt{L}} , \frac{1}{P\sqrt{nL}} \right\} \right).
  \end{equation}
  For any $\Delta > K_{a,\text{thre}}$, we have
  \begin{align}
    & \mathbb{P} \left[ \left. \left| {\rm K}'_a-K_a \right| \geq \Delta , {\rm K}'_a \in \{0,1,\ldots,K\} \right| {\rm K}_a = K_a \right] \notag\\
    & \leq \sum_{K'_a \in \left\{ K''_a : \left| K''_a-K_a \right| \geq \Delta , K''_a \in \{0,1,\ldots,K\}  \right\}} \mathbb{P} \left[ K_a \to K'_{a} \right] \label{eq:proof_noCSI_noKa_Ka_Kahat_neq_scalinglaw1_1}\\
    & \leq
    2 \exp \left\{ \ln K - \frac{c_2^2 L}{2c_1} \left( 1 + \frac{ n-1 }{ (1+nK_aP)^2 } \right) \right\},\label{eq:proof_noCSI_noKa_Ka_Kahat_neq_scalinglaw1_2}
  \end{align}
  where \eqref{eq:proof_noCSI_noKa_Ka_Kahat_neq_scalinglaw1_1} follows from the union bound and \eqref{eq:proof_noCSI_noKa_Ka_Kahat_neq_scalinglaw1_2} follows from \eqref{eq:proof_noCSI_noKa_Ka_Kahat_scalinglaw_cond7}. We can observe that \eqref{eq:proof_noCSI_noKa_Ka_Kahat_neq_scalinglaw1_2} tends to $0$ as $L\to\infty$ in the case of $\frac{\ln K}{L} = o(1)$. Likewise, for any $1 \leq \Delta \leq K_{a,\text{thre}}$, we have
  \begin{align}
    & \mathbb{P} \left[ \left. \left| {\rm K}'_a-K_a \right| \geq \Delta , {\rm K}'_a \in \{0,1,\ldots,K\} \right| {\rm K}_a = K_a \right] \notag\\
    & \leq
    2 \exp \left\{ \ln K - \frac{c_1 \left( n L P \Delta  \right)^2}{ L \left( n + (nK_aP)^2 \right) } \right\} + 2 \exp \left\{ \ln K - \frac{c_2^2 L}{2c_1} \left( 1 + \frac{ n-1 }{ (1+nK_aP)^2 } \right) \right\}.\label{eq:proof_noCSI_noKa_Ka_Kahat_neq_scalinglaw2_1}
  \end{align}
  Under the condition that $\frac{\ln K}{L} = o(1)$ and $ \sqrt{ \frac{ \alpha \ln K \left( n+(nK_aP)^2 \right) }{L n^2P^2} } \leq \Delta \leq K_{a,\text{thre}}$ for some constant $\alpha > 32$, the RHS of~\ref{eq:proof_noCSI_noKa_Ka_Kahat_neq_scalinglaw2_1} is in the order of $ \mathcal{O}\left( \frac{1}{K} \right)$. Combining the cases of $\Delta > K_{a,\text{thre}}$ and $1 \leq \Delta \leq K_{a,\text{thre}}$, we can obtain that under the assumption of $L,K\to\infty$ and $\frac{\ln K}{L} = o(1)$, the probability of the event that $K_a$ is wrongly estimated as ${\rm K}'_a$ which belongs to an interval with a distance from $K_a$ no less than $\sqrt{ \frac{ \alpha \ln K \left( n + (nK_aP)^2 \right) }{L n^2P^2} }$ goes to $0$. 

  Moreover, if the number of active users satisfies $K_a \leq \sqrt{\frac{L}{\alpha\ln K} - \frac{1}{nP^2} }$, the estimation error of $K_a$ vanishes, i.e., we have $\mathbb{P}[ \left. {\rm K}'_a  \neq K_a \right| {\rm K}_a  = K_a ] = \mathcal{O}\left( \frac{1}{K} \right) \to 0$. In a special case of $\sqrt{\frac{nL}{\ln K}}P \to \infty$, the number $K_a = \mathcal{O} \left(\sqrt{\frac{L}{\ln K}}\right)$ of active users can be estimated with vanishing error probability $ \mathcal{O}\left( \frac{1}{K} \right)$.

\section{Proof of Theorem~\ref{Theorem_achievability}} \label{Appendix_proof_achievability}

  In this appendix, we prove Theorem~\ref{Theorem_achievability} to establish an achievability bound on the minimum required energy-per-bit for the scenario in which the number $\mathrm{K}_a$ of active users is random and unknown. We use a random coding scheme to generate a common codebook $\mathbf{C} = \left[ \mathbf{c}_{1}, \ldots, \mathbf{c}_{M} \right] \in \mathbb{C}^{n\times M}$ with each codeword  drawn i.i.d. according to $\mathcal{CN} \left(0,P'\mathbf{I}_{n}\right)$ or drawn uniformly i.i.d. from a sphere of radius $\sqrt{nP'}$. As explained in Theorem~\ref{Theorem_achievability}, we perform a change of measure, and the total variation distance between the true measure and the new one is bounded by $p_0$ in~\eqref{eq_noCSI_p0}.

  The decoder first obtains an estimate $K'_a$ of the number of active users as introduced in Section~\ref{Section:Ka}. Then, based on the MAP criterion, the decoder produces a set of decoded messages of size $\hat{K}_a$ belonging to an interval around the estimated value $K'_a$. That is, it is satisfied that $\hat{K}_a \in [K'_{a,l}, K'_{a,u}]$, where $K'_{a,l} = \max\left\{ K_l , K'_{a} -r' \right\}$, $K'_{a,u} = \min\left\{ K_u , K'_{a} +r' \right\}$, and $r'$ denotes a nonnegative integer referred to as the decoding radius. The per-user probability of misdetection in~\eqref{eq:MD} can be upper-bounded as in \eqref{eq:Theorem_achi_proofsketch0}, where $p_1$ can be bounded as in \eqref{eq:Theorem_achi_proofsketch2}. Likewise, the per-user probability of false-alarm in~\eqref{eq:FA} can be upper-bounded as
  \begin{equation}
    P_{\mathrm{FA}}
    \leq    \sum_{K_a=K_l}^{K_u}    P_{{\rm{K}}_a}(K_a)    \sum_{K'_a=K_l}^{K_u}
    \sum_{t \in \mathcal{T}_{K'_a} }
    \sum_{t' \in \mathcal{T}_{K'_a,t} }
        \frac{t' +  ( K'_{a,l}-K_a )^{+}}{ \hat{K}_a  } \mathbb{P}   \left[  \mathcal{F}_{t,t'}  \cap \left\{ K_a \to K'_{a} \right\} \right]_{\text{new}}
    +  p_0 , \label{eq_PUPE_FA_upper_noCSI_noKa_r_weight}
  \end{equation}
  where $\mathcal{F}_{t,t'}$ denotes the event that there are exactly $t+(K_a-K'_{a,u})^{+}$ misdetected codewords and $t' + ( K'_{a,l} - K_a )^{+}$ falsely alarmed codewords; the integer $t'$ takes value in $\mathcal{T}_{K'_a}$ defined in~\eqref{eq_noCSI_noKa_estimate_Tset2} because: 1)~$\hat{K}_a$ must be in $[K'_{a,l} : K'_{a,u} ]$; 2)~the number of falsely alarmed codewords is lower-bounded by $( K'_{a,l} - K_a )^{+}$ and upper-bounded by $M-K_a$; and 3)~there exist falsely alarmed codewords only when $\hat{K}_a\geq 1$.

  Next, we omit the subscript ``new'' for the sake of brevity. The probability on the RHS of~\eqref{eq:Theorem_achi_proofsketch2} can be bounded as
  \begin{align}
    \mathbb{P} \left[  \mathcal{F}_{t} \cap \left\{ K_a \to K'_{a} \right\} \right]
    & \leq \min \left\{ \mathbb{P}  \left[  \mathcal{F}_{t}
    \cap \left\{ \hat{K}_{a}  \in  [K'_{a,l} , K'_{a,u} ] \right\} \right]
    , \mathbb{P} \left[ K_a  \to  K'_{a} \right] \right\} \label{eq_PUPE_MD_FA_upper_noCSI_noKa2_r_weight}\\
    & \leq \min \left\{ 1, \sum_{t' \in \bar{\mathcal{T}}_{K'_a,t} }   \mathbb{P}  \left[  \mathcal{F}_{t,t'}
    \left|  \hat{K}_{a}  \in  [K'_{a,l}:K'_{a,u} ] \right.\right]
    , \mathbb{P} \left[ K_a  \to  K'_{a}  \right] \right\} . \label{eq_PUPE_MD_FA_upper_noCSI_noKa3_r_weight}
  \end{align}
  Here, $\bar{\mathcal{T}}_{K'_a,t}$ is defined in \eqref{eq_noCSI_noKa_estimate_Tset2bar}, which is obtained similar to $\mathcal{T}_{K'_a,t}$ with the difference that the number $\hat{K}_a$ of detected codewords can be $0$; \eqref{eq_PUPE_MD_FA_upper_noCSI_noKa2_r_weight} follows because the event $K_a \to K'_{a}$ implies that $ \hat{K}_{a} \in [K'_{a,l} : K'_{a,u} ]$ and the joint probability is upper-bounded by each of the individual probabilities~\cite{noKa}. Similarly, the probability on the RHS of~\eqref{eq_PUPE_FA_upper_noCSI_noKa_r_weight} can be bounded as
  \begin{equation}
    \mathbb{P} \left[  \mathcal{F}_{t,t'} \cap \left\{ K_a \to K'_{a} \right\} \right] \leq \min \left\{ 1 , \mathbb{P} \left[  \mathcal{F}_{t,t'}
    \left| \hat{K}_{a} \in [ K'_{a,l} : K'_{a,u} ] \right.\right]
    , \mathbb{P} \left[ K_a \to K'_{a} \right] \right\} . \label{eq_PUPE_MD_FA_upper_noCSI_noKa_r}
  \end{equation}
  The upper bound on $\mathbb{P} \left[ K_a \to K'_{a} \right]$ is given in Theorem~\ref{Theorem_Ka}. Then, we proceed to bound the probability $\mathbb{P} \left[  \mathcal{F}_{t,t'} \left| \hat{K}_{a} \in [ K'_{a,l} : K'_{a,u} ] \right.\right]$.

  We use the MAP decoder shown in \eqref{eq:decoderoutput_noCSI_W}, \eqref{eq:decoderoutput_noCSI}, and \eqref{eq:g_noCSI}. Let the set $\mathcal{W}_1 \subset \mathcal{W}$ of size $t+(K_a-K'_{a,u})^{+}$ denote the set of misdecoded messages. The set $\mathcal{W}_1$ can be divided into two subsets $\mathcal{W}_{1,1}$ and $\mathcal{W}_{1,2}$ of size $(K_a-K'_{a,u})^{+}$ and $t$, respectively. Let the set $\mathcal{W}_2 \subset [M]\backslash \mathcal{W}$ of size $t'+(K'_{a,l}-K_a)^{+}$ denote the set of false-alarm codewords. Let $\mathcal{W}_{2,1}$ denote an arbitrary subset of $\mathcal{W}_2$ of size $(K'_{a,l}-K_a)^{+}$. For the sake of simplicity, we rewrite ``$\cup_{ \mathcal{W}_{1} \subset \mathcal{W}, \;\! \left| \mathcal{W}_{1} \right| = t+(K_a-K'_{a,u})^{+} }  $'' to ``$\cup_{\mathcal{W}_{1}}$'' and ``$\cup_{{\mathcal{W}_{2} \subset [M] \backslash \mathcal{W},\;\! \left| \mathcal{W}_{2} \right| = t'+(K'_{a,l}-K_a)^{+} } }$'' to ``$\cup_{\mathcal{W}_2}$''; similarly for $\sum$ and $\cap$. Then, we have
  \begin{align}
      \mathbb{P}  \left[  \mathcal{F}_{t,t'} \left|  \hat{K}_a \in [K'_{a,l}, K'_{a,u} ] \right.\right]
      & \leq \mathbb{P}   \left[ \left. \mathcal{G}_e
      \right| \hat{K}_a \in [{K}'_{a,l} ,{K}'_{a,u} ] \right] \label{eq:proof_noCSI_noKa_pftt1} \\
      & \leq  \min_{ 0 \leq \omega \leq 1 , \nu\geq0 }
      \left\{  \mathbb{P}  \left[ \left. \mathcal{G}_e
      \cap \mathcal{G}_{\omega,\nu}
      \right| \hat{K}_a  \in  [{K}'_{a,l} ,{K}'_{a,u} ] \right] 
       +  \mathbb{P}   \left[ \mathcal{G}_{\omega,\nu}^c  \right]
      \right\} ,  \label{eq:proof_noCSI_noKa_pftt}
  \end{align}
  where $\mathcal{G}_e = {\cup}_{\mathcal{W}_{1}}  {\cup}_{\mathcal{W}_{2}} \left\{ g  \left( {\boldsymbol{\Gamma}}_{ \mathcal{W}  \backslash \mathcal{W}_1 \cup \mathcal{W}_2 }   \right) \leq  g  \left( \boldsymbol{\Gamma}_{ \mathcal{W} \backslash \mathcal{W}_{1,1} \cup \mathcal{W}_{2,1} } \right)  \right\}$, and \eqref{eq:proof_noCSI_noKa_pftt} follows from Fano's bounding technique $\mathbb{P}\left[A\right] \leq \mathbb{P}\left[A\cap B\right] +\mathbb{P}\left[ B^c \right]$ given in~\cite{1961}. According to the properly selected region around the linear combination of the transmitted signals given in~\cite[Eq.~(14)]{TIT}, we define the event
  \begin{equation}
    \mathcal{G}_{\omega,\nu} = \bigcap_{\mathcal{W}_1}  \left\{ g\left( \boldsymbol{\Gamma}_{\mathcal{W}} \right) \leq \omega g\left( {\boldsymbol{\Gamma}}_{ \mathcal{W} \backslash \mathcal{W}_1 } \right) + nL\nu \right\}.
  \end{equation}

  Applying similar ideas in \cite[(110) and (111)]{TIT}, the first probability on the RHS of~\eqref{eq:proof_noCSI_noKa_pftt} can be bounded as
    \begin{align}
      & \mathbb{P}  \left[ \left. \mathcal{G}_e
      \cap \mathcal{G}_{\omega,\nu}
      \right| \hat{K}_a \in [{K}'_{a,l} ,{K}'_{a,u} ] \right] \notag\\ 
      & \leq   C_{K'_a,t,t'}
      \mathbb{E}_{ \mathbf{C}_{ \mathcal{W} \cup \mathcal{W}_2 } }   \left[ \min_{ {0\leq u,0\leq r} } 
      \mathbb{E}_{\mathbf{H},\mathbf{Z}}  \Big[  \exp  \left\{ rnL\nu  +
      u g  \left( \boldsymbol{\Gamma}_{ \mathcal{W} \backslash \mathcal{W}_{1,1} \cup \mathcal{W}_{2,1} } \right)
      - u g \left( {\boldsymbol{\Gamma}}_{ \mathcal{W}  \backslash \mathcal{W}_1 \cup \mathcal{W}_2 }   \right)
      - r g  \left(\boldsymbol{\Gamma}_{ \mathcal{W} } \right)
       \right. \right.  \notag\\
      & \;\;\;\;\;
      \left.\left.\left.
      + \;\! r\omega \;\! g \left(  {\boldsymbol{\Gamma}}_{ \mathcal{W} \backslash \mathcal{W}_1 } \right) \right\}       \right|  
      \mathbf{C}_{ \mathcal{W} \cup \mathcal{W}_2 } , \hat{K}_a \in [{K}'_{a,l} ,{K}'_{a,u} ] \right] \bigg],  \label{eq_noCSI_q1t_chernoff}
    \end{align}
  where the term $C_{K'_a,t,t'} = \binom{K_a}{ t+(K_a-K'_{a,u})^{+} } \binom{M-K_a}{t'+(K'_{a,l}-K_a)^{+}}$. For any subset $S \subset [M]$, the matrix $\mathbf{C}_S\in \mathbb{C}^{n\times |S|}$ denotes the concatenation of codewords $\left\{ \mathbf{c}_i : i\in S \right\}$. Exploiting symmetry, we assume w.l.o.g. that $\mathcal{W}=[K_a]$, $\mathcal{W}_{1,1} = [(K_a-K'_{a,u})^{+}]$, $\mathcal{W}_{1,2} = [(K_a-K'_{a,u})^{+}+1:(K_a-K'_{a,u})^{+}+t]$, $\mathcal{W}_{2,1} = [K_a+1 : K_a+(K'_{a,l}-K_a)^{+}]$, and $\mathcal{W}_{2,2} = [K_a+(K'_{a,l}-K_a)^{+}+1: K_a+(K'_{a,l}-K_a)^{+}+t']$. Define $\mathbf{F}$, $\mathbf{F}_1$, $\mathbf{F}'$, $\mathbf{F}''$, and $b_{u,r}$ as in Theorem~\ref{Theorem_achievability}.  
      
  Denote the conditional expectation over $\mathbf{H}$ and $\mathbf{Z}$ on the RHS of~\eqref{eq_noCSI_q1t_chernoff} as $E_1$, which can be calculated as
  \begin{equation}
      E_1
      = \exp  \left\{ L \left(
      u \ln  \left|\mathbf{F}''\right|
       -  u \ln  \left| {\mathbf{F}'} \right|
      - r \ln  \left|\mathbf{F}\right|
      + r\omega \ln  \left| \mathbf{F}_{1} \right|
      - \ln  \left| \mathbf{B} \right| \right)  + b_{u,r}
      \right\}   \label{eq_noCSI_q1t_exp_y}.
  \end{equation}
  Here, \eqref{eq_noCSI_q1t_exp_y} follows from Lemma \ref{expectation_bound} by taking the expectation over $\mathbf{H}$ and $\mathbf{Z}$ provided that the minimum eigenvalue of $\mathbf{B}$ satisfies  $\lambda_{\min}\left(\mathbf{B}\right) > 0$, where the matrix $\mathbf{B}$ is given in \eqref{eq_noCSI_noKa_estimation_B}. Substituting \eqref{eq_noCSI_q1t_exp_y} into \eqref{eq_noCSI_q1t_chernoff}, we can obtain an upper bound on $\mathbb{P}  \left[ \left. \mathcal{G}_e \cap \mathcal{G}_{\omega,\nu} \right| \hat{K}_a \in [{K}'_{a,l} ,{K}'_{a,u} ] \right]$, which is denoted as $q_{1,K'_a,t,t'} \left(\omega,\nu\right)$ given in~\eqref{eq_noCSI_noKa_estimation_q1t}.

  Define the event $\mathcal{G}_\delta =\bigcap_{\mathcal{W}_1} \left\{ \sum_{i=1}^{n}  {\chi_i^2(2L)}  \leq  2nL(1+\delta) \right\}$ for $\delta\geq0$. In the case of $t+(K_a-K'_{a,u})^{+} > 0$, we can bound the second probability $\mathbb{P}  \left[ \mathcal{G}_{\omega,\nu}^c  \right]$ on the RHS of~\eqref{eq:proof_noCSI_noKa_pftt} as
  \begin{equation} \label{eq:noCSI_q2t_goodregion}
    \mathbb{P} \left[ \mathcal{G}_{\omega,\nu}^c  \right]
    \leq \min_{\delta\geq 0}
    \left\{  \mathbb{P} \left[ \mathcal{G}_{\omega,\nu}^c \cap \mathcal{G}_\delta  \right]
    + \mathbb{P} \left[ \mathcal{G}_\delta^c \right]  \right\} , 
  \end{equation}
  where $\mathbb{P} \left[ \mathcal{G}_\delta^c \right] = \binom{K_a}{t+(K_a-K'_{a,u})^{+}} \left( 1 - \frac{\gamma\left( nL, nL \left( 1+\delta \right)\right)}{\Gamma\left( nL \right)} \right)$ and $\mathbb{P} \left[  \mathcal{G}_{\omega,\nu}^c \cap \mathcal{G}_\delta \right]$ can be bounded as
    \begin{equation}
      \mathbb{P}  \left[ \mathcal{G}_{\omega,\nu}^c \cap \mathcal{G}_{\delta} \right]
      \leq  \binom{K_a}{t+(K_a-K'_{a,u})^{+}} \mathbb{E}_{\mathbf{C}_{ \mathcal{W} }}   \left[
      \frac{\gamma  \left( Lm, \frac{nL(1+\delta)(1-\omega)- \bar{b} }{\omega \prod_{i=1}^{m} \lambda_i^{ {1}/{m}}  } \right)}{\Gamma\left( Lm \right)}  \right]  .\label{eq:noCSI_q2t_prod}
    \end{equation}
  Here, the term $\bar{b} = \omega L \ln \left|\mathbf{F}_1\right| - L \ln \left|\mathbf{F}\right| + n L \nu - \omega b_1 + b $, and $\lambda_1, \ldots, \lambda_n$ denote the eigenvalues of the matrix $\mathbf{F}_1^{-1} \mathbf{C}\boldsymbol{\Gamma}_{\mathcal{W}_1}\mathbf{C}^H$ of rank $m = \min \left\{ n, t+(K_a-K'_{a,u})^{+} \right\}$ in decreasing order. The upper bound in \eqref{eq:noCSI_q2t_prod} is obtained applying similar ideas in \cite[(197)-(202)]{TIT} with the following changes: 1)~the number of misdetections is changed from $t$ in \cite{TIT} to $t+(K_a-K'_{a,u})^{+}$ since there are additional $(K_a-K'_{a,u})^{+}$ misdetections due to the inaccuracy in the estimation of the number of active users; and 2)~since we consider the MAP criterion instead of the ML criterion used in \cite[(197)-(202)]{TIT}, $b$ and $b_1$ given in \eqref{eq_noCSI_noKa_estimation_b} and \eqref{eq_noCSI_noKa_estimation_b1}, respectively, are introduced in \eqref{eq:noCSI_q2t_prod}. Substituting \eqref{eq:noCSI_q2t_prod} into \eqref{eq:noCSI_q2t_goodregion}, we obtain an upper bound on $\mathbb{P}  \left[ \mathcal{G}_{\omega,\nu}^c \right]$ in the case of $t+(K_a-K'_{a,u})^{+} > 0$, which is denoted as $q_{2,K'_a,t}\left(\omega,\nu\right)$ given in~\eqref{eq_noCSI_noKa_estimation_q2t}. In the case of $t+(K_a-K'_{a,u})^{+} = 0$, we have
  \begin{align}
    \mathbb{P}  \left[ \mathcal{G}_{\omega,\nu}^c \right] 
    & = \mathbb{P} \left[ g\left(\boldsymbol{\Gamma}_{\mathcal{W}} \right) >  \frac{nL\nu}{1-\omega} \right] \\
    & = \mathbb{E}_{\mathbf{C}_{ \mathcal{W} }} \left[ 1 - \frac{\gamma\left( nL, \frac{nL\nu}{1-\omega} - L \ln \left| \mathbf{F} \right|  +  b \right)}{\Gamma\left( nL \right)}  \right] , \label{eq:proof_noCSI_noKa_pft_t0_rbig}
  \end{align}
  where the constant $b$ is given in~\eqref{eq_noCSI_noKa_estimation_b}. The RHS of \eqref{eq:proof_noCSI_noKa_pft_t0_rbig} is denoted as $q_{2,K'_a,t,0}$ as in~\eqref{eq_noCSI_noKa_estimation_q2t0}. This concludes the proof of Theorem~\ref{Theorem_achievability}.


\section{Proof of Theorem~\ref{Theorem_converse_single_noKa}} \label{Appendix_proof_converse_single_noKa}

  We assume there are $K$ users and each of them becomes active with probability $p_a$ independently. The converse bound for the scenario with knowledge of the activities of $K-1$ users and the transmitted codewords and channel coefficients of active users among them, is also converse in the case without this knowledge. If the remaining user is active, it equiprobably selects a message $W \in [M]$, and the transmitted codeword satisfies the maximal power constraint in~\eqref{eq:power_constraint}. Let $\mathcal{F} \subset \mathbb{C}^n$ be a set of channel inputs satisfying~\eqref{eq:power_constraint}. Denote the set of decoded message as $\hat{\mathcal{W}}$ of size $|\hat{\mathcal{W}}| \leq K$. The requirements on the per-user probabilities of misdetection and false-alarm are given by
    \begin{equation} \label{eqR:MD}
      P_{e,{\rm{MD}}}  =   \frac{p_a}{ M }  \sum_{m \in [M]}  \mathbb{P} \big[ m  \notin  \hat{\mathcal{W}}  | W  =  m \big]   \leq  p_a \epsilon_2  =  \epsilon_{\rm{MD}},
    \end{equation}
    \begin{equation} \label{eqR:FA2}
      P_{e,{\rm{FA}}}  \leq  (1 - p_a) \mathbb{P} \big[ \hat{\mathcal{W}} \neq \emptyset | W  =  0 \big] \leq (1-p_a) \epsilon_1 = \epsilon_{\rm{FA}}.
    \end{equation}

  In the case of $|\hat{\mathcal{W}}|\leq 1$, an upper bound on the number of codewords that are compatible with the requirements on the probabilities of false-alarm, misdetection, and inclusive error is provided in~\cite[Theorem~2]{On_joint}. This converse bound was derived based on \cite[Theorem 15]{yury_beta} and \cite[Theorem 27]{Channel_coding_rate}. By enlarging the list size from $|\hat{\mathcal{W}}|\leq 1$ to $|\hat{\mathcal{W}}|\leq K$, changing the error requirement in~\cite{On_joint} to~\eqref{eqR:MD} and~\eqref{eqR:FA2}, and considering MIMO fading channels, we can obtain the following proposition.
  \begin{prop} \label{R_Theorem_AWGN_singleUE_conv}
    Consider the single-user setup with active probability $p_a$. Let $Q_{ Y^{n\times L} }$ be an arbitrary distribution on $\mathbb{C}^{n\times L}$. The $(n,M,\epsilon_{\rm{MD}},\epsilon_{\rm{FA}},P)$ code satisfies
    \begin{equation} \label{eqR:beta_conv_AWGN_singleUE_1}
        \frac{M}{K} \leq   \sup_{  { {P_{X^{n}} : \mathbf{x} \in \mathcal{F} , \epsilon_{1} ,\epsilon_{2} \in [0,1]} } }
        \frac{ 1 - \beta_{1-\epsilon_{1}}  \left( P_{ Y^{n\times L} | X^{n} = \mathbf{0} } , Q_{ Y^{n\times L} } \right) }{ \beta_{ 1-\epsilon_{2} }
         \left( P_{ X^{n} } P_{Y^{n\times L}|X^{n}} , P_{X^{n}} Q_{Y^{n\times L}} \right)  } ,
      \end{equation}
    where
      \begin{equation}
        p_a \epsilon_{2} \leq \epsilon_{\rm{MD}} ,
      \end{equation}
      \begin{equation}
        (1-p_a) \epsilon_{1} \leq \epsilon_{\rm{FA}} .
      \end{equation}
    \begin{IEEEproof}
      Let $P_{X^{n}}$ be the distribution on $X^{n}$ induced by the encoder when the messages are uniform. Let $P_{Y^{n\times L}}$ be the output distribution. By assumption, we have
      \begin{equation}\label{eqR:proof_epsilon1}
        P_{ Y^{n\times L} \mid X^{n}=\bf{0} } \big[ \hat{\mathcal{W}} = \emptyset \big] \geq 1-\epsilon_1.
      \end{equation}
      Let $Z = 1\big[ \big\{ W\in\hat{\mathcal{W}} \big\} \cap \big\{ \hat{\mathcal{W}} \neq \emptyset \big\} \big]$. We have
      \begin{equation}\label{eqR:proof_PXY}
        P_{X^{n}, Y^{n\times L}}[Z=1]=P_{X^{n}, Y^{n\times L}}[ W\in\hat{\mathcal{W}} ] \geq 1-\epsilon_2,
      \end{equation}
      \begin{align}\label{eqR:proof_PXQY}
        P_{X^{n}} Q_{Y^{n\times L}}[Z=1]
        & = \sum_{W\in[M]} \frac{1}{M}
        \sum_{\hat{\mathcal{W}} \in \binom{[M]}{\leq K} } Q (\hat{\mathcal{W}}) 1\left[ \left\{ W \in \hat{\mathcal{W}} \right\} \cap \left\{ \hat{\mathcal{W}} \neq \emptyset \right\} \right] \notag\\
        & = \frac{1}{M} \sum_{\hat{\mathcal{W}} \in \binom{[M]}{\leq K} } Q (\hat{\mathcal{W}})
        \; |\hat{\mathcal{W}}| \notag\\
        & \leq \frac{K  Q_{Y^{n\times L}} \left[ \hat{\mathcal{W}} \neq \emptyset \right] }{M}.
      \end{align}
      It then follows that
      \begin{equation}\label{eqR:proof_beta_epsilon2}
        \beta_{1-\epsilon_2}  \left( P_{X^{n}, Y^{n\times L}}, P_{X^{n}} Q_{Y^{n\times L}} \right)
         \leq  \frac{ K}{M}  Q_{Y^{n\times L}}   \big[ \hat{\mathcal{W}} \neq \emptyset \big] .
      \end{equation}
      The term $Q_{Y^{n\times L}} \left[ \hat{\mathcal{W}} \neq \emptyset \right]$ can be further bounded as
      \begin{align}
        Q_{Y^{n\times L}}   \big[ \hat{\mathcal{W}} \neq \emptyset \big]
        & =  1-Q_{Y^{n\times L}} \big[ \hat{\mathcal{W}} = \emptyset \big] \\
        & \leq  1-\beta_{1-\epsilon_1} \left( P_{ Y^{n\times L} \mid X^{n}=\bf{0} }, Q_{Y^{n\times L}} \right) .\label{eqR:proof_beta_epsilon2_Q}
      \end{align}
      Substituting \eqref{eqR:proof_beta_epsilon2_Q} into \eqref{eqR:proof_beta_epsilon2}, \eqref{eqR:beta_conv_AWGN_singleUE_1} can be obtained.
    \end{IEEEproof}
  \end{prop}

  Next, we follow similar lines as in~\cite[Appendix~I]{TIT} to loosen Proposition~\ref{R_Theorem_AWGN_singleUE_conv} and obtain an easy-to-evaluate bound in~\eqref{eqR:beta_conv_AWGN_singleUE_1e}, which completes the proof of Theorem~\ref{Theorem_converse_single_noKa}.

%
\section{Proof of Theorem~\ref{Theorem_converse_noCSI_Gaussian_cKa}} \label{Appendix_proof_converse_noCSI_Gaussian_noKa}

  When there are $K_a$ active users, we assume w.l.o.g. that the active user set is $\mathcal{K}_a = [K_a]$. Let $\mathbf{X}\in\mathbb{C}^{n\times M}$ be the codebook matrix. Denote $\bar{\mathbf{X}}_{K_aM} = \left[\mathbf{X},\mathbf{X},\ldots,\mathbf{X}\right] \in \mathbb{C}^{n\times K_aM}$ and $\bar{\mathbf{X}} = \operatorname{diag} \left\{\bar{\mathbf{X}}_{K_aM}, \bar{\mathbf{X}}_{K_aM}, \ldots, \bar{\mathbf{X}}_{K_aM} \right\} \in \mathbb{C}^{nL\times K_aML}$. Let $\bar{\mathbf{H}}_{l}$ be a $K_a M  \times K_a M$ block diagonal matrix, whose block $k$ is a diagonal $M \times M$ matrix with diagonal entries equal to ${h}_{k,l}$. Let $\bar{\mathbf{H}} = \left[\bar{\mathbf{H}}_{1}, \ldots , \bar{\mathbf{H}}_{L} \right]^T$. The vector $\bar{\boldsymbol{\beta}} \in \left\{ 0,1 \right\}^{K_aM}$ has $K_a$ blocks, whose block $k$ denoted as $\bar{\boldsymbol{\beta}}_k$ is of size $M$ and includes one $1$. Assume $\bar{\mathbf{z}} \sim \mathcal{CN} ({\bf{0}} , \mathbf{I}_{nL} )$. We have
  \begin{equation}
    \bar{\mathbf{y}} = \bar{\mathbf{X}}  \bar{\mathbf{H}}  \bar{\boldsymbol{\beta}} + \bar{\mathbf{z}} \in \mathbb{C}^{nL\times 1}.
  \end{equation}

  For $k\in[K_a]$, define $\mathcal{M}_k = 1 [ W_k \notin \hat{\mathcal{W}}] $, where $\hat{\mathcal{W}}$ denotes the decoded list of size $| \hat{\mathcal{W}} | = \hat{K}_a \leq K$. Let $P_{e,k} = \mathbb{P} [ \mathcal{M}_k = 1 ]$, $\bar{P}_{e, K_a} = \frac{1}{K_a}\sum_{k\in [K_a]} P_{e,k}$ if $K_a \geq 1$, and $\bar{P}_{e, K_a} = 0$ if $K_a =0$. We have $P_{e} = \sum_{K_a \geq 1} P_{{\rm K}_a}(K_a) \bar{P}_{e,K_a} \leq \epsilon_{\rm MD}$. For any user $k\in[K_a]$, the standard Fano inequality gives that
  \begin{align}
    &H_2( W_k | \bar{\mathbf{X}} , {\rm{K}}_a = K_a )
    - H_2\left( \mathcal{M}_k | \bar{\mathbf{X}} , {\rm{K}}_a = K_a \right)
    - H_2 ( W_k | \mathcal{M}_k , \hat{\mathcal{W}} , \bar{\mathbf{X}} , {\rm{K}}_a = K_a ) \notag\\
    & \leq I_2  (   \mathcal{W}_{k};\hat{\mathcal{W}} \left| \bar{\mathbf{X}} , {\rm{K}}_a = K_a \right. ) , \label{eq:fano_noKa}
  \end{align}
  where the first three entropies are given by $H_2( W_k | \bar{\mathbf{X}} , {\rm{K}}_a = K_a ) = J$, $ H_2( \mathcal{M}_k | \bar{\mathbf{X}} , {\rm{K}}_a = K_a ) = h_2(P_{e,k})$, and
  \begin{align}
    H_2 ( W_k | \mathcal{M}_k , \hat{\mathcal{W}} , \bar{\mathbf{X}} , {\rm{K}}_a  =  K_a )
    & \leq (1 - P_{e,k})   \log_2  \hat{K}_a  +  P_{e,k} J \label{eq:fano_noCSI_entropy1} \\
    & \leq \log_2 \hat{K}_a + P_{e,k} J. \label{eq:fano_noCSI_entropy}
  \end{align}

  Then, by taking the average over $k\in[K_a]$ and $K_a \in S \subset \mathcal{K}$ on both sides of~\eqref{eq:fano_noKa}, and applying Jensen's inequality and $\sum_{k\in[K_a]}  I_2 (  \mathcal{W}_{k};\hat{\mathcal{W}}  \left| \bar{\mathbf{X}} , {\rm{K}}_a  =  K_a \right.  )  \leq  I_2 \big(  \mathcal{W}_{[K_a]};\hat{\mathcal{W}}  \left| \bar{\mathbf{X}} , {\rm{K}}_a  =  K_a  \right.\big)$, we can obtain
  \begin{align}
    & \left( \mathbb{P}[ {\rm K}_a \in S ] - \sum_{K_a \in S} P_{{\rm K}_a}(K_a) \bar{P}_{e,K_a} \right) J
    - \sum_{K_a \in S} P_{{\rm K}_a}(K_a) h_2 \left( \bar{P}_{e,K_a} \right)
    - \sum_{K_a \in S} P_{{\rm K}_a}(K_a) \log_2  \hat{K}_a \notag\\
    & \leq
    \sum_{K_a \in S} P_{{\rm K}_a}(K_a)  \frac{1}{K_a} I_2 \big(  \mathcal{W}_{[K_a]};\hat{\mathcal{W}} \left| \bar{\mathbf{X}} , {\rm{K}}_a  =  K_a \right.  \big) . \label{eq:fano_noCSI_Ka2}
  \end{align}
  It is satisfied that
  \begin{align}
    \sum_{K_a \in S} P_{{\rm K}_a}(K_a) \bar{P}_{e,K_a} J
    + \sum_{K_a \in S} P_{{\rm K}_a}(K_a) h_2\left( \bar{P}_{e,K_a} \right) 
    & \leq P_e J + h_2(P_e) \label{eq:fano_noCSI_Ka2_left1} \\
    & \leq \epsilon_{\rm MD} J + h_2(\epsilon_{\rm MD}), \label{eq:fano_noCSI_Ka2_left2}
  \end{align}
  where \eqref{eq:fano_noCSI_Ka2_left1} follows because $S \subset \mathcal{K}$ and by applying Jensen's inequality for $h_2(\cdot)$ and \eqref{eq:fano_noCSI_Ka2_left2} holds under the condition that $P_e \leq \epsilon_{\rm MD} \leq \frac{M}{1+M}$.

  The mutual information in~\eqref{eq:fano_noCSI_Ka2} can be upper-bounded as~\cite[Eq.~(149)]{finite_payloads_fading}
    \begin{equation}
      I_2 \big(  \mathcal{W}_{[K_a]};\hat{\mathcal{W}} \left| \bar{\mathbf{X}} , {\rm{K}}_a = K_a \right.  \big)  \leq
      I_2  \left( \bar{\mathbf{H}}\bar{\boldsymbol{\beta}}; \bar{\mathbf{y}} | \;\!  \bar{\mathbf{X}} , {\rm{K}}_a =  K_a  \right)
       -  I_2  \left( \bar{\mathbf{H}}\bar{\boldsymbol{\beta}}; \bar{\mathbf{y}} | \;\! \bar{\boldsymbol{\beta}}, \bar{\mathbf{X}} , {\rm{K}}_a   =  K_a  \right) .\label{eq_conv_fano_nocsi}
    \end{equation} 
  Next, we focus on the two terms on the RHS of~\eqref{eq_conv_fano_nocsi}. We have
    \begin{equation}
      I_2\left(\left. \bar{\mathbf{H}}\bar{\boldsymbol{\beta}};\bar{\mathbf{y}} \right| \bar{\mathbf{X}} = \bar{\mathbf{X}}^r , {\rm{K}}_a = K_a \right)
      \leq L \log_2 \left| \mathbf{I}_{n } + \frac{K_a}{M} \mathbf{X}^{r} \left( \mathbf{X}^{r}\right)^{H} \right|,\label{eq_conv_inf_sup3}
    \end{equation}
  which follows by applying similar ideas in \cite[(236)-(239)]{TIT} by allowing all users share the same codebook. Here, $ \mathbf{X}^r$ is a realization of $\mathbf{X}$ and $\bar{\mathbf{X}}^r$ is a realization of $\bar{\mathbf{X}}$. Then, we have
    \begin{align}
      I_2 \left( \left. \bar{\mathbf{H}}\bar{\boldsymbol{\beta}}; \bar{\mathbf{y} } \right| \bar{\mathbf{X}} , {\rm{K}}_a = K_a \right)
      & \leq L \mathbb{E}_{\mathbf{X}}   \left[  \left. \log_2 \left| \mathbf{I}_{n} + \frac{K_a}{M} \mathbf{X} \mathbf{X}^{ H}  \right|  \; \right|   {\rm{K}}_a = K_a \right] \\
      & \leq nL \log_2 \left( 1 + K_aP \right), \label{eq_conv_inf_term1}
    \end{align}
    where \eqref{eq_conv_inf_term1} follows from the concavity of the $\log_2 \left|\cdot\right|$ function and Jensen's inequality under the assumption that the entries of codebooks are i.i.d. with mean $0$ and variance $P$. Following from (43) and (44) in~\cite{letter}, we have 
    \begin{equation}
      I  \left( \bar{\mathbf{H}}\bar{\boldsymbol{\beta}};\bar{\mathbf{y}} | \bar{\boldsymbol{\beta}}, \bar{\mathbf{X}}, {\rm{K}}_a = K_a \right)
      \geq \left( 1- \frac{1}{M} {\binom{K_a}{2}} \right) L \mathbb{E} \left[  \log_2  \left| \mathbf{I}_{n} + \mathbf{X}_{K_a}  \mathbf{X}_{K_a}^{ H} \right|  \right] \label{eq_conv_inf_term2},
    \end{equation}
    where $\mathbf{X}_{K_a}$ includes $K_a$ different codewords.

    Substituting \eqref{eq:fano_noCSI_Ka2_left2}, \eqref{eq_conv_fano_nocsi}, \eqref{eq_conv_inf_term1}, and \eqref{eq_conv_inf_term2} into \eqref{eq:fano_noCSI_Ka2}, we can obtain~\eqref{P_tot_conv_noCSI} in Theorem~\ref{Theorem_converse_noCSI_Gaussian_cKa}. Moreover, we can obtain \eqref{P_tot_conv_noCSI_FA_Fano} considering that at least $\max \{0,\hat{K}_a-K_a\}$ messages are false-alarmed when the size of the decoded list is $\hat{K}_a$.

\section{ Proof of Theorem \ref{Theorem_scalinglaw_noCSI_Kaknown}  } \label{Proof_scalinglaw_noCSI_Kaknown}

  We first establish in Appendix \ref{Proof_scalinglaw_noCSI_achi_Kaknown} the achievability result of Theorem \ref{Theorem_scalinglaw_noCSI_Kaknown} and then Appendix~\ref{Proof_scalinglaw_noCSI_conv_Kaknown} shows the converse part.

\subsection{ Achievability  bound } \label{Proof_scalinglaw_noCSI_achi_Kaknown}

  Consider the case where the number $K_{a,n}$ active users is known in advance and the receiver is required to output a list of $K_{a,n}$ messages. All users share a common codebook $\mathbf{X} \in \mathbb{C}^{n\times M_n}$ with each codeword drawn uniformly i.i.d. from the sphere of radius $\sqrt{nP_n}$. Let $\mathcal{W}$ denote the set of transmitted messages of size no more than $K_{a,n}$. To derive an upper bound on the error probability, we change the measure to the new one under which the number of users who select the same message is assumed to be no more than a constant $D$. The total variation distance between the true measure and the new one can be bounded by $p_0$ as follows
  \begin{align}
    p_0
    & \leq \sum_{d=D}^{K_{a,n}} \frac{ \binom{K_{a,n}}{d} M_n (M_n-1)^{K_{a,n}-d} }{ M_n^{K_{a,n}} }   \\
    & \leq \sum_{d=D}^{K_{a,n}} \left( \frac{eK_{a,n}}{d} \right)^{d} \frac{1}{M_n^{d-1}}  \label{eq:p0_scalinglaw1_Kaknown} \\
    & \leq \left( \frac{e}{D} \right)^{D} \frac{K_{a,n}^{D+1}}{M_n^{D-1}}  , \label{eq:p0_scalinglaw_Kaknown}
  \end{align}
  where \eqref{eq:p0_scalinglaw1_Kaknown} holds due to $\binom {a } {b} \leq \left( \frac{ea}{b} \right)^{b}$ for $a\geq b>0$; \eqref{eq:p0_scalinglaw_Kaknown} follows because $\left( \frac{eK_{a,n}}{d} \right)^{d} \frac{1}{M_n^{d-1}}$ decreases with $d$ in the case of $d\geq1$ and $M_n\geq K_{a,n}$. Under the assumption of $M_n \geq K_{a,n}^{r_1}$ for a constant $r_1>1$, there exists a constant $D$ making sure $p_0 \to 0$ as $M_n,K_{a,n}\to\infty$.

  Then, we upper-bound the per-user probability of misdetection as
  \begin{align}
    P_{\rm MD}
    &\leq \frac{1}{K_{a,n}}  \sum_{k\in {\mathcal{K}_{a,n}}}   \mathbb{P}  \left[ W_{k}  \notin  \hat{\mathcal{W}}  \right]_{\text{new}}   +  p_0 \label{eq_PUPE_MD_upper_noCSI_noKa_scalinglaw0_Kaknown} \\
    & = \sum_{t = 0 }^{|\mathcal{W}|}
    \frac{t}{K_{a,n}} \mathbb{P} \left[  \mathcal{F}_{t,K_{a,n}-|\mathcal{W}|+t} \right]_{\text{new}}
    + p_0   \\
    & \leq \sum_{t = 1 }^{|\mathcal{W}|} \mathbb{P} \left[  \mathcal{F}_{t,K_{a,n}-|\mathcal{W}|+t} \right]_{\text{new}}
    + p_0, \label{eq_PUPE_MD_upper_noCSI_noKa_scalinglaw2_Kaknown}
  \end{align}
  where the subscript ``new'' indicates that the probability is measured under the aforementioned new measure; \eqref{eq_PUPE_MD_upper_noCSI_noKa_scalinglaw0_Kaknown} follows from~\eqref{eq:MD} and \eqref{eq:p0_scalinglaw_Kaknown}; and $\mathcal{F}_{t,K_{a,n}-|\mathcal{W}|+t}$ denotes the event that there are $t$ misdetected messages and $K_{a,n}-|\mathcal{W}|+t$ falsely alarmed messages. Likewise, $P_{\mathrm{FA}}$ can be upper-bounded as
  \begin{align}
    P_{\mathrm{FA}}
    & \leq \sum_{t = 0}^{|\mathcal{W}|}
    \frac{K_{a,n}-|\mathcal{W}|+t}{ K_{a,n} }
    \mathbb{P} \left[  \mathcal{F}_{t,K_{a,n}-|\mathcal{W}|+t} \right]_{\text{new}} + p_0 \\
    &\leq \sum_{t = 1 \left[ K_{a,n}-|\mathcal{W}| = 0 \right]}^{|\mathcal{W}|} \mathbb{P} \left[  \mathcal{F}_{t,K_{a,n}-|\mathcal{W}|+t} \right]_{\text{new}} +  p_0. \label{eq_PUPE_FA_upper_noCSI_noKa_scalinglaw4_Kaknown}
  \end{align}

  For a set $S\subset[M_n]$, the matrix ${\boldsymbol{\Gamma}}_{S} = \operatorname{diag} \left\{ {\boldsymbol{\gamma}}_{S} \right\}\in {\mathbb{N}}^{M_n\times M_n}$ satisfies $1 \leq \left[ {\boldsymbol{\gamma}}_{S} \right]_{i} \leq D$ if $i\in S$ and $\left[ {\boldsymbol{\gamma}}_{S} \right]_{i} = 0$ otherwise. Based on the ML criterion, the log-likelihood cost function is given by
  \begin{equation}
    g\left( {\boldsymbol{\Gamma}_{S}} \right)
     =  L_n \ln  \left| \mathbf{I}_{n}   +  \mathbf{X}\boldsymbol{\Gamma}_{S} \mathbf{X}^H \right|
    + \operatorname{tr} \left(  \left( \mathbf{I}_{n}   +  \mathbf{X}\boldsymbol{\Gamma}_{S} \mathbf{X}^H \right)^{ -1}  \mathbf{Y} \mathbf{Y}^H \right).
  \end{equation}
  Next, we omit the subscript ``new'' in~\eqref{eq_PUPE_MD_upper_noCSI_noKa_scalinglaw2_Kaknown} and \eqref{eq_PUPE_FA_upper_noCSI_noKa_scalinglaw4_Kaknown} and denote $t'=K_{a,n}-|\mathcal{W}|+t$ for simplicity. We can upper-bound $\mathbb{P} \left[ \mathcal{F}_{t,t'}  \right]$ as
  \begin{align}
      \mathbb{P} \left[ \mathcal{F}_{t,t'}   \right]
      &\leq  \mathbb{P}   \left[
      \cup_{ \stackrel{S_1\subset \mathcal{W}}{|S_{1}|=t} }  \cup_{ \stackrel{S_2 \subset [M_n] \backslash \mathcal{W}}{|S_2|=t'} }
        \left\{ g \left( {\boldsymbol{\Gamma}}_{  \mathcal{W} \backslash S_1 \cup S_2 , {\rm opt} } \right)
       \leq  g \left(\boldsymbol{\Gamma}_{ \mathcal{W} , {\rm true} } \right)  \right\}
       \right]  \\
      &\leq \binom {K_{a,n}} {t} \binom {M_n} {t'} \mathbb{P} \left[ g\left( {\boldsymbol{\Gamma}}_{ \mathcal{W} \backslash S_1 \cup S_2 , {\rm opt} } \right)
      \leq g \left(\boldsymbol{\Gamma}_{\mathcal{W} , {\rm true} } \right)
      \right] \label{eq_Pft_upper_noCSI_scalinglaw0_Kaknown} \\
      &\leq  \left( \max\{K_{a,n},1\} M_n \right)^{t+t'}
       \mathbb{P}  \left[ g \left( {\boldsymbol{\Gamma}}_{ \mathcal{W} \backslash S_1 \cup S_2 , {\rm opt} } \right)
      \leq g  \left(\boldsymbol{\Gamma}_{\mathcal{W} , {\rm true} } \right)
      \right] \label{eq_Pft_upper_noCSI_scalinglaw1_Kaknown} .
  \end{align}
  Here, the matrix $\boldsymbol{\Gamma}_{ \!\mathcal{W} , {\rm true} }$ indicates the true message selection pattern and the matrix ${\boldsymbol{\Gamma}}_{ \! \mathcal{W} \backslash  S_1 \cup S_2 , {\rm opt} }  \! =    \arg \min_{ {[ {\boldsymbol{\gamma}}_{ \mathcal{W}  \backslash  S_1 \cup S_2}]}_{ i}  \in[D] , i \in \mathcal{W} \backslash  S_1 \cup S_2 }  g  \left( {\boldsymbol{\Gamma}}_{  \mathcal{W} \backslash  S_1 \cup S_2}  \right) $; \eqref{eq_Pft_upper_noCSI_scalinglaw0_Kaknown} follows from the union bound; and \eqref{eq_Pft_upper_noCSI_scalinglaw1_Kaknown} follows by applying the inequality $\binom {K_{a,n}} {t} \leq \left( \max\{K_{a,n},1\} \right)^{t+t'}$ and the inequality $\binom {M_n} {t'} \leq M_n^{t+t'}$.

  Along similar lines as in (296) and (299) in~\cite{TIT}, we have
  \begin{equation}
    \mathbb{P}  \left[ g \left( {\boldsymbol{\Gamma}}_{  \mathcal{W} \backslash S_1 \cup S_2 , {\rm opt} } \right)
    \leq  g   \left(\boldsymbol{\Gamma}_{ \mathcal{W} , {\rm true} } \right) \right]
    \leq \mathbb{E}  \left[ \exp   \left\{  -
    \frac{ L_n\left\| \mathbf{F} - \mathbf{F}_2 \right\|_F^2 }
    {  8\sigma_{\rm max}^2  \left( \frac{\mathbf{F}+\mathbf{F}_2}{2}  \right) } \right\}  \right] ,
  \end{equation}
  where $\sigma_{max}\left( \frac{\mathbf{F}+\mathbf{F}_2}{2} \right)$ denotes the maximum singular value of $\frac{\mathbf{F}+\mathbf{F}_2}{2}$, $\mathbf{F} = \mathbf{I}_{n}  +  \mathbf{X} \boldsymbol{\Gamma}_{\mathcal{W} , {\rm true}} \mathbf{X}^H$, and $\mathbf{F}_2  = \mathbf{I}_{n}  +  \mathbf{X} {\boldsymbol{\Gamma}}_{ \mathcal{W} \backslash S_1 \cup S_2 , {\rm opt}}  \mathbf{X}^H$. Following from the restricted isometry property result in \cite[Theorems~2 and~5]{Caire1}, with probability exceeding $1-\exp\left\{ -c_{\delta} n  \right\}$ on a draw of the common codebook, we have
  \begin{equation}
    \left\| \mathbf{F}-\mathbf{F}_2 \right\|_F^2
    \geq c_2 (1-\delta) n^2 P_n^2 (t+t') ,\label{eq_RIP_lower_noCSI_scalinglaw_simple_Kaknown}
  \end{equation}
  provided that the conditions
  \begin{equation}\label{eq_noCSI_scalinglaw_RIP2_condition_simple_Kaknown}
     2n(n-1) < M_n < \frac{1}{4} \exp\left\{ c_1\sqrt{n(n-1)} \right\} ,
  \end{equation}
  \begin{equation}
    K_{a,n}  \leq C_{\delta} \frac{n(n-1)}{\ln^{2}\left( \frac{e M_n}{n(n-1)}\right)}, \label{eq_noCSI_scalinglaw_RIP1_condition_simple_Kaknown}
  \end{equation}
  are satisfied with positive constants $c_1,c_2,C_\delta$, and $\delta\in(0,1)$. Along similar lines as in~\cite[Appendix A]{Caire1}, we have
  \begin{equation}
    \sigma_{\max }  \left(  \frac{\mathbf{F}+\mathbf{F}_2}{2}  \right)
     \leq  P_n c_3 D
    \left( c_4 \ln  \frac{eM_n}{K_{a,n}} +  \frac{ c_{\epsilon} n + \ln2 }{ \max  \{ K_{a,n} , n  \} } \right) \max  \{ K_{a,n} , n  \}  +  1   \label{eq_daviation_lower_noCSI_scalinglaw_simple_Kaknown},
  \end{equation}
  independently of the set of transmitted codewords and that of decoded codewords with probability at least $1-\exp\left\{ -c_{\epsilon} n \right\}$, where $c_{\epsilon}$, $c_3$, and $c_4$ are positive constants. Denote by $\mathcal{G}_{t,t'}$ the event that the common codebook is such that the conditions in~\eqref{eq_RIP_lower_noCSI_scalinglaw_simple_Kaknown} and \eqref{eq_daviation_lower_noCSI_scalinglaw_simple_Kaknown} hold. If the event $\mathcal{G}_{t,t'}$ occurs, we can obtain that
  \begin{equation}
    \frac{ \left\| \mathbf{F}-\mathbf{F}_2 \right\|_F^2 }
    { \sigma_{max}^2\left( \frac{\mathbf{F}+\mathbf{F}_2}{2} \right) }
      \geq m^{*} (t+t')  ,  \label{eq_Dt_bound_noCSI_scalinglaw_Kaknown}
  \end{equation}
  \begin{equation}\label{eq_m_cond_noCSI_scalinglaw_Kaknown}
     m^{*}  =  \frac{ c_2(1-\delta)n^2 }{   \left(  c_3 D
    \left( c_4\ln \frac{eM_n}{K_{a,n}} + \frac{c_{\epsilon} n  + \ln2}{\max  \{ K_{a,n} , n  \} }  \right)
    \max  \{ K_{a,n} , n  \} + \frac{1}{P_n} \right)^{ 2}} .
  \end{equation}
  Then, we have
  \begin{align}
    \mathbb{P} \left[ \mathcal{F}_{t,t'} \right]
    & \leq \mathbb{P} \left[  \left. \mathcal{F}_{t,t'}  \right|  \mathcal{G}_{t,t'}  \right]
    + \mathbb{P} \left[  \mathcal{G}_{t,t'}^{c}  \right] \label{eq_pfttG_upper_noCSI_scalinglaw1_Kaknown} \\
    & \leq \exp\left\{ - (t+t')  \left(  \frac{L_n m^{*} }{8}  -   \ln \left( \max\{K_{a,n},1\} M_n \right) \right) \right\} + \exp\left\{ -c_{\delta} n  \right\} + \exp\left\{ -c_{\epsilon} n \right\}  \label{eq_pfttG_upper_noCSI_scalinglaw3_Kaknown} \\
    & \leq b^{- (t+t')}
    + 2\exp\left\{ -c' n  \right\}  , \label{eq_pfttG_upper_noCSI_scalinglaw4_Kaknown}
  \end{align}
  where \eqref{eq_pfttG_upper_noCSI_scalinglaw3_Kaknown} follows from \eqref{eq_Pft_upper_noCSI_scalinglaw1_Kaknown} and \eqref{eq_Dt_bound_noCSI_scalinglaw_Kaknown}, and \eqref{eq_pfttG_upper_noCSI_scalinglaw4_Kaknown} holds for some positive constant $c'\leq \min\{c_{\delta} , c_{\epsilon}\}$ in the case of
  \begin{equation}\label{eq_L_cond_noCSI_scalinglaw_Kaknown}
    1 < b  \leq  \exp\left\{ \frac{L_n m^{*} }{8}  -   \ln \left( K_{a,n} M_n \right) \right\} .
  \end{equation}

  Then, we can bound the misdetection probability as
  \begin{align}
    P_{\rm MD}
    & \leq \sum_{t = 1 }^{K_{a,n}} \left(  b^{-t}   +  2\exp \left\{ -c' n \right\}  \right)  +  p_0 \label{eq_PUPE_MD_upper_noCSI_noKa_scalinglaw_ultimate2_Kaknown} \\
    & \leq \frac{1}{b-1}  + \exp \left\{ \ln(2K_{a,n}) - c' n \right\}   +  p_0  , \label{eq_PUPE_MD_upper_noCSI_noKa_scalinglaw_ultimate4_Kaknown}
  \end{align}
  where \eqref{eq_PUPE_MD_upper_noCSI_noKa_scalinglaw_ultimate2_Kaknown} follows by substituting \eqref{eq_pfttG_upper_noCSI_scalinglaw4_Kaknown} into \eqref{eq_PUPE_MD_upper_noCSI_noKa_scalinglaw2_Kaknown} and applying $b^{-(t+t')}   \leq  b^{-t}$ and $|\mathcal{W}| \leq K_{a,n}$, and \eqref{eq_PUPE_MD_upper_noCSI_noKa_scalinglaw_ultimate4_Kaknown} follows from the geometric sum formula $\sum_{t=1}^{K_{a,n}}  b^{-t}  =  \frac{b^{-1}\left(1-b^{-K_{a,n}}\right)}{1-b^{-1}}  \leq  \frac{1}{b-1}$. The false-alarm probability can be bounded as follows
  \begin{align}
    P_{\mathrm{FA}}
    & \leq \sum_{t = 1 \left[ K_{a,n}-|\mathcal{W}| = 0 \right]}^{|\mathcal{W}|} \left(  b^{-\left(t+t'\right)}   +  2\exp \left\{ -c' n \right\}  \right) +  p_0 \label{eq_PUPE_FA_upper_noCSI_noKa_scalinglaw_ultimate4_Kaknown0}\\
    & \leq \frac{1}{b-1}  + \exp \left\{ \ln(2K_{a,n}) - c' n \right\}   +  p_0  , \label{eq_PUPE_FA_upper_noCSI_noKa_scalinglaw_ultimate4_Kaknown}
  \end{align}
  where \eqref{eq_PUPE_FA_upper_noCSI_noKa_scalinglaw_ultimate4_Kaknown0} follows by substituting \eqref{eq_pfttG_upper_noCSI_scalinglaw4_Kaknown} into \eqref{eq_PUPE_FA_upper_noCSI_noKa_scalinglaw4_Kaknown}, and \eqref{eq_PUPE_FA_upper_noCSI_noKa_scalinglaw_ultimate4_Kaknown} follows by applying the inequality $\sum_{t = 1 \left[ K_{a,n}-|\mathcal{W}| = 0 \right]}^{|\mathcal{W}|}  b^{-\left(t+t'\right)} \leq \sum_{t = 1 \left[ K_{a,n}-|\mathcal{W}| = 0 \right]}^{|\mathcal{W}|} b^{-\left(t+K_{a,n}-|\mathcal{W}|\right)} \leq \sum_{t=1}^{K_{a,n}} b^{-t} \leq \frac{1}{b-1}$.

  As aforementioned, we have $p_0 \to 0$ under the assumption that $M_n \geq K_{a,n}^{r_1}$ for a constant $r_1>1$ and $M_n,K_{a,n}\to\infty$. Together with \eqref{eq_PUPE_MD_upper_noCSI_noKa_scalinglaw_ultimate4_Kaknown} and \eqref{eq_PUPE_FA_upper_noCSI_noKa_scalinglaw_ultimate4_Kaknown}, it concludes that in the case of $b = \omega\left(1\right)$, we have $P_{\mathrm{MD}} \to 0$ and $P_{\mathrm{FA}} \to 0$ as $n\to\infty$. When $M_n = \Theta  \left( n^{c} \right)$ for any constant $c > 2$, the condition in~\eqref{eq_noCSI_scalinglaw_RIP2_condition_simple_Kaknown} is satisfied. In this case, following from \eqref{eq_noCSI_scalinglaw_RIP1_condition_simple_Kaknown}, the maximum number of reliably supported active users satisfies $K_{a,n} = \mathcal{O}  \left( \frac{n^2}{ \ln^2 n }  \right)$. Then, combining \eqref{eq_L_cond_noCSI_scalinglaw_Kaknown} and the condition $b = \omega\left(1\right)$, it concludes that $L_n = \Theta  \left( \max  \left\{ \frac{n^{2}}{\ln n} , \frac{ \ln n }{n^2 P_n^2} \right\} \right)$ receive antennas are needed to make the error probability as small as desired. This condition can be rewritten as $n^2P_n^2L_n = \Theta\left( \ln n \right)$ in the case of $\frac{n^2P_n}{\ln n} = \mathcal{O}(1)$. Likewise, we can obtain that when both $P_{\mathrm{MD}}$ and $P_{\mathrm{FA}}$ are required to be less than a constant, it should also be satisfied that $L_n = \Theta  \left( \max  \left\{ \frac{n^{2}}{\ln n} , \frac{ \ln n }{n^2 P_n^2} \right\} \right)$.

\subsection{ Converse bound } \label{Proof_scalinglaw_noCSI_conv_Kaknown}

  According to Theorem 2 in \cite{letter}, when the number $K_{a,n}$ of active users is fixed and known in advance and the receiver is required to output a list of $K_{a,n}$ messages, the single-user type converse bound on the minimum required energy-per-bit is shown in Lemma~\ref{Theorem_converse_single_Kaknown}.
  \begin{Lemma}[Theorem 2 in \cite{letter}] \label{Theorem_converse_single_Kaknown}
    Assume that there are $K_{a,n}$ active users, which is fixed and known beforehand. The receiver is required to output a list of $K_{a,n}$ messages. The minimum required energy-per-bit for URA in MIMO quasi-static Rayleigh fading channels is lower-bounded as
    \begin{equation} \label{eq:P_tot_conv_singleUE_EbN0_Kaknown}
      E^{*}_{b}(n,M_n,\epsilon_{\rm MD},\epsilon_{\rm FA}) \geq \inf \frac{nP_n}{\log_2 M_n}.
    \end{equation}
    Here, the $\inf$ is taken over all $P_n > 0$ satisfying that
      \begin{equation} \label{eq:P_tot_conv_singleUE1_Kaknown}
        \log_2 M_n - \log_2 {K_{a,n}} \leq
        - \log_2 { \mathbb{P}\left[ \chi^2(2L_n) \geq (1+(n+1)P_n)r
        \right] } ,
      \end{equation}
    where $r_n$ is the solution of
      \begin{equation}\label{eqR:beta_conv_AWGN_singleUE_e_constraint1_Kaknown}
        \mathbb{P} \left[ \chi^2(2L_n) \leq r_n \right] = \epsilon_{\rm{MD}} .
      \end{equation}
  \end{Lemma}

  In the case of $K_{a,n} = \mathcal{O}\left( \frac{n^2}{\ln^2 n} \right)$ and $M_n = \Theta \left( n^{c} \right)$ with constant $c > 2$, the LHS of~\eqref{eq:P_tot_conv_singleUE1_Kaknown} is on the order of $\Theta(\ln n)$. Next, we proceed to figure out the minimum required number of receive antennas and transmitting power for satisfying~\eqref{eq:P_tot_conv_singleUE1_Kaknown}. Following from \cite[Theorem~A]{chi_bound}, for every $L_n\geq 1$ and $\epsilon_{\rm{MD}} \in (0,1)$, $r_n$ satisfying \eqref{eqR:beta_conv_AWGN_singleUE_e_constraint1_Kaknown} can be upper-bounded as
   \begin{equation}
     r_n \leq 2L_n + 2\ln\left( \frac{1}{1-\epsilon_{\rm{MD}}} \right) + 2 \sqrt{2L_n \ln\left( \frac{1}{1-\epsilon_{\rm{MD}}} \right)} .
   \end{equation}
  When the target misdetection error probability is required to satisfy $\epsilon_{\rm{MD}} \leq \epsilon$ with constant $\epsilon\in(0,1)$, we have $\ln\left( \frac{1}{1-\epsilon_{\rm{MD}}} \right) \leq C_{\epsilon} = \ln\left( \frac{1}{1-\epsilon} \right)$. The term $\ln\left( \frac{1}{1-\epsilon_{\rm{MD}}} \right)$ is also bounded by the constant $C_{\epsilon}$ under the condition of $\epsilon_{\rm{MD}} \to 0$. Thus, the upper bound of $r_n$ is given by
   \begin{equation} \label{eq:beta_conv_AWGN_singleUE_rmax_Kaknown}
     r_n \leq r_{{\rm max},n} = 2L_n + 2 C_{\epsilon} + 2 \sqrt{2C_{\epsilon}L_n} .
   \end{equation}

  The probability in~\eqref{eq:P_tot_conv_singleUE1_Kaknown} can be lower-bounded as follows
  \begin{align}
     & \mathbb{P} \left[ \chi^2(2L_n) \geq (1+(n+1)P_n) r_n \right] \notag\\
     & \geq \mathbb{P} \left[ \chi^2(2L_n) \geq (1+(n+1)P_n) r_{{\rm max},n} \right] \\
     & \geq \frac{1-e^{-2}}{2} \frac{1}{\sqrt{2L_n}} \frac{(1+(n+1)P_n) r_{{\rm max},n}}{ (1+(n+1)P_n) r_{{\rm max},n} - 2L_n + 2\sqrt{2L_n} } \notag\\
     & \;\;\;\; \cdot \exp\left\{ -\frac{1}{2} \left( (1+(n+1)P_n) r_{{\rm max},n} - 2L_n    - (2L_n-2)\ln\frac{(1+(n+1)P_n) r_{{\rm max},n}}{2L_n}  \right) \right\},  \label{eq:P_chi_bound}
   \end{align}
  where \eqref{eq:P_chi_bound} follows from Lemma~\ref{lemma:P_chi_bound} shown below.
  \begin{Lemma}[Proposition 3.1 in \cite{chi_bound}] \label{lemma:P_chi_bound}
    For all $k\geq 2$ and all $u \geq k-1$ we have
    \begin{equation}
      \mathbb{P} \left[ \chi^{2}\left(k\right) \geq u \right] \geq
      \frac{1-e^{-2}}{2} \frac{1}{\sqrt{k}} \frac{u}{ u - k + 2\sqrt{k} }   \exp\left\{ -\frac{1}{2} \left( u - k - (k-2)\ln\frac{u}{k}  \right) \right\} .
    \end{equation}
  \end{Lemma}

  The term $\frac{(1+(n+1)P_n) r_{{\rm max},n}}{ (1+(n+1)P_n) r_{{\rm max},n} - 2L_n + 2\sqrt{2L_n} } $ on the RHS of~\eqref{eq:P_chi_bound} can be further bounded as
   \begin{align}
     & \frac{(1+(n+1)P_n) r_{{\rm max},n}}{ (1+(n+1)P_n) r_{{\rm max},n} - 2L_n + 2\sqrt{2L_n} } \notag\\
     & \geq \frac{ r_{{\rm max},n}}{ (1+(n+1)P_n) r_{{\rm max},n} - 2L_n + 2\sqrt{2L_n} } \\
     & = \frac{ L_n + \sqrt{2C_{\epsilon}L_n} + C_{\epsilon}  }{ (n+1)P_n \left( L_n+ C_{\epsilon}+ \sqrt{2C_{\epsilon}L_n}\right) + \sqrt{2C_{\epsilon}L_n } + \sqrt{2L_n}  + C_{\epsilon} } . \label{eq:single_conv_term1} 
   \end{align}

  We can upper-bound the term on the exponent of the RHS of~\eqref{eq:P_chi_bound} as follows
   \begin{align}
     & \frac{1}{2}  \left(  (1 + (n + 1)P_n) r_{{\rm max},n}  -  2L_n - (2L_n - 2)\ln \frac{(1 + (n + 1)P_n) r_{{\rm max},n} }{2L_n}   \right) \notag\\
     & = \frac{1}{2} \left( (n+1)P_n r_{{\rm max},n} + 2C_{\epsilon} + 2 \sqrt{2C_{\epsilon}L_n }
     - (2L_n-2)\ln\left( 1 + T_{n,P_n,L_n} \right)  \right)  \\
     & = L_n\left( T_{n,P_n,L_n}
     - \ln \left( 1 +  T_{n,P_n,L_n} \right) \right)
     + \ln \left( 1 +  T_{n,P_n,L_n} \right)  \\
     & \leq \frac{L_nT_{n,P_n,L_n}^2}{2}
     + T_{n,P_n,L_n} ,  \label{eq:x_ln}
   \end{align}
   where \eqref{eq:x_ln} follows from the inequality $x-\ln(1+x) \leq \frac{x^2}{2}$ for $x\geq 0$ and the inequality $\ln(1+x) \leq x$ for $x> -1$, and $T_{n,P_n,L_n}$ is given by
   \begin{align}
     T_{n,P_n,L_n}
     & = (n+1)P_n \frac{r_{{\rm max},n}}{2L_n} + \frac{C_{\epsilon}}{L_n} + \sqrt{\frac{2C_{\epsilon}}{L_n}} \\
     & = (n+1)P_n\left( 1 + \sqrt{\frac{2C_{\epsilon}}{L_n}} + \frac{C_{\epsilon}}{L_n} \right) + \frac{C_{\epsilon}}{L_n} + \sqrt{\frac{2C_{\epsilon}}{L_n}}. \label{eq:single_conv_term_T}
   \end{align}

   In the case of $L_n\to \infty$ and $n^2P_n^2L_n = \mathcal{O}(1)$, following from~\eqref{eq:single_conv_term1}, we have
   \begin{equation}
     \frac{(1+(n+1)P_n) r_{{\rm max},n}}{ (1+(n+1)P_n) r_{{\rm max},n} - 2L_n + 2\sqrt{2L_n} }
     \geq \frac{ \sqrt{L_n} + o(\sqrt{L_n})  }{ nP_n\sqrt{L_n} + \mathcal{O}(1) } = \Omega\left( \sqrt{L_n} \right).
   \end{equation}
   Likewise, following from~\eqref{eq:x_ln} and \eqref{eq:single_conv_term_T}, we have
   \begin{align}
     & \frac{1}{2}  \left(  (1 + (n + 1)P_n) r_{{\rm max},n}  -  2L_n - (2L_n - 2)\ln \frac{(1 + (n + 1)P_n) r_{{\rm max},n} }{2L_n}   \right) \notag\\
     & \leq  \frac{ \left( (n+1)P_n \left( \sqrt{L_n} + \sqrt{2C_{\epsilon}} + \frac{C_{\epsilon}}{\sqrt{L_n}} \right) +  \sqrt{2C_{\epsilon}} + \frac{C_{\epsilon}}{\sqrt{L_n}} \right)^2}{2}
     +  o(1)\\
     & = \mathcal{O}(1) . 
   \end{align}
   Thus, we can obtain that $\mathbb{P} \left[ \chi^2(2L_n) \geq (1+(n+1)P_n) r_n \right] = \Omega(1)$ in the case of $L_n\to \infty$ and $n^2P_n^2L_n = \mathcal{O}(1)$. Considering that the LHS of~\eqref{eq:P_tot_conv_singleUE1_Kaknown} is on the order of $\Theta(\ln n)$, the inequality in~\eqref{eq:P_tot_conv_singleUE1_Kaknown} cannot be satisfied in this case.

   In the following, we consider the case with larger $n^2P_n^2L_n$. We assume $nP_n=o(1)$ and $n^2P_n^2L_n \to \infty$. Following from~\eqref{eq:single_conv_term1}, we have
   \begin{align}
     & \frac{(1+(n+1)P_n) r_{{\rm max},n}}{ (1+(n+1)P_n) r_{{\rm max},n} - 2L_n + 2\sqrt{2L_n} } \notag\\
     & \geq \frac{ L_n + C_{\epsilon} + \sqrt{2C_{\epsilon}L_n } }{ (n+1)P_n \left( L_n+ C_{\epsilon}+ \sqrt{2C_{\epsilon}L_n}\right) + C_{\epsilon} + \sqrt{2C_{\epsilon}L_n } + \sqrt{2L_n} } \\
     & = \frac{ \sqrt{L_n} + o(\sqrt{L_n})  }{ nP_n\sqrt{L_n} + o(nP_n\sqrt{L_n}) } \\
     & = \frac{1}{nP_n} + o\left( \frac{1}{nP_n} \right). \label{eq:single_conv_term1_asym}
   \end{align}
   Likewise, following from~\eqref{eq:x_ln} and \eqref{eq:single_conv_term_T}, we have
   \begin{align}
     & \frac{1}{2}  \left(  (1 + (n + 1)P_n) r_{{\rm max},n}  -  2L_n - (2L_n - 2)\ln \frac{(1 + (n + 1)P_n) r_{{\rm max},n} }{2L_n}   \right) \notag\\
     & \leq  \frac{ \left( (n+1)P_n \left( \sqrt{L_n} + \sqrt{2C_{\epsilon}} + \frac{C_{\epsilon}}{\sqrt{L_n}} \right) +  \sqrt{2C_{\epsilon}} + \frac{C_{\epsilon}}{\sqrt{L_n}} \right)^2}{2}
     +  o(1)\\
     & = \frac{n^2P_n^2L_n}{2} + o\left( n^2P_n^2L_n \right) . \label{eq:single_conv_term2_asym}
   \end{align}
   Substituting \eqref{eq:single_conv_term1_asym} and \eqref{eq:single_conv_term2_asym} into \eqref{eq:P_chi_bound}, we have
   \begin{equation}
     \mathbb{P} \left[ \chi^2(2L_n) \geq (1+(n+1)P_n) r_n \right] \geq \sqrt{\frac{1}{2L_n}} \left( \frac{1}{nP_n} + o\left( \frac{1}{nP_n} \right) \right) \exp\left\{ - \frac{n^2P_n^2L_n}{2} + o\left( n^2P_n^2L_n \right) \right\}.
   \end{equation}
   It indicates that
   \begin{equation}
     - \log_2 \mathbb{P} \left[ \chi^2(2L_n) \geq (1+(n+1)P_n) r_n \right] \leq  \frac{n^2P_n^2L_n \log_2 e}{2} + o \left( n^2P_n^2L_n \right) + \mathcal{O} \left( \log_2  \left( nP_n\sqrt{L_n} \right) \right).
   \end{equation}
  Thus, the condition in~\eqref{eq:P_tot_conv_singleUE1_Kaknown} is satisfied in the case of $n^2P_n^2L_n = \Theta\left( \ln n \right)$ and $nP_n=o(1)$.

\section{ Proof of Theorem \ref{Theorem_scalinglaw_noCSI}  } \label{Proof_scalinglaw_noCSI}
  In this appendix, we consider the case with a random and unknown number of active users. We first establish in Appendix \ref{Proof_scalinglaw_noCSI_achi} the achievability result of Theorem \ref{Theorem_scalinglaw_noCSI} and then Appendix \ref{Proof_scalinglaw_noCSI_conv} shows the converse part.

\subsection{ Achievability  bound } \label{Proof_scalinglaw_noCSI_achi}

  We adopt a two-stage approach to derive the achievability bound. First, active users transmit a common sequence of length $n_0$ and power $P_n$ and the BS outputs the estimate of the number of active users. Second, users share a common codebook $\mathbf{X} \in \mathbb{C}^{n_1\times M_n}$ with each codeword drawn uniformly i.i.d. from the sphere of radius $\sqrt{n_1P_n}$. Let $n_0 = C_0 n$ and $n_1 = (1-C_0)n$ for some constant $C_0 \in(0,1)$. We change the measure to the new one under which: 1)~the number of users who select the same message is no more than a constant $D$; and 2)~there are at least $K_{l,n}$ and at most $K_{u,n}$ active users, where $K_{l,n} = 0$ and $K_{u,n} = \min\left\{ \mathbb{E}\left[{\rm{K}}_{a,n}\right] + C_1 \frac{n^2}{\ln^2 n}   ,  K_n \right\}$ for some constant $C_1>0$. The total variation distance between the true measure and the new one can be bounded by $p_0$ as follows: 
  \begin{align}
    p_0
    & \leq \mathbb{E}_{{\rm{K}}_{a,n}}   \left[
    \sum_{d=D}^{{\rm{K}}_{a,n}} \frac{ \binom{{\rm{K}}_{a,n}}{d} M_n (M_n-1)^{{\rm{K}}_{a,n}-d} }{ M_n^{{\rm{K}}_{a,n}} } \right]
    + \mathbb{P} \left[ {\rm{K}}_{a,n} \notin [K_{l,n},K_{u,n}] \right]  \\
    & \leq \mathbb{E}_{{\rm{K}}_{a,n}}   \left[  \sum_{d=D}^{{\rm{K}}_{a,n}} \left( \frac{e{\rm{K}}_{a,n}}{d} \right)^{d} \frac{1}{M_n^{d-1}}  \right]
    + \mathbb{P} \left[ {\rm{K}}_{a,n} > K_{u,n} \right] \label{eq:p0_scalinglaw1} \\
    & \leq \left( \frac{e}{D} \right)^{D} \frac{K_n^{D+1}}{M_n^{D-1}}
    +  \mathbb{P} \left[ {\rm{K}}_{a,n} > K_{u,n} \right] , \label{eq:p0_scalinglaw}
  \end{align}
  where \eqref{eq:p0_scalinglaw1} holds due to $\binom {a } {b} \leq \left( \frac{ea}{b} \right)^{b}$ for $a\geq b>0$, and \eqref{eq:p0_scalinglaw} follows because $\left( \frac{e{\rm{K}}_{a,n}}{d} \right)^{d} \frac{1}{M_n^{d-1}}$ decreases with $d$ in the case of $d\geq1$ and $M_n\geq K_n$. Under the assumption of $M_n \geq K_n^{r_1}$ for a constant $r_1>1$, there exists a constant $D$ making sure $\left( \frac{e}{D} \right)^{D} \frac{K_n^{D+1}}{M_n^{D-1}} \to 0$ as $M_n\to \infty$. In the case of $K_n \leq \mathbb{E}\left[ {\rm{K}}_{a,n}\right] + C_1 \frac{n^2}{\ln^2 n} $, we have $\mathbb{P} \left[ {\rm{K}}_{a,n} > K_{u,n} \right] = 0$; when $K_n = \omega \left( \mathbb{E}\left[ {\rm{K}}_{a,n}\right] + C_1 \frac{n^2}{\ln^2 n} \right)$, we have 
  \begin{align}
    \mathbb{P} \left[ {\rm{K}}_{a,n} > K_{u,n} \right]
    & \leq \mathbb{P} \left[ \left| {\rm{K}}_{a,n} - \mathbb{E}\left[ {\rm{K}}_{a,n}\right] \right| > C_1 \frac{n^2}{\ln^2 n} \right]   \\
    & \leq \frac{{\rm var}({\rm{K}}_{a,n})}{ C_1^2 \left( \frac{n^2}{\ln^2 n} \right)^{2} } , \label{eq:p0_Cheby_scalinglaw}
  \end{align}
  where \eqref{eq:p0_Cheby_scalinglaw} follows by applying Chebyshev's inequality. The RHS of \eqref{eq:p0_Cheby_scalinglaw} tends to $0$ under the condition of $\frac{{\rm var}({\rm{K}}_{a,n})}{ \left( \frac{n^2}{\ln^2 n} \right)^{2} } = o(1)$.

  Then, we upper-bound the per-user probability of misdetection as
  \begin{align}
    P_{\rm MD}
    &\leq   p_0 + \sum_{K_{a,n}=\max\{K_{l,n},1\}}^{K_{u,n}}
            \frac{P_{{\rm{K}}_{a,n}} (K_{a,n})}{K_{a,n}}  \sum_{k\in {\mathcal{K}_{a,n}}}   \mathbb{P}  \left[ W_{k}  \notin  \hat{\mathcal{W}}  \right]_{\text{new}}    \label{eq_PUPE_MD_upper_noCSI_noKa_scalinglaw0} \\
    & \leq  p_0 + \underbrace{\sum_{K_{a,n}=K_{l,n}}^{K_{u,n}}   P_{{\rm{K}}_{a,n}} (K_{a,n}) \mathbb{P}  \left[ \left.  {\rm K}'_{a,n} \geq K'_{a,n,u} , {\rm K}'_{a,n}  \in [0:K_n]  \right|  {\rm K}_{a,n} = K_{a,n} \right]_{\text{new}} }_{p_1}    \notag \\
    & \;\;  + \!\!\!\underbrace{ \sum_{K_{a,n} =\max\{ K_{l,n},1 \}}^{K_{u,n}} \!\!\!
            \frac{P_{{\rm{K}}_{a,n}} \!(K_{a,n})}{K_{a,n}}  \!  \sum_{k\in {\mathcal{K}_{a,n}}}  \! \mathbb{P} \! \left[ \left.  W_{k}  \notin  \hat{\mathcal{W}}  \right|
              {\rm K}'_{a,n}  \leq  K'_{a,n,u}  , {\rm K}'_{a,n}  \in  [0: K_n] \right]_{\text{new}} }_{p_2}  , \label{eq_PUPE_MD_upper_noCSI_noKa_scalinglaw2}
  \end{align}
  where the random variable ${\rm K}_{a,n}$ denotes the number of active users; the random variable ${\rm K}'_{a,n}$ denotes the estimated number of active users; \eqref{eq_PUPE_MD_upper_noCSI_noKa_scalinglaw0} follows from~\eqref{eq:MD} and \eqref{eq:p0_scalinglaw}; \eqref{eq_PUPE_MD_upper_noCSI_noKa_scalinglaw2} follows because $\mathbb{P}\left[A\right] \leq \mathbb{P}\left[A | B\right] +\mathbb{P}\left[ B^c \right]$ for events $A$ and $B$~\cite{1961}; and $K'_{a,n,u} = K_{a,n} + \Theta \left(\max\left\{ K_{a,n} , n_0^{\frac{1}{2} } \right\}\right) $. The term $p_1$ in~\eqref{eq_PUPE_MD_upper_noCSI_noKa_scalinglaw2} can be bounded as
  \begin{align}
    p_1 
    & \leq \max_{K_{l,n}\leq K_{a,n}\leq K_{u,n}} 
    \mathbb{P} \left[ \left.  {\rm K}'_{a,n} \geq K'_{a,n,u} , {\rm K}'_{a,n}  \in [0:K_n]  \right|  {\rm K}_{a,n} = K_{a,n} \right]_{\text{new}} \\
    & = \max_{K_{l,n}\leq K_{a,n}\leq K_{u,n}} \mathbb{P} \left[ \left. {\rm K}'_{a,n} - K_{a,n} \geq \Theta \left(\max\left\{ K_{a,n} , n_0^{\frac{1}{2} } \right\}\right)  , {\rm K}'_{a,n}  \in [0:K_n] \right|  {\rm K}_{a,n} = K_{a,n}  \right]_{\text{new}} .  \label{eq:p1_scalinglaw}
  \end{align}
  
  Considering that \eqref{eq:proof_noCSI_noKa_Ka_Kahat_neq_scalinglaw1_2} and \eqref{eq:proof_noCSI_noKa_Ka_Kahat_neq_scalinglaw2_1} hold no matter whether active users transmit a common sequence or different sequences, in the case of $\ln K_n = \mathcal{O}\left( \ln n_0 \right)$, we can obtain from \eqref{eq:proof_noCSI_noKa_Ka_Kahat_neq_scalinglaw1_2} and \eqref{eq:proof_noCSI_noKa_Ka_Kahat_neq_scalinglaw2_1} that for some positive constant $\alpha$,
  \begin{equation}
    \mathbb{P} \left[ \left.   \left| {\rm K}'_{a,n}  - K_{a,n} \right|   \geq   \sqrt{  \frac{\alpha\ln n_0}{L_n}  \left(  K_{a,n}^2  +  \frac{ 1 }{ n_0P_n^2} \right) } , {\rm K}'_{a,n}  \in [0:K_n]  \right|  {\rm K}_{a,n} = K_{a,n}   \right]  \to  0, \label{eq:p1_scalinglaw2}
  \end{equation}
  as $L_n\to \infty$ while satisfying $\frac{\ln n_0}{L_n} = o(1)$. Combining \eqref{eq:p1_scalinglaw} and \eqref{eq:p1_scalinglaw2}, we can obtain that $p_1 \to 0$ as $n_0 \to \infty$ in the case of $n_0^2P_n^2L_n = \Theta\left( \ln n_0 \right)$ and $\frac{\ln n_0}{L_n} = o(1)$.

  The term $p_2$ in~\eqref{eq_PUPE_MD_upper_noCSI_noKa_scalinglaw2} can be bounded as
  \begin{align}
    p_2
    & \leq    \sum_{K_{a,n}=\max\{K_{l,n},1\}}^{K_{u,n}}     P_{{\rm{K}}_{a,n}} (K_{a,n})
    \sum_{t = 0 }^{|\mathcal{W}|}
    \sum_{t' = 0}^{K'_{a,n,u}  - |\mathcal{W}| + t }
       \frac{t}{K_{a,n}}
    \mathbb{P} \left[  \mathcal{F}_{t,t'} \right]_{\text{new}} \label{eq_PUPE_MD_upper_noCSI_noKa_scalinglaw3}\\
    & \leq    \sum_{K_{a,n}=\max\{K_{l,n},1\}}^{K_{u,n}}    P_{{\rm{K}}_{a,n}}(K_{a,n})
    \sum_{t = 1 }^{|\mathcal{W}|}
    \sum_{t' = 0}^{K'_{a,n,u}  - |\mathcal{W}| + t }
      \mathbb{P} \left[  \mathcal{F}_{t,t'} \right]_{\text{new}} , \label{eq_PUPE_MD_upper_noCSI_noKa_scalinglaw4}
  \end{align}
  where \eqref{eq_PUPE_MD_upper_noCSI_noKa_scalinglaw3} holds because $M_n \gg K_{u,n}$ and the size of the decoded list satisfies $0\leq |\hat{\mathcal{W}}| \leq K'_{a,n,u}$, and $\mathcal{F}_{t,t'}$ denotes the event that there are $t$ misdetected messages and $t'$ falsely alarmed messages. Likewise, $P_{\mathrm{FA}}$ can be upper-bounded as
  \begin{equation}
    P_{\mathrm{FA}}
    \leq    \sum_{K_{a,n}=K_{l,n}}^{K_{u,n}}     P_{{\rm{K}}_{a,n}} (K_{a,n})
    \sum_{t = 0}^{|\mathcal{W}|}
    \sum_{t' = 1 }^{K'_{a,n,u}  - |\mathcal{W}| + t}
       \mathbb{P}  \left[  \mathcal{F}_{t,t'} \right]_{\text{new}}
     + p_1 +  p_0. \label{eq_PUPE_FA_upper_noCSI_noKa_scalinglaw4}
  \end{equation}
  Next, we omit the subscript ``new'' and proceed to upper-bound $\mathbb{P} \left[ \mathcal{F}_{t,t'} \right]$. Along similar lines as in Appendix~\ref{Proof_scalinglaw_noCSI_achi_Kaknown}, we have
  \begin{equation}
    \mathbb{P} \left[ \mathcal{F}_{t,t'} \right]
    \leq b^{- (t+t')} + 2\exp\left\{ -c' n_1 \right\}  , \label{eq_pfttG_upper_noCSI_scalinglaw4}
  \end{equation}
  under the conditions that
  \begin{equation}\label{eq_noCSI_scalinglaw_RIP2_condition_simple}
     2n_1(n_1-1) < M_n < \frac{1}{4} \exp\left\{ c_1\sqrt{n_1(n_1-1)} \right\} ,
  \end{equation}
  \begin{equation}
    K_{a,n} + K'_{a,n,u}  \leq C_{\delta} \frac{n_1(n_1-1)}{\ln^{2}\left( \frac{e M_n}{n_1(n_1-1)}\right)}, \label{eq_noCSI_scalinglaw_RIP1_condition_simple}
  \end{equation}
  \begin{equation}\label{eq_L_cond_noCSI_scalinglaw}
    1 < b  \leq  \exp\left\{ \frac{L_n m^{*} }{8}  -   \ln \left( 
    K_{u,n} M_n \right) \right\} ,
  \end{equation}
  \begin{equation}\label{eq_m_cond_noCSI_scalinglaw}
     m^{*}    \geq  \frac{ c_2(1-\delta)n_1^2 }{   \left(  c_3 D
    \left( c_4\ln \left(eM_n\right) + \frac{c_{\epsilon} n_1 + \ln2}{\max  \{ c_5 K_{u,n} , n_1  \} }  \right)
    \max  \{ c_5 K_{u,n} , n_1  \} + \frac{1}{P_n} \right)^{ 2}} ,
  \end{equation}
  where $c',c_1,c_2,c_3,c_4,c_5,c_{\epsilon},C_{\delta}$, and $D$ are positive constants, and the constant $\delta\in(0,1)$.

  Then, we can bound the misdetection probability as
  \begin{align}
    P_{\rm MD}
    & \leq \sum_{K_{a,n}=\max\{K_{l,n},1\}}^{K_{u,n}}   \left( P_{{\rm{K}}_{a,n}} (K_{a,n})
    \left( K'_{a,n,u}  +  1 \right)
    \sum_{t = 1 }^{K_{a,n}}
     \left(  b^{-t}   +  2\exp \left\{ -c' n_1 \right\}  \right)  \right) + p_1  +  p_0 \label{eq_PUPE_MD_upper_noCSI_noKa_scalinglaw_ultimate2} \\
    & \leq \frac{1}{b-1}\Theta \left(\max \left\{  K_{u,n} , n_0^{\frac{1}{2} } \right\}\right)
    + \exp \left\{ \Theta \left( \max  \left\{ \ln  K_{u,n} , \ln  n_0 \right\} \right)  -  c' n_1 \right\}  +  p_1   +   p_0  , \label{eq_PUPE_MD_upper_noCSI_noKa_scalinglaw_ultimate4}
  \end{align}
  where \eqref{eq_PUPE_MD_upper_noCSI_noKa_scalinglaw_ultimate2} follows by substituting \eqref{eq_pfttG_upper_noCSI_scalinglaw4} into \eqref{eq_PUPE_MD_upper_noCSI_noKa_scalinglaw4} and applying $b^{-(t+t')}   \leq  b^{-t}$; \eqref{eq_PUPE_MD_upper_noCSI_noKa_scalinglaw_ultimate4} follows from the geometric sum formula $\sum_{t=1}^{K_{a,n}} b^{-t}  =  \frac{b^{-1}\left(1-b^{-K_{a,n}}\right)}{1-b^{-1}}  \leq  \frac{1}{b-1}$ and the fact that the term $\left( K'_{a,n,u}  +  1 \right)   \sum_{t = 1 }^{K_{a,n}}  \left(  b^{-t}    +  2\exp \left\{ -c' n_1 \right\}  \right)$ increases with $K_{a,n}$. Likewise, the false-alarm probability also can be bounded by~\eqref{eq_PUPE_MD_upper_noCSI_noKa_scalinglaw_ultimate4}.

  As aforementioned, we have $p_0\to 0$ as $M_n\to \infty$ under the conditions that $M_n \geq K_n^{r_1}$ for some constant $r_1>1$ and $\frac{{\rm var}({\rm{K}}_{a,n})}{ \left( \frac{n^2}{\ln^2 n} \right)^{2} } = o(1)$ in the case of $K = \omega \left( \mathbb{E}\left[ {\rm{K}}_{a,n}\right] + C_1 \frac{n^2}{\ln^2 n} \right)$; and we have $p_1\to 0$ in the case of $n_0^2P_n^2L_n = \Theta\left( \ln n_0 \right)$, $\frac{\ln n_0}{L_n} = o(1)$, $\ln K_n = \mathcal{O}\left( \ln n_0 \right)$, and $n_0, L_n \to \infty$. Together with \eqref{eq_PUPE_MD_upper_noCSI_noKa_scalinglaw_ultimate4}, it concludes that in the case of $n_1 = \Omega\left( \max \left\{ \ln K_{u,n} , \ln n_0 \right\} \right)$ and $b = \Omega\left(\max\left\{ K_{u,n} , n_0^{\frac{1}{2} } \right\}\right)$, the misdetction error probability and false-alarm error probability is bounded by positive constants in $(0,1)$ or even vanish. When $\mathbb{E} \left[{\rm{K}}_{a,n}\right] = \mathcal{O}  \left( \frac{n^2}{ \ln^2 n }  \right)$ and $M_n = \Theta \left( n^{r_2} \right)$ for any constant $r_2 > 2$, the conditions \eqref{eq_noCSI_scalinglaw_RIP2_condition_simple} and \eqref{eq_noCSI_scalinglaw_RIP1_condition_simple} are satisfied. Combining $b = \Omega\left(\max\left\{ K_{u,n} , n_0^{\frac{1}{2} } \right\}\right)$ and the conditions in \eqref{eq_L_cond_noCSI_scalinglaw} and \eqref{eq_m_cond_noCSI_scalinglaw}, it is required that $L_n = \Theta  \left( \max  \left\{ \frac{n^{2}}{\ln n} , \frac{ \ln n }{n^2 P_n^2} \right\} \right)$, which can be rewritten as $n^2P_n^2L_n = \Theta\left( \ln n \right)$ in the case of $\frac{n^2P_n}{\ln n} = \mathcal{O}(1)$.

\subsection{ Converse bound } \label{Proof_scalinglaw_noCSI_conv}

  Assuming $S = \mathcal{K}$, the condition \eqref{eq:P_tot_conv_singleUE3} in Theorem~\ref{Theorem_converse_single} becomes
      \begin{equation} \label{eq:P_tot_conv_singleUE3_scalinglaw}
        \sum_{K_{a,n} \in \mathcal{K}}  P_{{\rm K}_{a,n}}(K_{a,n}) \; \epsilon_{K_a} \leq \epsilon_{\mathrm{MD}}.
      \end{equation}
  When the target misdetection probability $\epsilon_{\mathrm{MD}}$ is a constant in $(0,1)$ or tend to $0$, we can observe from~\eqref{eq:P_tot_conv_singleUE3_scalinglaw} that there exists at least an element in the set $\left\{ \epsilon_{K_a}: K_{a,n} \in \mathcal{K} \right\}$ being less than a constant in $(0,1)$. We denote this element as $\epsilon_{k}$ without loss of generality. Following from~\eqref{eq:P_tot_conv_singleUE2} and applying similar lines as in~\eqref{eq:beta_conv_AWGN_singleUE_rmax_Kaknown}, for $L_n\geq 1$, we have
   \begin{equation} \label{eq:beta_conv_AWGN_singleUE_rmax_k}
     r_k \leq r_{k,max} = 2L_n + 2\ln\left( \frac{1}{1-\epsilon_{k}} \right) + 2 \sqrt{2L_n \ln\left( \frac{1}{1-\epsilon_{k}} \right)} .
   \end{equation}

  Then, we enlarge the size of the decoded list to $K$ and loose the condition \eqref{eq:P_tot_conv_singleUE1} in Theorem~\ref{Theorem_converse_single} as
      \begin{equation} \label{eq:P_tot_conv_singleUE1_scalinglaw}
        \log_2  M_n - \log_2  K_n   \leq
        - \log_2  { \mathbb{P} \left[ \chi^2(2L_n) \geq (1+(n+1)P_n) \;\! r_k \right] } .
      \end{equation}
  Under the assumption of $K_n^{r_1} \leq M_n = \Theta \left( n^{r_2} \right)$ with constants $r_1>1$ and $r_2> 2$, the LHS of~\eqref{eq:P_tot_conv_singleUE1_scalinglaw} is on the order of $\Theta(\ln n)$. Along similar ideas as in Appendix~\ref{Proof_scalinglaw_noCSI_conv_Kaknown}, we can obtain that the condition in~\eqref{eq:P_tot_conv_singleUE1_scalinglaw} is satisfied in the case of $n^2P_n^2L_n = \Theta\left( \ln n \right)$ and $nP_n=o(1)$.

%
\section{Proof of Theorem~\ref{Theorem_converse_noCSI_Gaussian_cKa_joint_fixed}} \label{Appendix_proof_converse_noCSI_Gaussian_noKa_joint_fixed}

  Assume that there are $K_a$ active users, which is fixed and known in advance. We assume w.l.o.g. that the active user set is $\mathcal{K}_a = [K_a]$. The set of messages transmitted by active users is denoted as $\mathcal{W}_{\mathcal{K}_a}$, and the set of decoded messages is denoted as $\hat{\mathcal{W}}_{\mathcal{K}_a}$. Following from \cite[(113)]{noKa}, we have
  \begin{align}
    P_{e,J} & = \mathbb{P}  \left[  \mathcal{W}_{\mathcal{K}_a} \neq \hat{\mathcal{W}}_{\mathcal{K}_a}  \right]  \\
    & = \mathbb{P}  \left[ \left. \mathcal{W}_{\mathcal{K}_a} \neq \hat{\mathcal{W}}_{\mathcal{K}_a} \right|  |\mathcal{W}_{\mathcal{K}_a}| = K_a   \right]  \mathbb{P} \left[ |\mathcal{W}_{\mathcal{K}_a}| = K_a \right]  \notag\\
    & \;\;\;\;\; + \mathbb{P}  \left[ \left. \mathcal{W}_{\mathcal{K}_a} \neq \hat{\mathcal{W}}_{\mathcal{K}_a} \right|  |\mathcal{W}_{\mathcal{K}_a}| < K_a   \right]  \mathbb{P} \left[ |\mathcal{W}_{\mathcal{K}_a}| < K_a \right] \\
    & \geq \mathbb{P}  \left[ \left. \mathcal{W}_{\mathcal{K}_a} \neq \hat{\mathcal{W}}_{\mathcal{K}_a} \right|  |\mathcal{W}_{\mathcal{K}_a}| = K_a  \right]  \frac{M !}{M^{K_a} (M-K_a)! } . 
  \end{align}
  To derive the converse bound, we loosen the error requirement $ P_{e,J} = \mathbb{P}  \left[  \mathcal{W}_{\mathcal{K}_a} \neq \hat{\mathcal{W}}_{\mathcal{K}_a}  \right] \leq \epsilon_J$ in~\eqref{eq:joint_error} to $\bar{P}_{e,J} = \mathbb{P}  \left[ \left. \mathcal{W}_{\mathcal{K}_a} \neq \hat{\mathcal{W}}_{\mathcal{K}_a} \right|  |\mathcal{W}_{\mathcal{K}_a}| = K_a   \right]  \leq \frac{\epsilon_J M^{K_a} (M-K_a)! }{M !}$.

  Define $\mathcal{M}_{\mathcal{K}_a} = 1 \left[ \mathcal{W}_{\mathcal{K}_a} \neq \hat{\mathcal{W}}_{\mathcal{K}_a} \right]$. We have $\bar{P}_{e,J} = \mathbb{P} \left[  \mathcal{M}_{\mathcal{K}_a} = 1 \left| |\mathcal{W}_{\mathcal{K}_a}| = K_a \right. \right]$. The standard Fano inequality gives that 
  \begin{align}
    &H_2( \mathcal{W}_{\mathcal{K}_a} | \bar{\mathbf{X}} , |\mathcal{W}_{\mathcal{K}_a}| = K_a )
    - H_2\left( \mathcal{M}_{\mathcal{K}_a} | \bar{\mathbf{X}} , |\mathcal{W}_{\mathcal{K}_a}| = K_a \right)
    - H_2 ( \mathcal{W}_{\mathcal{K}_a} | \mathcal{M}_{\mathcal{K}_a} , \hat{\mathcal{W}}_{\mathcal{K}_a} , \bar{\mathbf{X}} , |\mathcal{W}_{\mathcal{K}_a}| = K_a ) \notag\\
    & \leq I_2  \left(  \mathcal{W}_{\mathcal{K}_a} ; \hat{\mathcal{W}}_{\mathcal{K}_a} \left| \bar{\mathbf{X}} , |\mathcal{W}_{\mathcal{K}_a}| = K_a \right. \right) , \label{eq:fano_noKa_joint_fixed}
  \end{align}  
  where the entropies are given by $H_2\left( \mathcal{W}_{\mathcal{K}_a} | \bar{\mathbf{X}} , |\mathcal{W}_{\mathcal{K}_a}| = K_a \right) = \log_2 \binom{M}{K_a}$, $ H_2\left( \mathcal{M}_{\mathcal{K}_a} | \bar{\mathbf{X}} ,|\mathcal{W}_{\mathcal{K}_a}| = K_a \right) = h_2(\bar{P}_{e,J})$, and $H_2 \left( \mathcal{W}_{\mathcal{K}_a} | \mathcal{M}_{\mathcal{K}_a} , \hat{\mathcal{W}}_{\mathcal{K}_a} , \bar{\mathbf{X}} , |\mathcal{W}_{\mathcal{K}_a}| =  K_a \right) \leq \bar{P}_{e,J} \log_2 \binom{M}{K_a}$. Thus, we have
  \begin{equation}
    (1 - \bar{P}_{e,J})\log_2 \binom{M}{K_a} - h_2(\bar{P}_{e,J}) 
    \leq I_2  \left(  \mathcal{W}_{\mathcal{K}_a} ; \hat{\mathcal{W}}_{\mathcal{K}_a} \left| \bar{\mathbf{X}} , |\mathcal{W}_{\mathcal{K}_a}| = K_a \right. \right) .  \label{eq:fano_noKa_joint1_fixed}
  \end{equation}  
  In the case of $\bar{P}_{e,J} \leq \frac{\epsilon_J M^{K_a} (M-K_a)! }{ M !  } \leq \frac{ \binom{M}{K_a}}{1+\binom{M}{K_a}}$, we have
  \begin{equation}
    \bar{P}_{e,J} \log_2 \binom{M}{K_a} + h_2(\bar{P}_{e,J})  
    \leq\frac{\epsilon_J M^{K_a} (M-K_a)! }{ M !  } \log_2 \binom{M}{K_a} + 
    h_2\left( \frac{\epsilon_J M^{K_a} (M-K_a)! }{ M !  } \right), \label{eq:fano_noCSI_Ka2_left2_joint_fixed}
  \end{equation} 
  
  The mutual information in~\eqref{eq:fano_noKa_joint1_fixed} can be upper-bounded by the RHS of~\ref{eq_conv_fano_nocsi} by allowing $|\mathcal{W}_{\mathcal{K}_a}| = K_a$. Thus, we can obtain \eqref{P_tot_conv_noCSI_joint_fixed} in Theorem~\ref{Theorem_converse_noCSI_Gaussian_cKa_joint_fixed}.


\end{document}